\documentclass[floatfix,letterpaper,longbibliography,nofootinbib,aps,pre,preprint,superscriptaddress,showkeys,showpacs,12pt]{revtex4-1}
\usepackage{caption}
\usepackage[caption=false]{subfig}
\usepackage{amsfonts}
\usepackage{amsmath, amsthm, amssymb}
\usepackage{mathtools}
\usepackage{graphicx}
\usepackage{float}
\usepackage{breqn}
\usepackage{hyperref}
\usepackage{bm}
\usepackage{url}
\usepackage{placeins}
\usepackage{multirow}
\newcommand\numberthis{\addtocounter{equation}{1}\tag{\theequation}}

\numberwithin{equation}{section}

\begin{document}
\title{Spectral-Domain-Based Scattering Analysis of Fields Radiated by Distributed Sources in Planar-Stratified Environments with Arbitrarily Anisotropic Layers}
\date{\today}
\author{Kamalesh Sainath}
\email{sainath.1@osu.edu}
\author{Fernando L. Teixeira}
\email{teixeira@ece.osu.edu}
\affiliation{The Ohio State University: ElectroScience Laboratory}
\altaffiliation[Address: ]{1330 Kinnear Road, Columbus, Ohio, USA 43212}

\begin{abstract}
\noindent
We discuss the numerically stable, spectral-domain computation and extraction of the scattered electromagnetic field excited by distributed sources embedded in planar-layered environments, where each layer may exhibit arbitrary and independent electrical and magnetic anisotropic response and loss profiles. This stands in contrast to many standard spectral-domain algorithms that are restricted to computing the fields radiated by Hertzian dipole sources in planar-layered environments where the media possess azimuthal-symmetric material tensors (i.e., isotropic, and certain classes of uniaxial, media). Although computing the scattered field, particularly when due to distributed sources, appears (from the analytical perspective, at least) relatively straightforward, different procedures within the computation chain, if not treated carefully, are inherently susceptible to numerical instabilities and (or) accuracy limitations due to the potential manifestation of numerically overflown and (or) numerically unbalanced terms entering the chain. Therefore, primary emphasis herein is given to effecting these tasks in a numerically stable and robust manner for all ranges of physical parameters. After discussing the causes behind, and means to mitigate, these sources of numerical instability, we validate the algorithm's performance against closed-form solutions. Finally, we validate and illustrate the applicability of the proposed algorithm in case studies concerning active remote sensing of marine hydrocarbon reserves embedded deep within lossy, planar-layered media.
\end{abstract}
\pacs{02.70.-c,02.70.Hm,95.75.Pq}

\keywords{Sommerfeld integral; anisotropic media; integral acceleration; Green's function; stratified media}
\maketitle
\section{\label{intro}Introduction}
Spectral-domain based computation and analysis of electromagnetic (EM) fields radiated by current distributions, embedded within planar-stratified environments with generally anisotropic media characterized by arbitrary (diagonalizable\footnote{The diagonalizability constraint ensures completeness of the plane wave basis; naturally-occurring media are always characterizable by diagonalizable material tensors, however.}) 3 $\times$ 3 relative permeability and permittivity tensors $\boldsymbol{\bar{\mu}}_r$ and $\boldsymbol{\bar{\epsilon}}_r$ (resp.), finds application in myriad areas. Some examples are geophysical prospection in subterranean \cite{anderson1,chen,teixeira9,teixeira10} and sub-oceanic \cite{macgregor1,key2,constable2,constable3,um1} environments, microstrip antennas~\cite{teixeira1,pozar,pozar2,balanis2}, planar waveguides~\cite{paulus}, transionospheric EM propagation studies~\cite{jehle}, ground penetrating radar~\cite{spagnolini1}, and so on. To facilitate field computation in such problems, which can possess domains exhibiting length scales on the order of hundreds or even thousands of wavelengths, spectral/Fourier-domain based EM field calculation methods exhibit both robustness and speed as defining virtues, making them oftentimes indispensable~\cite{sainath,sainath2,sainath3,mich1,mosig1,mich2}. For example, as demonstrated in \cite{sainath3}, through use of Complex-Plane Gauss-Laguerre Quadrature (CGLQ) and adaptive \emph{hp} refinement one can rapidly and accurately evaluate, without analytical-stage\footnote{As opposed to when using, for example, discrete image methods~\cite{mich5} which can involve approximating the spectral integrand as a sum of analytically invertible ``images" possessing closed-form integral solutions.} approximations, the radiated EM field via direct numerical integration without major concern about slow integrand decay (and hence slow convergence) or rapid integrand oscillation (necessitating fine sampling and high computational cost). It is upon these and other previous works \cite{sainath,sainath2} that we build to create a robust, error-controllable, and rapid direct integration algorithm directed specifically at achieving two objectives concerning EM radiation and scattering in planar-layered, generally anisotropic media, which comprise the main contributions of this paper: (1) An ``in-situ" scattered EM field extraction method, applicable to both point-like/Hertzian and distributed radiators (e.g., wire and aperture antennas), and (2) Direct Fourier-domain evaluation of the radiation integral used to compute fields radiated by distributed sources. Elaborated upon in detail below, these contributions add to the extensive body of work concerning spectral-domain based calculation of EM fields in planar-layered media which dominantly focus on radiation of Hertzian dipole sources in planar-layered media where the layers possess azimuthal-symmetric material tensors \cite{mich1,lambot1,mich1,mosig1,mich2}. In contrast, in the spirit of previous work \cite{sainath,sainath2,sainath3}, our proposed scattered-field and distributed radiator computation algorithms are applicable to planar-layered media where the layers can possess \emph{arbitrary} (diagonalizable) material tensors.

It is often the case that the time-harmonic scattered field $\bm{\mathcal{E}}_s(\bold{r})=\bm{\mathcal{E}}(\bold{r})-\delta_{L,M}\bm{\mathcal{E}}_d(\bold{r})$\footnote{Rather than the time-harmonic homogeneous medium/``direct" field $\bm{\mathcal{E}}_d(\bold{r})$ or time-harmonic total field $\bm{\mathcal{E}}(\bold{r})$. The Kronecker delta $\delta_{L,M}$ equals either one or zero when the source and observation layers ($M$ and $L$, resp.) either coincide or differ, respectively.} (or, via Fourier synthesis, the time domain scattered field) constitutes the signal of interest as it carries information about the inhomogeneity of the medium under interrogation. For example, in geophysical borehole prospection it is well known that planar inhomogeneity can contribute to erroneous extraction of the resistivity tensor of the local earth formation in which the sonde is embedded \cite{chen}. Therefore, being able to extract and analyze only the scattered field contribution may facilitate mitigating formation inhomogeneity effects in induction sonde measurements. Similarly concerning radars, one is usually only interested in the scattered field as it carries information about the surrounding environment's parameter(s) of interest~\cite{fiori,zhan}. Two straightforward ways to effect scattered field extraction are (1) \emph{a-posteriori} subtraction of $\bm{\mathcal{E}}_d(\bold{r})$ (computed in closed form) from $\bm{\mathcal{E}}(\bold{r})$ (numerically evaluated with spectral methods) and (2) temporal discrimination between the time domain (TD) direct and scattered field signals.

There are important drawbacks with each of these two methods, however. The subtraction method suffers from lack of general applicability when the source is embedded in generally anisotropic media wherein the time-harmonic space domain tensor Green's functions may not be available in closed form. Furthermore, even when $\bm{\mathcal{E}}_d(\bold{r})$ is available in closed form, \emph{a posteriori} direct field subtraction lacks robustness in the numerical evaluation of $\bm{\mathcal{E}}(\bold{r})$ since $\{|\bm{\mathcal{E}}(\bold{r})|,|\bm{\mathcal{E}}_d(\bold{r})| \}\to \infty$ as the observation point $\bold{r}=(x,y,z)$ approaches a source point $\bold{r}'=(x',y',z')$ (e.g., time-harmonic scattered field received at a mono-static radar), leading to the subtraction of two numerically overflown results. Time-gating, on the other hand, is feasible subject to temporal resolvability between the direct and scattered fields; this is fundamentally absent in \emph{time-harmonic} fields, however, which are oftentimes the quantities of interest. Furthermore, the time-gating method also suffers from the same numerical instability issue when simulating ``mono-static"-like scenarios. Indeed when the TD signal, in such eigenfunction expansion techniques, is synthesized through a superposition of frequency domain signals, obviously one requires here too a numerically stable and robust scheme to compute the total (frequency domain) field $\bm{\mathcal{E}}(\bold{r})$ at each desired frequency to facilitate TD windowing of the synthesized TD signal. Therefore, both the subtraction and TD windowing techniques return us, in general, back to the question of how to compute the frequency-domain field across a wide range of source distribution and observer position scenarios.

In contrast to the above two techniques, the proposed scattered field extraction approach relies upon ``in-situ" subtraction of the direct field during the modal field synthesis (i.e., spectral integration) process itself. This ``in-situ" subtraction approach, constituting the first of our two contributions, sports the following advantages:
\begin{enumerate}
\item Applicability to time-harmonic fields and, through Fourier temporal harmonic synthesis, TD fields.
\item Does not require the space-domain tensor Green's functions (either in the frequency or time domain) in closed form.
\item Robustness and numerical stability even as $|\bold{r}-\bold{r}'|\to 0$, rendering it applicable even to ``mono-static"-like radiation and reception scenarios.
\item Imposes no additional computational burden versus computing $\bm{\mathcal{E}}(\bold{r})$ \cite{sainath,sainath2,sainath3}.
\item Imparts added exponential decay to the spectral integrand that further accelerates convergence of the field solution.
\item Automatically and rigorously effects the time-windowing function ordinarily performed after synthesis of the TD total field signal, removing any need for additional processing to discriminate between the TD direct and scattered field signals.
\item Applicability to general source geometries possessing a closed form Fourier domain representation.
\end{enumerate}
Beyond extracting the scattered field, we propose a rapid, robust algorithm to compute the spectral domain integral representation of the field produced by \emph{distributed} sources embedded in planar-layered, generally anisotropic media. This strategy is based on the spectral representation of compactly-supported, otherwise arbitrary distributed sources in terms of spatial (sinusoidal) current harmonics and finds applicability where realistic modeling of current sources (whether they be physical antennas or equivalent current distributions) is otherwise prohibitive due to the computationally expensive task of either repeatedly computing the (space-domain) tensor Green's function and/or having to perform spatial discretization of the source distribution. One such example includes computing the received scattered field at a spaceborne radar platform in such a way that captures the effects of an inhomogeneous atmosphere and (or) subsurface environment. This approach may also prove desirable in aperture field synthesis, where it can separately compute the field pattern of (orthogonal) Fourier current modes radiating in a given inhomogeneous, anisotropic environment, and thus constitute an efficient forward engine for aperture synthesis-based optimization algorithms seeking to solve the inverse problem of procuring an aperture current distribution leading to a desired, pre-defined field pattern.

The relative efficiency of the spectral domain method, concerning distributed sources, arises primarily from two factors. The first factor is the lower sampling requirement needed in the spectral domain as compared to the spatial domain (e.g., spatial sampling using a Hertzian dipole/``pulse" basis) to represent a given harmonic current distribution and its radiated field. That is to say, for each harmonic current (requiring, self-evidently, only one spectral domain sample) and its radiated field that is simulated, one must use (based on our numerical experiments) approximately ten Hertzian dipole samplings per half-cycle variation of current amplitude. This sampling efficiency in turn amounts to approximately one order of magnitude solution speed acceleration for one-dimensional, wire-like source distributions and approximately two orders of magnitude solution speed acceleration for two-dimensional, aperture-like source distributions. The second contributing factor towards efficiency is the \emph{sparse} representation of many commonly encountered current distributions in terms of mutually orthogonal spatial harmonic current distributions.\footnote{As mathematically exhibited in Section \ref{ArbAnt} a secondary phenomenon, relating to the ``tapering" of the distributed source's field spectrum (as compared to the spectrum of the fields from a Hertzian dipole), also imparts added efficiency in evaluating the fields in the spectral domain.} Although in the spectral domain the distributed characteristic of the source enters into the field spectrum as a (deceptively simple) multiplicative factor augmenting the Hertzian dipole field spectrum, serious numerical instability issues can arise in a practical implementation due to the manifestation of exponentially rising field terms. This issue must be addressed to realize the computational efficiency benefits of the spectral domain evaluation of distributed source fields.

We discuss the above-mentioned stability and robustness issues, along with the proposed solutions to them, in Sections~\ref{DirField} and~\ref{ArbAnt}. First for convenience, we briefly summarize some fundamentals behind the underlying formulation; notation information and details of the underlying formulation can be found in other references \cite{sainath,sainath2,sainath3}.
\section{Formulation Fundamentals: Overview}
Initially assume a homogeneous medium possessing material tensors\footnote{$\epsilon_0$, $c$, and $\mu_0=1/(\epsilon_0c^2)$ are the vacuum permittivity, speed of light, and permeability, respectively. Furthermore, $\omega=2\pi f$, $k_0=\omega/c$, and $\eta_0=\sqrt{\mu_0/\epsilon_0}$ are the angular radiation frequency, vacuum wavenumber, and intrinsic impedance~\cite{balanis1,chew}, respectively, while $i$ denotes the unit imaginary number.} $\boldsymbol{\bar{\epsilon}}_c=\epsilon_0\boldsymbol{\bar{\epsilon}}_r$ (permittivity, including losses) and $\boldsymbol{\bar{\mu}}_c=\mu_0\boldsymbol{\bar{\mu}}_r$ (permeability, including losses) exhibiting arbitrary and independent anisotropy and loss, in which there are impressed (i.e., causative) electric and (equivalent) magnetic current densities $\bm{\mathcal{J}}(\bold{r})$ and $\bm{\mathcal{M}}(\bold{r})$ (resp.), as well as impressed volumetric electric and (equivalent) magnetic charge densities $\rho_v$ and $\rho_m$ (resp.). From Maxwell's equations in the frequency domain, one obtains~\cite{sainath,chew}:
\begin{equation}\label{WaveOp} \bm{\mathcal{\bar{A}}}=\nabla \times \boldsymbol{\bar{\mu}}_r^{-1} \cdot \nabla \times - k_0^2 \boldsymbol{\bar{\epsilon}}_r \cdot \end{equation}
\begin{equation} \label{ME5B}  \bm{\mathcal{\bar{A}}} \cdot \bm{\mathcal{E}} = ik_0\eta_0\bm{\mathcal{J}}-\nabla \times \boldsymbol{\bar{\mu}}_r^{-1} \cdot \bm{\mathcal{M}} \end{equation}
where the exp($-i\omega t$) convention is assumed and suppressed. Subsequently defining the three-dimensional Fourier Transform (FT) pair, for some generic vector field $\bm{\mathcal{L}}$, as \cite{sainath}
\begin{align}\label{FT3D} \ \bold{\tilde{L}}(\bold{k}) \ &= \ \iiint\limits_{-\infty}^{+\infty} \bm{\mathcal{L}}(\bold{r}) \, \mathrm{e}^{-i\bold{k} \cdot \bold{r}} \, \mathrm{d}x \, \mathrm{d}y \, \mathrm{d}z \\
\label{IFT3D} \ \bm{\mathcal{L}}(\bold{r}) \ &= \ \left(\frac{1}{2\pi}\right)^3\iiint\limits_{-\infty}^{+\infty} \bold{\tilde{L}}(\bold{k}) \, \mathrm{e}^{i\bold{k} \cdot \bold{r}} \, \mathrm{d}k_x \,  \mathrm{d}k_y \, \mathrm{d}k_z \end{align}
with \(\bold{r}=(x,y,z) \ \mathrm{and} \ \bold{k}=(k_x,k_y,k_z) \), one can take the FT of \eqref{ME5B} to yield its Fourier domain version followed by multiplying $\bold{\tilde{\bar{A}}}^{-1}$ on both sides of the resultant Fourier-domain expression. Further manipulations, upon assuming a single Hertzian dipole source at $\bold{r}'$ and denoting the observation point as $\bold{r}$, leads to the following expression for the time-harmonic direct electric field $\bm{\mathcal{E}}_d(\bold{r})$ radiated by said distribution in this layer possessing the material properties of (what is, in the multi-layered medium scenario, defined as) layer $M$~\cite{sainath}:
\begin{multline}\label{EHM2a} \bm{\mathcal{E}}_d(\bold{r})=\frac{i}{(2\pi)^{2}} \iint \limits_{-\infty}^{+\infty}
\left[u(z-z')\sum_{n=1}^2{\tilde{a}_{M,n}\bold{\tilde{e}}_{M,n}\mathrm{e}^{i\tilde{k}_{M,nz}\Delta z}}+ u(z'-z)\sum_{n=3}^4{\tilde{a}_{M,n}\bold{\tilde{e}}_{M,n}\mathrm{e}^{i\tilde{k}_{M,nz}\Delta z}}
\right] \times \\ \mathrm{e}^{ik_x\Delta x+ik_y\Delta y} \, \mathrm{d}k_x \, \mathrm{d}k_y \end{multline}
where $\Delta z=z-z'$, $\Delta x=x-x'=\Delta y=y-y' \geq 0$,\footnote{An azimuthal coordinate rotation is assumed to have been performed such that $\Delta x=\Delta y \geq 0$ \cite{sainath2,sainath3}.} and $\{\bold{\tilde{e}}_{P,n},\tilde{k}_{P,nz},\tilde{a}_{P,n} \}$ stand for the modal electric field vector, longitudinal propagation constant, and (source dependent) direct field amplitude of the $P$th layer's $n$th mode ($1\leq P \leq N$) (resp.); furthermore, $u(\cdot)$ denotes the Heaviside step function. Similarly, the time-harmonic scattered electric field $\bm{\mathcal{E}}_s(\bold{r})$ writes as \cite{sainath}:
\begin{multline}\label{Espace1}   \bm{\mathcal{E}}_s(\bold{r})  =
\frac{i}{(2\pi)^{2}} \iint\limits_{-\infty}^{+\infty}\left[ (1-\delta_{L,N})\sum_{n=1}^2{\tilde{a}^s_{L,n}\bold{\tilde{e}}_{L,n}\mathrm{e}^{i\tilde{k}_{L,nz}z}}+
(1-\delta_{L,1})\sum_{n=3}^4{\tilde{a}^s_{L,n}\bold{\tilde{e}}_{L,n}\mathrm{e}^{i\tilde{k}_{L,nz}z}}\right] \times \\ \mathrm{e}^{ik_x\Delta x+ik_y\Delta y} \, \mathrm{d}k_x \, \mathrm{d}k_y\end{multline}
 where $\tilde{a}^s_{P,n}$ is the scattered field amplitude in layer $P$.

Before proceeding, we note that when referring to the $n$th modal field in layer $P$ being ``phase-referenced" to a particular $z=z_o$ plane, this means that its longitudinal propagator has been cast in the form $\mathrm{e}^{i\tilde{k}_{P,nz}(z-z_o)}$.
\section{\label{DirField}Direct Field Subtraction}
\subsection{Modal Field Representation Modifications}
We now exhibit the formulation to extract the scattered electric field observed at $\bold{r}$ in layer $L$ due to a source at $\bold{r}'$ in layer $M=L$ for $1\leq (M=L) \leq N$. We use here the same notation and nomeclature as \cite{sainath} concerning scattered fields, whose expressions we briefly review next. First define $\bold{\tilde{a}}_D^{+}$ ($\bold{\tilde{a}}_D^{-}$) as the direct field 2$\times$1 modal amplitude vector associated with up-going (down-going) characteristic modes in layer $M$ phase-referenced to the top (bottom) bounding interface at depth $z=z_{M-1}$ ($z=z_{M}$) \cite{sainath}. Second, define the 2$\times$1 vectors $\bold{\tilde{a}}_{S1}^+$ and $\bold{\tilde{a}}_{S1}^-$ as the up-going and down-going scattered field modal amplitudes (resp.) in layer $M$ whose respective modal fields are phase-referenced to the interface at $z=z_{M-1}$; likewise, $\{\bold{\tilde{a}}_{S2}^+,\bold{\tilde{a}}_{S2}^-\}$ are the amplitudes for scattered modal fields that are phase-referenced to the interface $z=z_M$ \cite{sainath}. Third, denote the 2$\times$2 generalized reflection matrix from layer $P$ to adjacent layer $P'$\footnote{That is, $P'$ equals either $P+1$ or $P-1$ when $P'$ corresponds to the layer below or above layer $P$, respectively, where layer $P$ is the layer containing the incident modal fields.} as $\bold{\tilde{\bar{R}}}_{P,P'}$. One then obtains the standard formulae below for the scattered field amplitudes in the source-containing layer $M$ as a function of the direct field amplitudes \cite{sainath}\footnote{$\bold{\bar{I}}_{\nu}$ is the $\nu \times \nu$ identity matrix.}:
\begin{align}
\bold{\bar{\Lambda}}_M^{+}(z_o)&=\mathrm{Diag}\left[\mathrm{e}^{i\tilde{k}_{M,1z}z_o} \ \mathrm{e}^{i\tilde{k}_{M,2z}z_o}\right], \ \bold{\bar{\Lambda}}_M^{-}(z_o)=\mathrm{Diag}\left[\mathrm{e}^{i\tilde{k}_{M,3z}z_o} \ \mathrm{e}^{i\tilde{k}_{M,4z}z_o}\right] \\ \bold{\tilde{\bar{M}}}_1  &=   \bold{\bar{\Lambda}}_M^+(z_{M-1}-z_{M}) \cdot \bold{\tilde{\bar{R}}}_{M,M+1}, \ \bold{\tilde{\bar{M}}}_2  =  \bold{\bar{\Lambda}}_M^-(z_{M}-z_{M-1}) \cdot \bold{\tilde{\bar{R}}}_{M,M-1} \\
\bold{\tilde{a}}_{S1}^-  &=  \left[\bold{\bar{I}}_2-\bold{\tilde{\bar{R}}}_{M,M-1} \cdot \bold{\tilde{\bar{M}}}_1 \cdot \bold{\bar{\Lambda}}_M^-(z_M-z_{M-1})\right]^{-1} \cdot \bold{\tilde{\bar{R}}}_{M,M-1} \cdot \left[\bold{\tilde{a}}_D^{+}+\bold{\tilde{\bar{M}}}_1 \cdot \bold{\tilde{a}}_D^{-} \right ] \numberthis \label{RFS1a} \\
\bold{\tilde{a}}_{S2}^+  &= \left[\bold{\bar{I}}_2-\bold{\tilde{\bar{R}}}_{M,M+1} \cdot \bold{\tilde{\bar{M}}}_2 \cdot \bold{\bar{\Lambda}}_M^+(z_{M-1}-z_M)\right]^{-1} \cdot \bold{\tilde{\bar{R}}}_{M,M+1} \cdot \left[\bold{\tilde{a}}_D^{-}+\bold{\tilde{\bar{M}}}_2 \cdot \bold{\tilde{a}}_D^{+}\right] \numberthis \label{RFS1b}
 \end{align}
 which are required when $M=L$ (i.e., when the observation and source layers coincide). Then the observed scattered field amplitudes, when $M=L$, write as
\begin{equation}\label{Case3b} \bold{\tilde{a}}_L^+=\bold{\bar{\Lambda}}_L^+(z-z_L) \cdot \bold{\tilde{a}}^+_{S2}, \ \bold{\tilde{a}}_L^-=\bold{\bar{\Lambda}}_L^-(z-z_{L-1}) \cdot \bold{\tilde{a}}^-_{S1} \end{equation}
 with the subsequent scattered amplitude-weighted superposition of the observed scattered modal fields following the prescription described in \cite{sainath}. Note that if $M=L=1$ or $M=L=N$, \eqref{RFS1a} or \eqref{RFS1b} (resp.) reduce to $\bold{0}$ and derive from the fact that no down-going or up-going reflected fields are present in layer $L$ (resp.) \cite{chew}[Ch. 2].

Now that the direct fields have served their purpose of exciting the scattered fields, their subtraction from the total field solution enters via coercion of the Kronecker delta $\delta_{L,M}$ in the expression $\bm{\mathcal{E}}(\bold{r})=\delta_{L,M}\bm{\mathcal{E}}_d(\bold{r})+\bm{\mathcal{E}}_s(\bold{r})$ to zero. Indeed, one uses the direct fields to excite the scattered fields, but does \emph{not} include the direct field contributions themselves when assembling the total observed modal field for some $(k_x,k_y)$ doublet, as evidenced by the expressions for $\bold{\tilde{a}}_L^+$ and $\bold{\tilde{a}}_L^-$ in \eqref{Case3b} being devoid of \emph{explicit} dependence on $\bold{\tilde{a}}_D^+$ and $\bold{\tilde{a}}_D^-$.\footnote{Implicitly, of course, $\bold{\tilde{a}}_L^+$ and $\bold{\tilde{a}}_L^-$ do depend on the direct field excitation.} Such a scattered-field extraction procedure is independent of the type of excitation involved; that is to say, this procedure features applicability to electric and (equivalent) magnetic currents of arbitrary polarization, (bounded) amplitude profile, and (compact) spatial support region subject to possessing a valid Fourier (wave-number) domain representation. Finally, we remark that due to the concept of a ``scattered" field becoming more ambiguous when $M \neq L$, the observed modal field amplitudes are computed identically to the procedure used in \cite{sainath} to compute the total field $\bm{\mathcal{E}}(\bold{r})$.
  \subsection{Additional Remarks}
The spectral integral in \eqref{Espace1} is evaluated along properly chosen integration contours in the $k_x$ and $k_y$ complex planes. For details, the reader is referred to~\cite{sainath2,sainath3}; at present it suffices to recall from \cite{sainath2,sainath3} that for the semi-infinite $k_x$ and $k_y$ ``tail" integrals one detours into the upper-half plane, parameterized by the detour angle $\gamma=\mathrm{tan}^{-1}\left(|\Delta x/\Delta z|\right)$ \cite{sainath2}, where one now replaces $\Delta z=z-z'$  with $\Delta z_{\mathrm{eff}}$:
\begin{equation}\label{zeff}
\Delta z_{\mathrm{eff}}=
\begin{cases}
|z-z'| &, M \neq L \\
(z-z_1)+(z'-z_1) &, M=L=1 \\
(z_{N-1}-z)+(z_{N-1}-z') &, M=L=N \\
\mathrm{min}\left[(z_{M-1}-z)+(z_{M-1}-z'),(z-z_M)+(z'-z_M)\right] &, 1<(M=L)<N
\end{cases}
\end{equation}
In the first and fourth cases above, for which $M \neq L$ and $1<(M=L)<N$ (resp.), one typically encounters (excepting when $M \neq L$, with $L=1$ or $N$) both up-going and down-going scattered fields. Therefore, we are obliged to make conservative (small) assumptions for $\Delta z_{\mathrm{eff}}$ to minimize the residual\footnote{$r_x$ is the real-valued variable in \cite{sainath3} in terms of which the $k_x$ plane integration contour path ``tail" is parameterized. The residual factor exp($r_x \cos{\gamma_x}\Delta z_{\mathrm{eff}}$) arises from using Complex-Plane Gauss-Laguerre Quadrature (CGLQ).} $\mathrm{exp}(r_x \cos{\gamma_x}\Delta z_{\mathrm{eff}})$ that we multiply back into the integrand (and similarly for the $k_y$ integration). Indeed, this is especially important due to the asymptotic Constant Phase Path (CPP), in general, not being well defined due to anisotropy and planar inhomogeneity \cite{sainath2,sainath3}. Therefore, conservatively assigning $\Delta z_{\mathrm{eff}}$ avoids situations where (for example) $1<(M=L)<N$, the source and observation points are both very close to the interface at $z=z_{M-1}$, and one uses an alternative effective propagation distance such as $\Delta z_{\mathrm{eff}}=[(2z_{M-1}-z-z')+(z+z'-2z_M)]/2$ that may over-estimate the effective longitudinal propagation distance of (the dominant contribution to) the down-going scattered fields (``single-bounce reflection term"). This may lead to exponential kernels of the form\footnote{$\Delta z'$ loosely denotes the correct mode-dependent effective longitudinal distance. Of course, in reality $\Delta z'$ is elusive to accurately quantify due to anisotropy and/or, when finite-thickness slabs are present, internal ``multi-bounce" effects.} $\mathrm{exp}(i\tilde{k}^-_z\Delta z'+\Delta z_{\mathrm{eff}}r_x\cos{\gamma_x})$, corresponding to the down-going scattered fields whose actual effective longitudinal propagation distance $\Delta z'$ has been overestimated as $\Delta z_{\mathrm{eff}}$. Such exponential residuals may lead to unbounded solutions for increasing $|r_x|$, rather than asymptotically tending to a constant magnitude and contributing towards a numerically stable computation process \cite{sainath3}.
\subsection{\label{ScatVal}Validation Results: Scattered Field Extraction}
To validate the algorithm's ability to accurately extract the scattered field, we use the algorithm to verify the following well-known results concerning the effect of placing Hertzian dipole radiators infinitesimally close to a perfectly conducting ground plane of infinite lateral extent~\cite{balanis1}:
\begin{enumerate}
\item The direct EM field of a vertical electric dipole (VED) will be reinforced by the field scattered off the ground. That is, the scattered and direct fields should be equal.
\item The direct EM field of a vertical magnetic dipole (VMD) will be canceled by the ground-scattered field. That is, the scattered and direct contributions to any given field component should be equal in magnitude and opposite in sign.
\end{enumerate}
To avoid (1) numerical instability due to entering an infinite conductivity for the ground plane and (2) inaccuracy stemming from a ground plane with finite conductivity, the presence of a perfectly reflecting ground plane is equivalently effected via manually coercing, within the code, the ground plane's intrinsic reflection coefficients for the incident $\mathrm{TE}_z$ and $\mathrm{TM}_z$ modes \cite{sainath}. We emphasize that this coercion is done only to facilitate the present image theory study and does not fundamentally alter any of the other computations.

Prior to discussing results, we note the following conventions used for all numerical results discussed in the paper: (1) All errors are displayed as field component-wise relative error 10$\mathrm{log}_{10}|(L_{\mathrm{num}}-L_{\mathrm{exact}})/L_{\mathrm{exact}}|$ (dB units); (2) all computations are performed in double precision; (3) any relative errors below -150dB are coerced to -150dB; (4) An adaptive integration tolerance of $1.2 \times 10^{-t}$ denotes a precision goal of approximately $t$ digits \cite{datta}; and (5) the error is coerced to -150dB whenever the computed and reference solution magnitudes are (within machine precision equal to) zero.

\begin{figure}[H]
\centering
\subfloat[\label{EzScat}]{\includegraphics[width=2.5in]{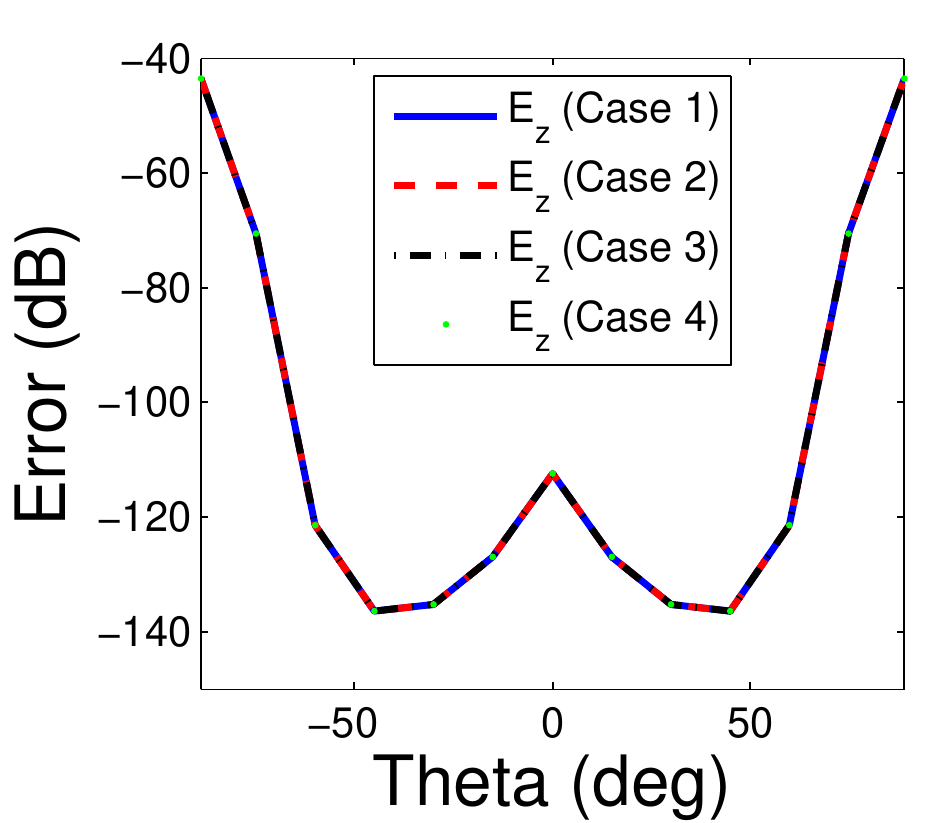}}
\subfloat[\label{HzScat}]{\includegraphics[width=2.5in]{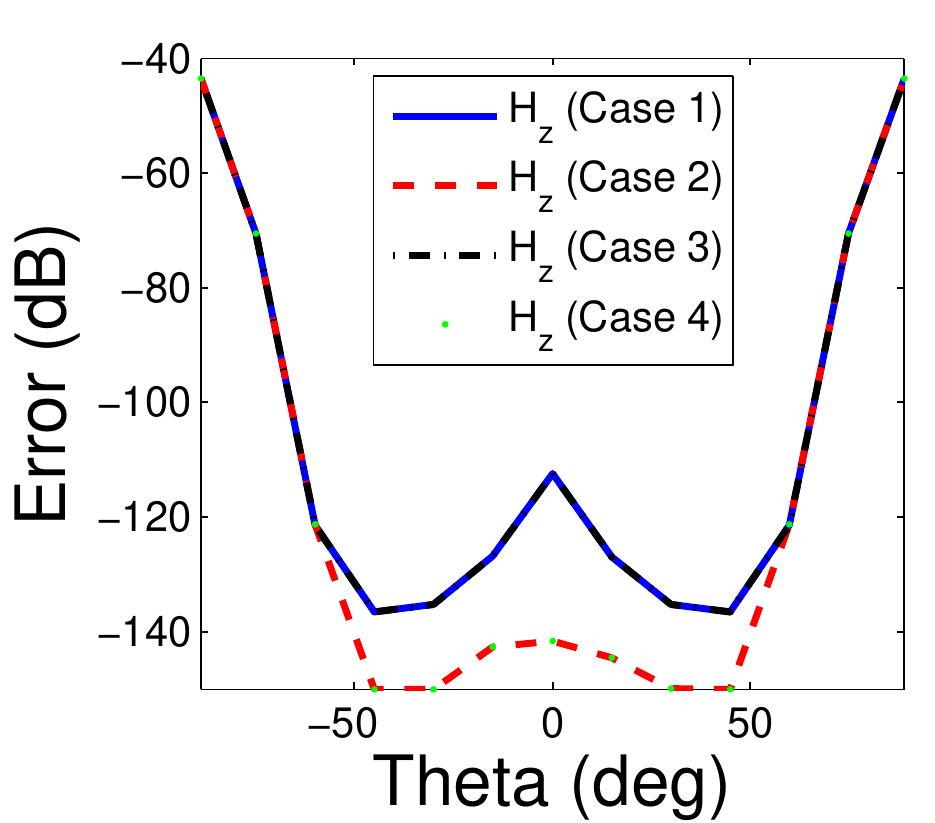}}
\caption{\label{ScatField} \small (Color online) Error in computing the field reflected off of the ground plane. For the VED and VMD cases, the reference field results are $E_z$ and $-H_z$ in homogeneous vacuum (resp.).}
\end{figure}

Figure \ref{EzScat} shows the error in computing the reflected electric field $E_z$ due to a VED, while Figure \ref{HzScat} shows the error in computing the reflected magnetic field $H_z$ due to a VMD, where both sources are radiating at $f$=2MHz. The observation point is kept at a fixed radial distance $|\bold{r}-\bold{r}'|$=10m from the source, the observation angle in azimuth is set to $\phi=0^{\circ}$, and the polar angle $\theta$ is swept from $-89^{\circ}\leq \theta \leq 89^{\circ}$. To test the scattered-field extraction for all possible scenarios concerning the source and observation points being in the same layer,\footnote{Recall that if $L\neq M$, the scattered-field extraction algorithm reduces to computing the total field.} we perform the scattered-field extraction in the following four cases referred to in the legends of Figures \ref{EzScat}-\ref{HzScat}. Case 1: Vacuum half-space above perfect electric conductor (PEC) half-space, with the source placed infinitesimally above the PEC ground. Case 2: Vacuum half-space below PEC half-space, with the source placed infinitesimally below the PEC ground. Case 3: Vacuum half-space, fictitiously partitioned into two layers such that both source and observer reside in a ``slab" of vacuum, above PEC half-space (source placed infinitesimally above the PEC ground). Case 4: Vacuum half-space, fictitiously partitioned into two layers such that both source and observer reside in a ``slab" of vacuum, below PEC half-space (source placed infinitesimally below the PEC ground). By being placed ``infinitesimally" above or below the ground plane, we mean to say that the longitudinal distance between the source and ground plane is set to $1.0 \times 10^{-15}$m. Needless to say, this distance (nuclear scale) is many orders of magnitude below the length scales of this problem; rather, it is simply used as a numerical means to test the proposed scattered-field extraction algorithm's accuracy.

We observe that approximately between the angles $-60^{\circ}\leq \theta \leq 60^{\circ}$ the algorithm delivers at least eleven digits of accuracy, which is consistent with the adaptive integration tolerance set as $1.2 \times 10^{-12}$. However, accuracy declines to approximately four digits as the observation point tends toward the surface of the ground plane. The cause behind this degradation of accuracy as the polar observation angle tends towards horizon, which is also evident in the wire and aperture antenna studies in Sections \ref{WireAntVal} and \ref{ApAntVal} below, is a topic of ongoing investigation.
\section{\label{ArbAnt}Distributed-Source Field Computation}
\subsection{Introduction}
The process of evaluating fields from distributed sources traditionally involves discretization of the space-domain radiation integrals concerning the (for example) electric field produced by either an electric or (equivalent) magnetic current source distribution (resp.) \cite{chew,felsen}:
\begin{align}\bm{\mathcal{E}}(\bold{r})&= ik_0 \eta_0 \iiint \limits_{V'} \bm{\mathcal{\bar{G}}}_{ee}(\bold{r};\bold{r}')\cdot \bm{\mathcal{J}}(\bold{r}')\mathrm{d}V' \numberthis \label{radint1} \\
\bm{\mathcal{E}}(\bold{r})&= -\iiint \limits_{V'} \bm{\mathcal{\bar{G}}}_{em}(\bold{r};\bold{r}')\cdot \boldsymbol{\bar{ \mu }}_r^{-1}(\bold{r}') \cdot \bm{\mathcal{M}}(\bold{r}')\mathrm{d}V'\numberthis  \label{radint2} \end{align}
where $\bm{\mathcal{\bar{G}}}_{ee}(\bold{r};\bold{r}')$ and $\bm{\mathcal{\bar{G}}}_{em}(\bold{r};\bold{r}')$ are the space domain tensor Green's functions describing electric fields radiated by Hertzian electric and magnetic dipole sources (resp.), and $\mathrm{d}V'=\mathrm{d}x'\mathrm{d}y'\mathrm{d}z'$ is the differential volume element on the emitter antenna manifold occupying volume $V'$ in layer $M$. Now admit either an electric \emph{or} magnetic source distribution and assume it is contained in one layer for simplicity. Furthermore, let $N_{avg}$ denote the average number of points on $V'$ whose equivalent Hertzian dipole contributions, to the electric field at observation point $\bold{r}$, need to be sampled to re-construct the observed field with some pre-prescribed accuracy level. In this case, one must (in general) evaluate a total of $N_{avg}$ two-dimensional Fourier integrals due to the space domain tensor Green's functions being translation-\emph{variant} along the longitudinal direction. In addition, one must then evaluate the space domain radiation integrals \eqref{radint1}-\eqref{radint2} themselves, which for an electrically large radiator with rapid variation in the current amplitude and/or polarization profile may itself also be a non-trivial task.

It turns out that for simple antenna geometries whose space-domain Fourier transforms are readily available in closed form, one can feasibly eliminate the intermediate step of evaluating the space-domain radiation integrals in \eqref{radint1}-\eqref{radint2} by directly computing the radiation integrals in the spectral domain itself. Indeed, recall that for a homogeneous medium \eqref{radint1}-\eqref{radint2} reduce to three-dimensional convolution integrals which can equivalently be computed in the Fourier domain \cite{chew}[Ch. 7]:
\begin{align}
\bm{\mathcal{E}}(\bold{r})&= ik_0 \eta_0\left(\frac{1}{2\pi}\right)^3\iiint \limits_{-\infty}^{\infty} \bold{\tilde{\bar{G}}}_{ee}(\bold{k};\bold{r}')\cdot \bold{\tilde{J}}(\bold{k})\mathrm{e}^{ik_xx+ik_yy+ik_zz}\mathrm{d}k_x\mathrm{d}k_y\mathrm{d}k_z \numberthis \label{radint1b} \\
\bm{\mathcal{E}}(\bold{r})&= -\left(\frac{1}{2\pi}\right)^3\iiint \limits_{-\infty}^{\infty} \bold{\tilde{\bar{G}}}_{em}(\bold{k};\bold{r}')\cdot \boldsymbol{\bar{ \mu }}_r^{-1} \cdot \bold{\tilde{M}}(\bold{k})\mathrm{e}^{ik_xx+ik_yy+ik_zz}\mathrm{d}k_x\mathrm{d}k_y\mathrm{d}k_z \numberthis  \label{radint2b}
\end{align}
whose more generalized manifestation, in the case of planar-layered media, writes as shown in \eqref{EHM2a}-\eqref{Espace1}. Of course, in the case of homogeneous isotropic media, the spectral-domain implementation may not be advantageous since the space-domain tensor Green's functions are available in closed form~\cite[Ch.\ 1,7]{chew}. However, in either homogeneous media exhibiting arbitrary anisotropy and/or planar-stratified media as considered here, wherein space-domain tensor Green's functions are typically unavailable in closed form, the spectral domain evaluation of the radiation integrals can offer a significant advantage in terms of solution speed and computational resource demand.
\subsection{Generalized Source Distribution: Formulation and Analyticity Considerations}
First we start with the vector wave equation \eqref{ME5B} under the assumption of a homogeneous medium, as usual. Assuming the source distribution to have a valid FT, i.e., $\bold{\tilde{J}}(\bold{k})$ and $\bold{\tilde{M}}(\bold{k})$ are well-defined spectral quantities, one can exhibit the spectral-domain version of \eqref{ME5B} as
\begin{equation}
   \label{ME5BFT}  \bold{\tilde{\bar{A}}}(\bold{k}) \cdot \bold{\tilde{E}}(\bold{k}) = ik_0\eta_0\bold{\tilde{J}}(\bold{k})-\tilde{\nabla} \times \boldsymbol{\bar{\mu}}_r^{-1} \cdot \bold{\tilde{M}}(\bold{k})
\end{equation}
 Inverting $ \bold{\tilde{\bar{A}}}(\bold{k})$ and taking the three-dimensional inverse Fourier integral on both sides of \eqref{ME5BFT} yields the space domain electric field $\bm{\mathcal{E}}(\bold{r})$. For a homogeneous medium, we note that \eqref{EHM2a} (with $\bm{\mathcal{L}}$ and $\bold{\tilde{L}}$ replaced by $\bm{\mathcal{E}}$ and $\bold{\tilde{E}}$ in \eqref{IFT3D}, resp.) is just the spectral-domain implementation of convolution for a general source distribution \eqref{radint1b}-\eqref{radint2b}. Despite the conceptually straightforward task of computing the radiation integral in the spectral domain, the presence of a \emph{distributed} source presents a practical challenge to the numerically robust and stable evaluation of the EM field. This is because the analyticity properties of the spectral integrand in regards to the $k_x$, $k_y$, and $k_z$ spectral variables in \eqref{IFT3D} are now obfuscated. That is, one often encounters the following scenario with distributed radiators: Regardless of whether one deforms the integration path into the upper- or lower-half section of the $k_x$ or $k_y$ complex plane,\footnote{This deformation is performed mainly to minimize integrand oscillation and accelerate integrand decay along the deformed Fourier tail integral paths, thereby resulting in rapid convergence of the evanescent spectrum field contribution \cite{sainath2,sainath3}.} one may encounter a numerically overflown result.

\begin{figure}[H]
\centering
\subfloat{\includegraphics[width=2.75in]{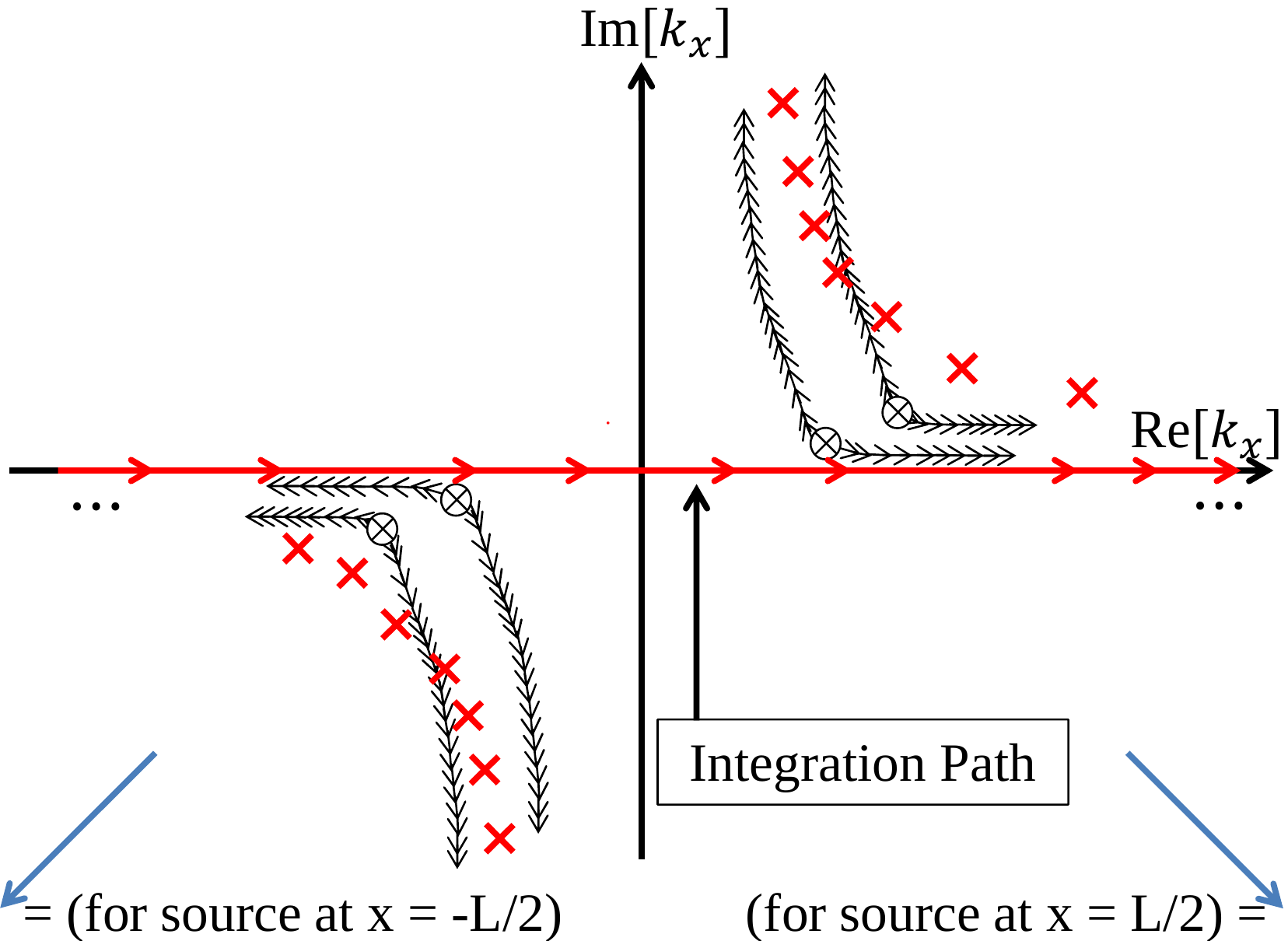}}

\subfloat{\includegraphics[width=2.75in]{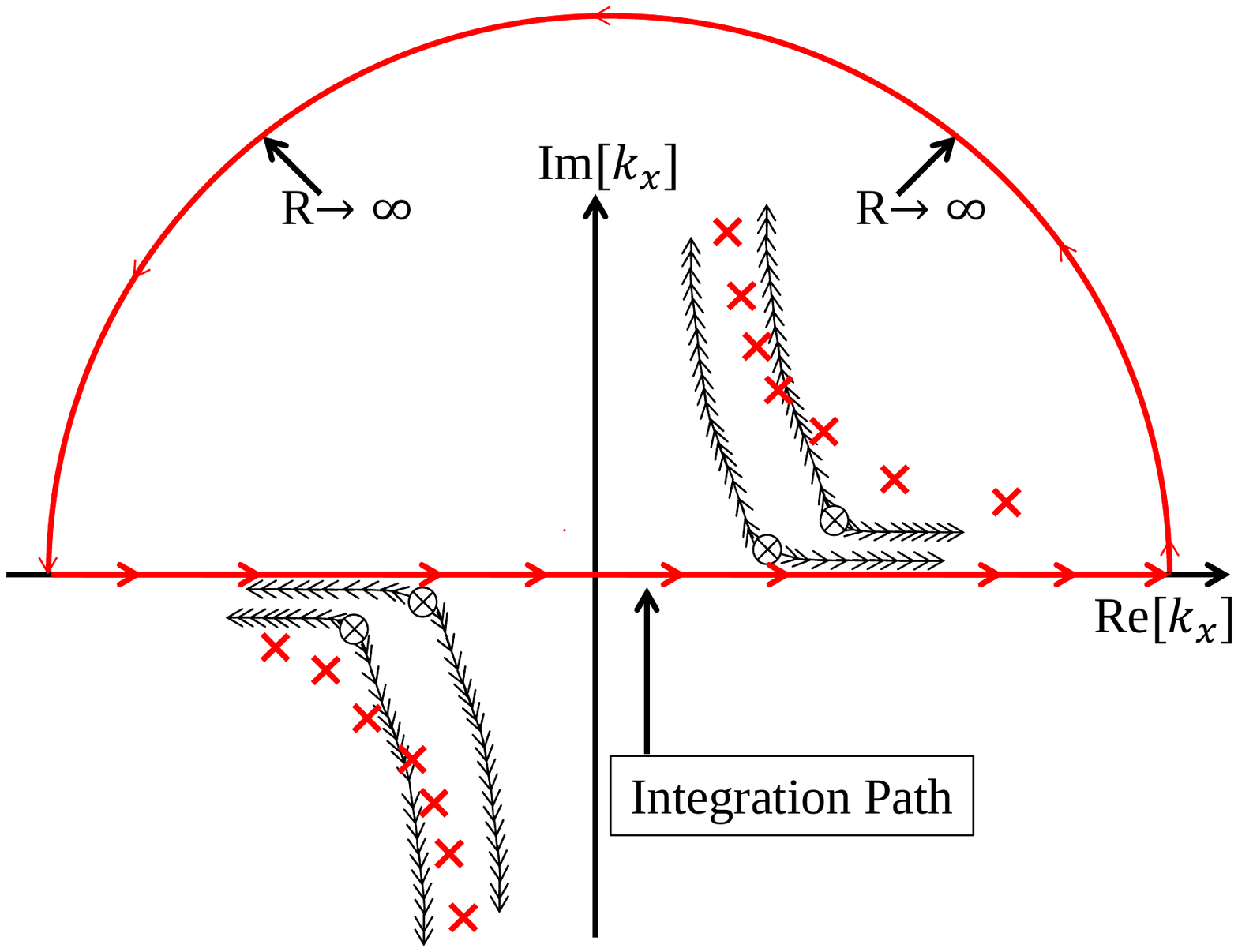}}
\subfloat{\includegraphics[width=2.75in]{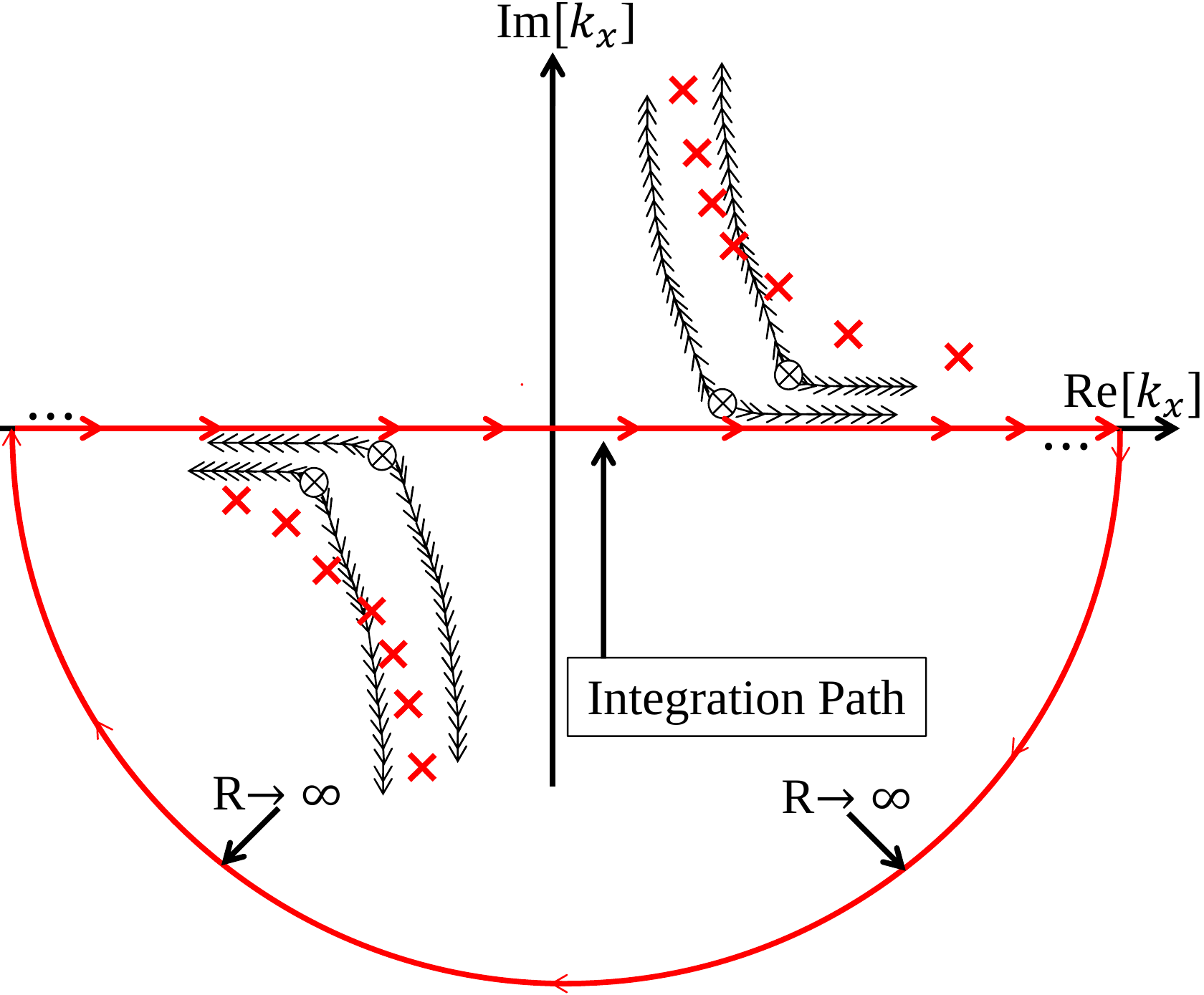}}
\caption{\label{JordLemma} \small (Color online) Source-location-dependent region of analyticity of the spectral EM field in the $k_x$ plane regarding the discussed example of two dipole sources. When $|x| < L/2$, the real-axis path is equivalent to enclosing either the upper-half or lower-half Im$[k_x]$ plane for the source located at $x'=-L/2$ or $x'=+L/2$ (resp.).}
\end{figure}

 To illustrate this, consider the modified FT pair below for the electric source current distribution $\bm{\mathcal{J}}(\bold{r})$:
 \begin{align}
\tilde{\bold{J}}(\bold{k})&=\bm{\mathcal{F}}\Bigg \{\bm{\mathcal{J}}(\bold{r}) \Bigg \}=\iiint \limits_{-\infty}^{\infty}\bm{\mathcal{J}}(\bold{r}')\mathrm{e}^{-i \bold{k}\cdot \bold{r}'}\mathrm{d}x'\mathrm{d}y'\mathrm{d}z'  \\
\bm{\mathcal{J}}(\bold{r})&=\bm{\mathcal{F}}^{-1}\Bigg \{\tilde{\bold{J}}(\bold{k})\Bigg \}=\left(\frac{1}{2\pi}\right)^3\iiint \limits_{-\infty}^{\infty} \iiint \limits_{-\infty}^{\infty}\bm{\mathcal{J}}(\bold{r}')\mathrm{e}^{i \bold{k}(\bold{r}-\bold{r}')}\mathrm{d}k_x\mathrm{d}k_y\mathrm{d}k_z\mathrm{d}x'\mathrm{d}y'\mathrm{d}z'
\end{align}
where one can draw a physical association between $\bold{r}$ and the observation point, as well as between $\bold{r}'$ and an equivalent Hertzian dipole source belonging to $\bm{\mathcal{J}}(\bold{r})$. Now, assume fixed $x$ and $x'$ values for which one evaluates the inner (i.e., along $k_x$) spectral integral. One promptly realizes that depending on whether $(x-x')>0$ or $(x-x')<0$ the spectral-domain integrand's region of analyticity, and hence the region in which one can apply Jordan's lemma in the $k_x$ plane \cite{chew}[Ch. 2], depends on the sign of $x-x'$. Physically, this corresponds to a situation of the observer witnessing incoming radiation from sources placed on either side of the observer along $x$. An analogous observation can be made with respect to more general volumetric sources distributed along $x$, $y$, and $z$. To better understand the analyticity issue, consider the example of the current distribution $\bm{\mathcal{J}}(\bold{r})=\left[\delta(x-L/2)+\delta(x+L/2)\right]\delta(y)\delta(z)\bold{\hat{z}}$,\footnote{$\delta (x)$ is the Dirac delta function.} possessing FT $\bold{\tilde{J}}(\bold{k})=\left[\mathrm{e}^{-ik_xL/2}+\mathrm{e}^{ik_xL/2}\right]\bold{\hat{z}}=2\bold{\hat{z}}\cos{(k_xL/2)}$, which radiates in free space. Evaluating (analytically) the $k_z$ integral of the spectral field and detouring in the $k_x$-plane integration path's ``tail" section~\cite{sainath2,sainath3} results in $|\bold{\tilde{J}}(\bold{k})| \to \infty$ as $|\mathrm{Im}(k_x)| \to \infty$ whenever $|x| < L/2 $; see Figure \ref{JordLemma}.
To mitigate the risk of numerical overflow for arbitrary $\bold{r}\neq \bold{r}'$, we make the conservative judgment to only allow observation of $\bm{\mathcal{E}}_s(\bold{r})$.\footnote{Restriction to calculating only the scattered field is done to lend exponential damping to the spectral integrand, helping to offset exponentially rising terms due to distributed sources.} To further suppress any exponentially rising terms, we purposefully incorporate the real-valued, numerical Laguerre-Gauss quadrature weights\footnote{The constant, complex valued factors $l^{+}$ and $l^{-}$ manifest in the expressions $k_x= l^{\pm}r_x \pm \xi_1$ used to parameterize the linear path deformation, appearing in Eqs. (2.10)-(2.11) of \cite{sainath3}, are placed outside the double Fourier integral and thus allow the weights to be real-valued. See \cite{sainath3} for details.} directly into the power of the complex exponentials prior to evaluating the exponentials themselves. It warrants pointing out that the importance of incorporating these quadrature weights directly into the exponentials should not be underestimated in comparison to the importance of restricting calculation to only the scattered field when it comes to distributed sources. Indeed, the weights themselves rapidly decay with respect to the real-valued variable $r_x$ used to parameterize the deformed path along which we evaluate the Fourier ``tail" integrals in the $k_x$ plane \cite{sainath3}. As a result, they serve to mitigate $r_x$-dependent exponential increase due to the presence of a distributed source and its exp($i k_x L/2$)-like terms. This is illustrated below in Figure \ref{weight}, where we plot $\mathrm{ln}(w_x)$ versus the ``normalized" quadrature node number $N^{\prime\prime}=n'/N'$ for various Laguerre-Gauss quadrature rules ($n'$ is the actual quadrature node number and $N'$ is the quadrature rule order); we indeed observe a rapid decrease in the weights as $n'$, and hence the $k_x$ plane integration path-parameterizing variable $r_x$, increases. An analogous discussion likewise holds for integration within the $k_y$ plane and the corresponding Laguerre-Gauss quadrature weights $w_y$.
\begin{figure}[H]
\centering
\includegraphics[width=2.75in]{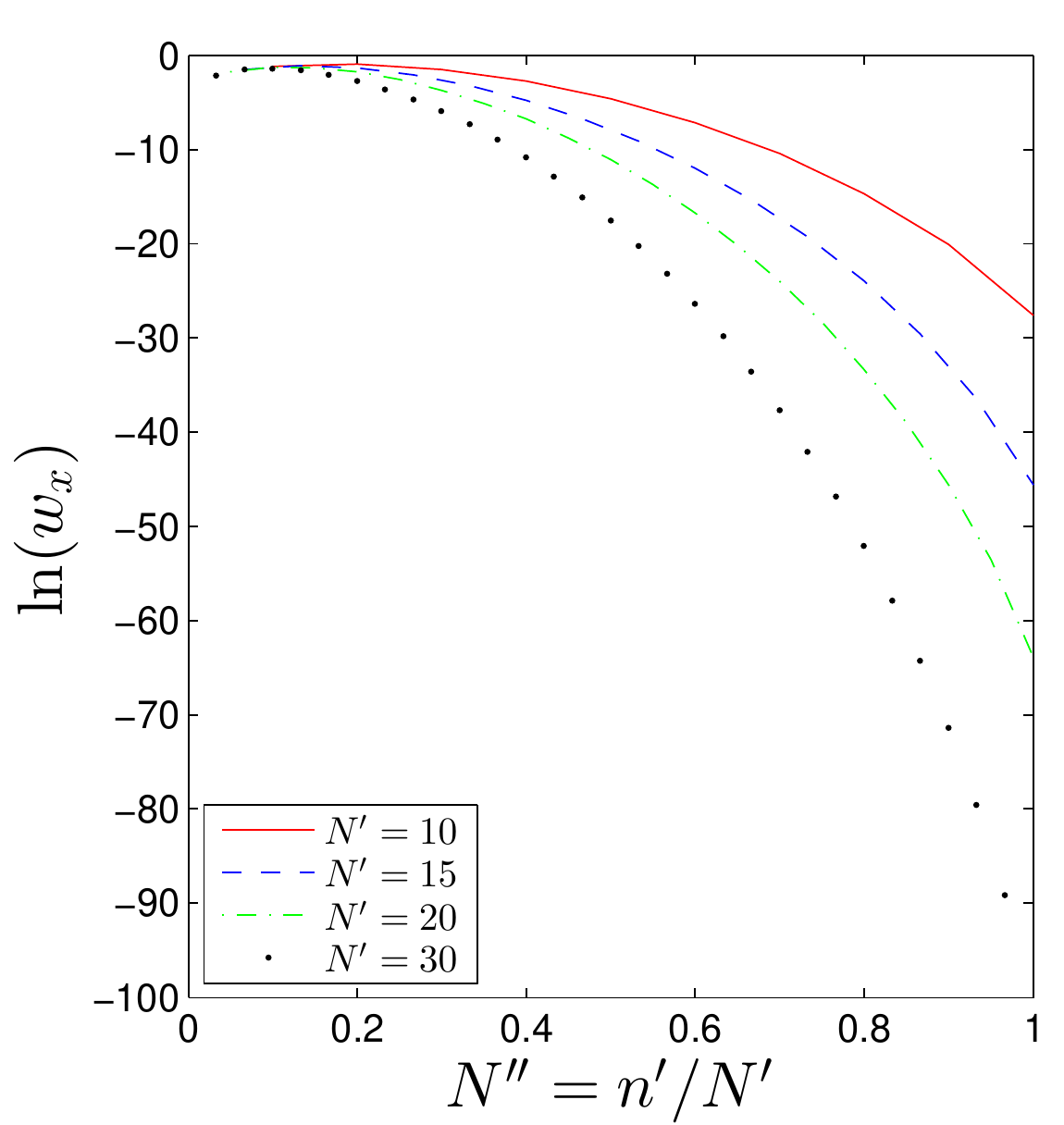}
\caption{\label{weight} \small (Color online) Natural logarithm of the Laguerre-Gauss quadrature weights (ln$[w_x]$) as a function of the normalized position $N^{\prime\prime}=n'/N'$ along the $k_x$ integration path for various quadrature rules. Note the rapid decline of the weights ln$(w_x)$ versus $N^{\prime\prime}$.}
\end{figure}
Referring again to the notation and terminology used in \cite{sainath3}, letting $w_x$ be the Laguerre-Gauss quadrature weight multiplying into the evaluation of the integrand for some ($k_x$, $k_y$) doublet ($k_{x0}=\xi_1+l^+r_{x0}$, $k_{y0}$), and (for the sake of illustration) assuming one is presently integrating in the intersection of the a) evanescent spectrum zone of the $k_x$ plane (let Re[$k_x]>$0) and b) propagation spectrum zone of the $k_y$ plane, then upon defining (let $\tilde{k}_z^+=\tilde{k}_{M,1z}=\tilde{k}_{M,2z}$)
\begin{equation}
\tau=i\tilde{k}_z^{+}\Delta z'+r_{x0}\cos(\gamma_x)(i\Delta x+\Delta z_{\mathrm{eff}})+ik_{y0}\Delta y
\end{equation}
for our twin vertical electric dipole radiation example, one can simply set
\begin{equation} \label{expcomp} \left(\mathrm{e}^{ik_{x0}L/2}+\mathrm{e}^{-ik_{x0}L/2}\right)\mathrm{e}^{\tau}w_x w_y \to
\Big(\mathrm{e}^{ik_{x0}L/2+\tau+\mathrm{ln}(w_x)}+\mathrm{e}^{-ik_{x0}L/2+\tau+\mathrm{ln}(w_x)}\Big) w_y
\end{equation}
Similarly, if one detours in the upper-half $k_y$ plane to evaluate the plane wave spectra evanescent with respect to $k_y$, then place the weights $\{\mathrm{ln}(w_y)\}$ into the exponentials. Detouring into both the $k_x$ and $k_y$ upper-half planes, by extension, mandates placing the weights $\{\mathrm{ln}(w_x)\}$ and $\{\mathrm{ln}(w_y)\}$ into both exponentials.

Now we exhibit the explicit spectral-domain representation of two commonly encountered distributed source geometries, the linear (wire) and rectangular aperture antennas.
\subsection{\label{dipole}Linear Antennas}
Consider a linear wire antenna of length $L$  centered at $\bold{r}_o'=\bold{0}$ whose current distribution can be written, without loss of generality, as a superposition of harmonic current modes \cite{balanis1}:
\begin{align}
 \bm{\mathcal{J}}(\bold{r})&=\bold{\hat{z}}\delta(x)\delta(y)\mathrm{rect}\left(\frac{z}{L}\right)\sum_{r=1}^{\infty}\left[J_c'(2r-1)\cos{\left(\frac{(2r-1) \pi z}{L}\right)}+J_s'(2r)\sin{\left(\frac{2 \pi r z}{L}\right)}\right] \label{SD1}
\end{align}
where $J_c'$ and $J_s'$ are the complex-valued modal current amplitudes for cosinusoidal and sinusoidal spatial current variation (resp.). The unit pulse function is defined as $\mathrm{rect}\left(u\right)=1$ for $|u| < 1/2$ and zero otherwise. It is a simple exercise to show that
\begin{equation}\label{FT2}
\tilde{\bold{J}}(\bold{k})=\frac{-r \bold{\hat{z}}}{L}\sum_{r=1}^{\infty}\left[\frac{\xi_r J_c'(2r-1)\cos{\left(\frac{k_zL}{2}\right)}+i\xi_r' J_s'(2r)\sin{\left(\frac{k_zL}{2}\right)}}{\left(k_z-\frac{r\pi}{L}\right)\left(k_z+\frac{r\pi}{L}\right)}\right]
\end{equation}
with
\begin{equation} \xi_r=\begin{cases}1,& r=1,5,9,... \\ -1,& r=3,7,11,... \\ 0,& \mathrm{else} \end{cases}, \\ \xi_r'=\begin{cases}-1,& r=2,4,6,... \\ 1,& r=4,8,12,... \\ 0,& \mathrm{else} \end{cases} \end{equation}
Now let the wire antenna be oriented along an \emph{arbitrary} direction $\bold{\hat{a}}$ relative to the $(x,y,z)$ system (i.e., $\bold{\hat{\dot{z}}}$ relative to the now-rotated antenna system) and with $J_c=J_c'\delta(\dot{x})\delta(\dot{y})$ and $J_s=J_s'\delta(\dot{x})\delta(\dot{y})$, where $\dot{\bold{r}}=(\dot{x},\dot{y},\dot{z})$ and $\dot{\bold{k}}=(\dot{k}_x,\dot{k}_y,\dot{k}_z)$ represent the position and wave vectors (resp.) in the antenna's local coordinates. To effect the spectral-domain current's representation in the original $(x,y,z)$ system, one can use polar and azimuthal rotation angles $\alpha$ and $\beta$ (resp.), along with their respective individual rotation matrices
\begin{equation}
\bold{\bar{U}}_{\alpha}=\begin{pmatrix} \cos{\alpha} & 0 & \sin{\alpha} \\ 0 & 1 & 0 \\ -\sin{\alpha} & 0 & \cos{\alpha}  \end{pmatrix}, \\ \,\,\, \,\,\,\,\, \bold{\bar{U}}_{\beta}=\begin{pmatrix} \cos{\alpha} & -\sin{\alpha} & 0  \\ \sin{\alpha} & \cos{\alpha} & 0 \\ 0 & 0 & 1  \end{pmatrix}
\end{equation}
and the composite rotation matrix $\bold{\bar{U}}=\bold{\bar{U}}_{\beta}\cdot \bold{\bar{U}}_{\alpha}$. Having defined these rotation matrices, we observe that $\bold{\dot{k}}$ and $\bold{\dot{r}}$ transform as $\bold{k}=\bold{\bar{U}} \cdot \bold{\dot{k}}$ and $\bold{r}=\bold{\bar{U}} \cdot \bold{\dot{r}}$, respectively.

Now we comment on some features of the resulting spectral-domain integral solution. Using Cauchy's Integral Theorem, the $n$th modal direct field residue ($n$=1,2,3,4) due to the source in layer $M$ writes as\footnote{Note that Adj($\bold{\bar{M}}$) is the adjugate (not adjoint) of matrix $\bold{\bar{M}}$ \cite{adjugate}, and Det($\bold{\bar{M}}$) is the determinant.}
\begin{equation}\label{res1}
\tilde{a}_{M,n}\bold{\tilde{e}}_{M,n}=2\pi i \left[\left(k_z-\tilde{k}_{M,nz}\right)ik_0\eta_0\mathrm{Adj}\left(\bold{\tilde{\bar{A}}}\right)\cdot \bold{\tilde{J}}(\bold{k})\mathrm{e}^{ik_zz}/\mathrm{Det}\left(\bold{\tilde{\bar{A}}}\right)\right]\Bigg|_{k_z=\tilde{k}_{M,nz}}
\end{equation}
for an arbitrary electric source distribution. The particular case of a vertically-oriented electric linear antenna centered at $\bold{r}_o'=(x_o,y_o,z_o)$ writes as
\begin{multline}\label{res2}
\tilde{a}_{M,n}\bold{\tilde{e}}_{M,n}=2\pi k_0\eta_0 \mathrm{Adj}\left(\bold{\tilde{\bar{A}}}\right)\cdot \bold{\hat{z}}\mathrm{e}^{ik_zz-i\bold{k}\cdot\bold{r}_o'} \left(k_z-\tilde{k}_{M,nz}\right) \times \\
\sum_{r=1}^{\infty}\left[r\frac{\xi_r J_c'(2r-1)\cos{\left(\frac{k_z L}{2}\right)}+i\xi_r' J_s'(2r)\sin{\left(\frac{k_z L}{2}\right)}}{L\left(k_z-\frac{r\pi}{L}\right)\left(k_z+\frac{r\pi}{L}\right)}\right] \Bigg|_{k_z=\tilde{k}_{M,nz}}
\end{multline}
where the entire expression for $\tilde{a}_{M,n}\bold{\tilde{e}}_{M,n}$ (not just the bracketed portion) is subject to evaluation at a particular eigenvalue $\tilde{k}_{M,nz}$. As can be observed in \eqref{res1}-\eqref{res2}, the distributed source spectrum manifests as a multiplicative sinc function-like ``taper" augmenting (and accelerating, versus $\{k_x,k_y\}$, the decay of) the computed direct field modal amplitudes of a Hertzian dipole \cite{sainath}. Finally, we should note that despite the factor $(k_z-r\pi/L)(k_z+r\pi/L)$ in the denominator of $\bold{\tilde{J}}(\bold{k})$ in \eqref{FT2} and \eqref{res2}, the values $k_z=\pm r\pi/L$ are \emph{not} poles because of the zeros in the numerator at those same points~\cite{clemmow}.
\subsection{\label{aper}Aperture Antennas}
Now consider a rectangular aperture possessing a tangential EM field distribution on its plane that one recasts, via the equivalence theorem, as a tangentially-polarized, surface-confined, magneto-electric current distribution \cite{balanis1}. Assuming the aperture's (1) principal axes are parallel with the $x$ and $y$ axes with principal lengths $L_x$ and $L_y$ (resp.) and (2) central location is $\bold{r}_o'=\bold{0}$, its spectral-domain representation readily follows from \eqref{FT2} upon letting $r$ and $q$ be the modal current indices describing current amplitude oscillation along $x$ and $y$ (resp.). However, unlike the linear antenna supporting a physical current that must vanish at the wire's ends, there are no such restrictions on the aperture's equivalent currents. Therefore, there may be both sinusoidal and cosinusoidal modal variations for each modal index $r > 0$ and modal index $q > 0$ in addition to a ``DC" term comprising a constant current amplitude sheet. Letting $J_{a,p',q'}'(r,q)$ stand for the (possibly complex-valued) Fourier coefficient of a current mode with constant current direction $\bold{\hat{a}}$ ($a$=$x$ or $y$), either sinusoidal ($p'$=$s$) or cosinusoidal ($p'$=$c$) current variation along the $x$ direction, and either sinusoidal ($q'$=$s$) or cosinusoidal ($q'$=$c$) current variation along the $y$ direction, then upon defining the sub-expressions
\begin{align}
\zeta(\bold{r})&=\mathrm{rect}\left(\frac{x}{L_x}\right)\mathrm{rect}\left(\frac{y}{L_y}\right)\delta(z) \\
\bm{\mathcal{J}}_1(r,q,\bold{r})&=\zeta(\bold{r})\left[\bold{\hat{x}}J_{x,s,s}'(r,q)\sin{\frac{r\pi x}{L_x}}\sin{\frac{q\pi y}{L_y}}+\bold{\hat{y}}J_{y,s,s}'(r,q)\sin{\frac{r\pi x}{L_x}}\sin{\frac{q\pi y}{L_y}}\right] \\ \bm{\mathcal{J}}_2(r,q,\bold{r})&=\zeta(\bold{r})\left[\bold{\hat{x}}J_{x,s,c}'(r,q)\sin{\frac{r\pi x}{L_x}}\cos{\frac{q\pi y}{L_y}}+\bold{\hat{y}}J_{y,s,c}'(r,q)\sin{\frac{r\pi x}{L_x}}\cos{\frac{q\pi y}{L_y}}\right]\\
\bm{\mathcal{J}}_3(r,q,\bold{r})&=\zeta(\bold{r})\left[\bold{\hat{x}}J_{x,c,s}'(r,q)\cos{\frac{r\pi x}{L_x}}\sin{\frac{q\pi y}{L_y}}+\bold{\hat{y}}J_{y,c,s}'(r,q)\cos{\frac{r\pi x}{L_x}}\sin{\frac{q\pi y}{L_y}}\right]\\ \bm{\mathcal{J}}_4(r,q,\bold{r})&=\zeta(\bold{r})\left[\bold{\hat{x}}J_{x,c,c}'(r,q)\cos{\frac{r\pi x}{L_x}}\cos{\frac{q\pi y}{L_y}}+\bold{\hat{y}}J_{y,c,c}'(r,q)\cos{\frac{r\pi x}{L_x}}\cos{\frac{q\pi y}{L_y}} \right]\end{align}
one has the following expression for the equivalent aperture currents in the space domain:
\begin{equation}\label{FT3a}
\bm{\mathcal{J}}(\bold{r})=\bm{\mathcal{J}}_4(0,0,\bold{r})+\sum_{r=1}^{\infty}\sum_{q=1}^{\infty}\sum_{p=1}^{4} \bm{\mathcal{J}}_{p}(r,q,\bold{r})
\end{equation}
with associated spectral-domain representation
\begin{equation}\label{FT3b}
\tilde{\bold{J}}(\bold{k})=\bold{\tilde{J}}_4(0,0,\bold{k})+\sum_{r=1}^{\infty}\sum_{q=1}^{\infty}\sum_{p=1}^{4}\bold{\tilde{J}}_p(r,q,\bold{k})
\end{equation}
 Analogous to the linear antenna, a more general aperture plane orientation can be effected using appropriate rotation matrices to represent arbitrarily-oriented rectangular aperture antennas. Further akin to the wire antenna case, we observe again the manifestation of a tapering in the field's Fourier spectrum (except now along both $k_x$ and $k_y$), the property of the distributed field computation imparting a (deceptively simple) multiplicative factor into the computed Hertzian dipole direct field modal amplitudes, the presence of (now four) fictitious poles, and the vulnerability of numerical instability when the observation point lies within the region $(|x| < L_x/2) \cup (|y| < L_y/2)$. The latter instability aspect, when one is detouring into the upper-half $k_x$ and $k_y$ planes, is mitigated in the same manner to that shown concerning linear antennas, i.e., via placing the natural logarithm of one or both of the Laguerre-Gauss quadrature weights into the exponentials prior to evaluating them, as well as only evaluating the scattered fields. Now, however, due to the multiplication of two sinusoid-type functions in the spectral domain one will have for each current mode functional dependance (i.e., cosinusoidal along both $x$ and $y$, etc.) four exponentials into which one places the (natural logarithm of the) quadrature weights, rather than two exponentials in the case of wire antennas.
\subsection{\label{WireAntVal}Validation Results: Linear Antennas}
In this subsection we first show results concerning the fields radiated by an infinitesimally thin linear (wire) antenna radiating at $f$=30MHz in unbounded free space. We set the wire antenna's length at half the free space wavelength (wire length $L=\lambda_0/2\sim$5m), partition free space into three fictitious layers, place the antenna in the 5m-thick central layer, position the observation point always either in the top or bottom layer to compute the total field, and restrict attention to an electric current distribution with mode index $r=1$. The radial distance between the antenna's center and all observation points is held fixed at $|\bold{r}-\bold{r}'|=50$m, while the adaptive integration error tolerance was set to $1.2 \times 10^{-4}$.

\begin{figure}[H]
\centering
\subfloat[\label{E2a}]{\includegraphics[width=2.8in]{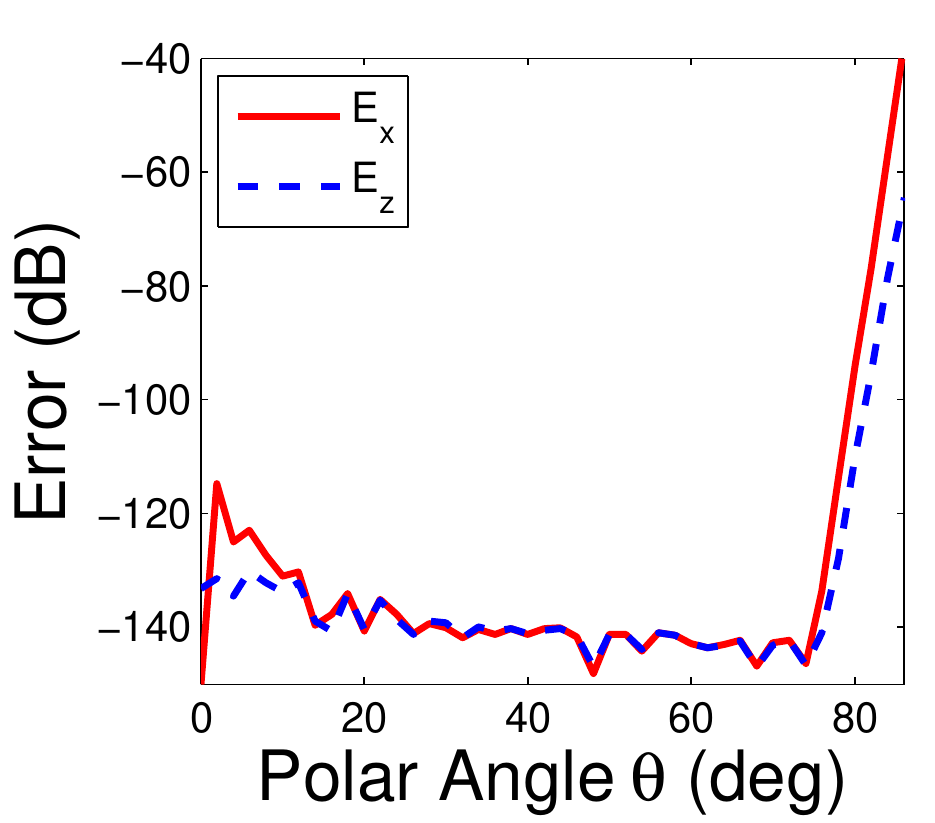}}
\subfloat[\label{E2b}]{\includegraphics[width=2.8in]{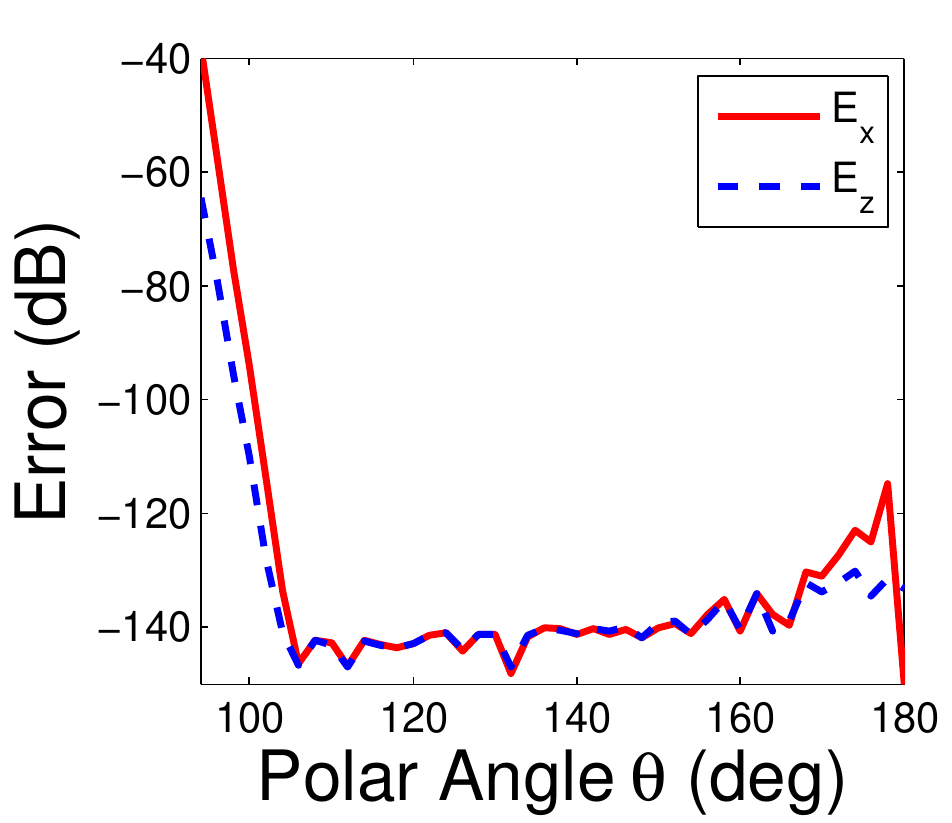}}
\caption{\small \label{E2}(Color online) Accuracy of $E_x$ and $E_z$, versus polar angle $\theta$, for the vertical wire antenna. Reference results computed using expressions from \cite{balanis1}.}
\end{figure}

Figure \ref{E2} shows the accuracy of the electric field, radiated by a vertically-oriented wire antenna, versus polar angle $\theta$ for a fixed azimuthal observation angle $\phi=0^{\circ}$. Note that for $|\bold{r}-\bold{r}'|=50$m the sampled polar angles $\theta=$$88^{\circ}$, $90^{\circ}$, and $92^{\circ}$ correspond to observation points lying within the central free space layer and thus zero scattered-field result. Thus, the polar angle sweep data is shown sub-divided into two plots to remove the artificial discontinuity in the data (versus $\theta$). We see that between $\theta=0^{\circ}-76^{\circ}$ and $\theta=104^{\circ}-180^{\circ}$, one realizes an accuracy of between thirteen to fourteen digits in $E_z$. An analogous statement applies for the error in $E_x$ except at $\theta=0^{\circ}$ and $\theta=180^{\circ}$, where the algorithm's computed solution (to within machine precision) and closed form solution yield answers for $E_x$ having magnitude equal to zero (hence the error's coercion to -150dB). We notice that the accuracy degrades as the polar observation angle tends towards horizon, but the algorithm still manages to deliver results accurate to approximately four digits. This trend is qualitatively consistent with the results in Section \ref{ScatVal}, where instead the field and observation points were in the same layer. Further extensive error studies (not shown herein) were also performed to better characterize the algorithm's performance, which consisted of all the following parameter permutations\footnote{The dot over the field component directions denotes components expressed with respect to the antenna's local (rotated) coordinate system.}: $(E_{\dot{x}},E_{\dot{y}},E_{\dot{z}},H_{\dot{x}},H_{\dot{y}})\times(\alpha=0^{\circ},\alpha=45^{\circ},\alpha=90^{\circ})\times (\theta=45^{\circ},\theta=135^{\circ})\times (\phi=0^{\circ},\phi=45^{\circ},\phi=90^{\circ},\phi=135^{\circ},\phi=180^{\circ},\phi=225^{\circ},\phi=270^{\circ},\phi=315^{\circ})$. The results in all these permutations indicated error ranging between -130dB to -140dB.
To gain a better understanding of the computational efficiency realized with our proposed distributed radiator simulation approach, we examined the time required to compute the total field $E_z$ radiated by a vertically-oriented, half-wavelength wire antenna ($L=\lambda_0/2$) which radiates at $f=$30MHz, is centered at the origin in free space, and has current distribution $\bm{\mathcal{J}}(\bold{r})=\bold{\hat{z}}\mathrm{cos}(\pi z/L)\mathrm{rect}\left(z/L\right)\delta(x)\delta(y)$ (i.e., the current variation is characterized by the harmonic $r=1$); the observation points examined were $\bold{r}=(0,0,50)$ [m] and $\bold{r}=50(\mathrm{cos}80^{\circ},0,\mathrm{sin}80^{\circ})$ [m]. In particular, we compared the time required to obtain the field solution from using the proposed spectral-domain approach versus performing space-domain radiation integral evaluation via Legendre-Gauss quadrature.\footnote{That is, with the Gauss quadrature nodes being the equivalent Hertzian dipole locations on $V'$.} Indeed, our study revealed that a factor of approximately \emph{one order of magnitude} in acceleration can be realized.\footnote{We found $N_{\mathrm{avg}}\sim10$ Hertzian dipole sampling points were required to achieve at most -100dB error in our study.} Furthermore, by the sampling theorem one reasons that when $r$ significant spatial current harmonics are required to adequately capture the spatial variation of current on $V'$, for a comparable accuracy with our proposed approach one would require approximately $r \times N_{\mathrm{avg}}$ Hertzian dipole sampling points (as compared to just $r$ evaluations with our approach).

\begin{figure}[H]
\centering
\subfloat[\label{geoA2}]{\includegraphics[width=2.9in]{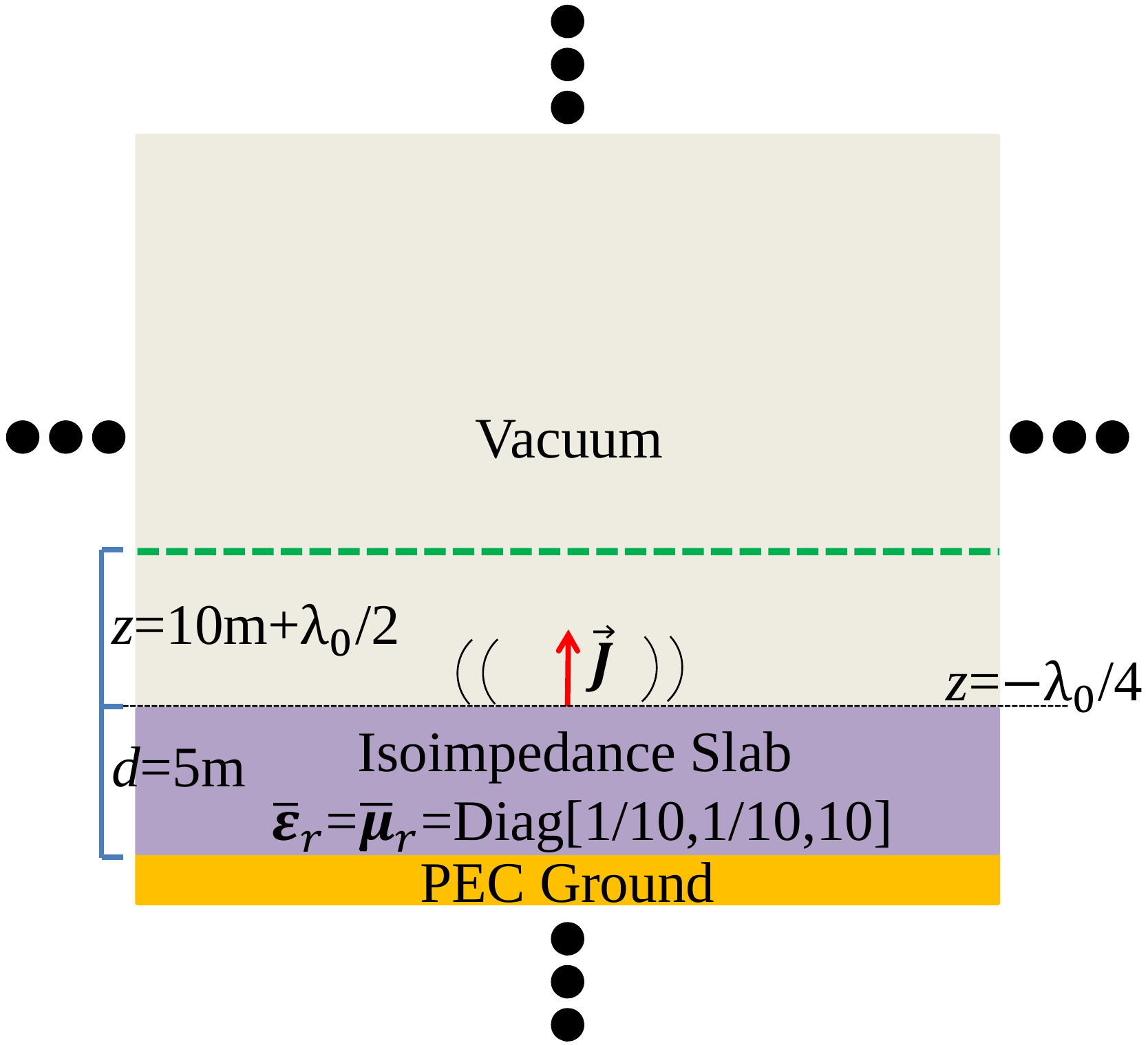}}
\subfloat[\label{geoB2}]{\includegraphics[width=2.9in]{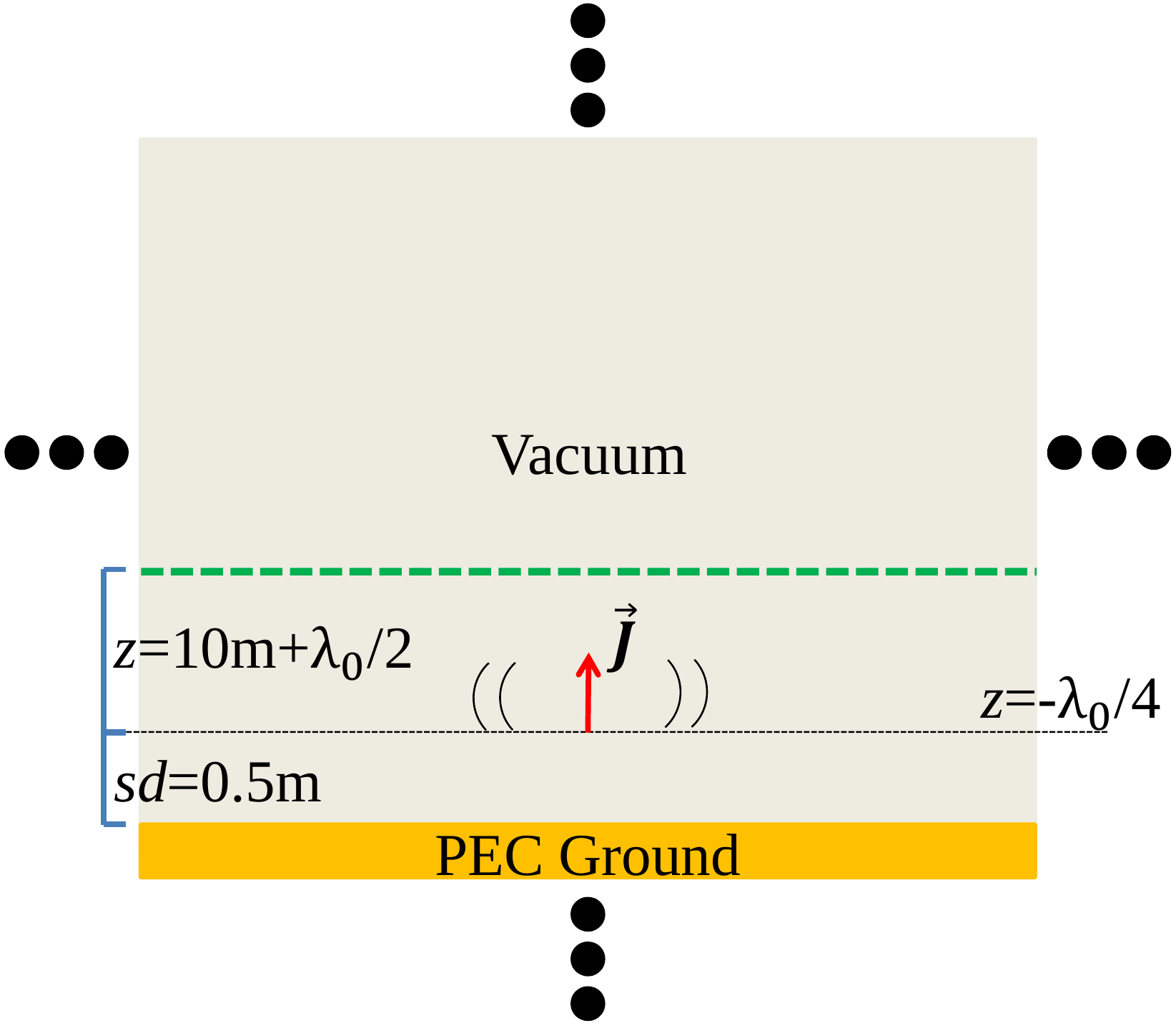}}

\subfloat[\label{geoC2}]{\includegraphics[width=2.9in]{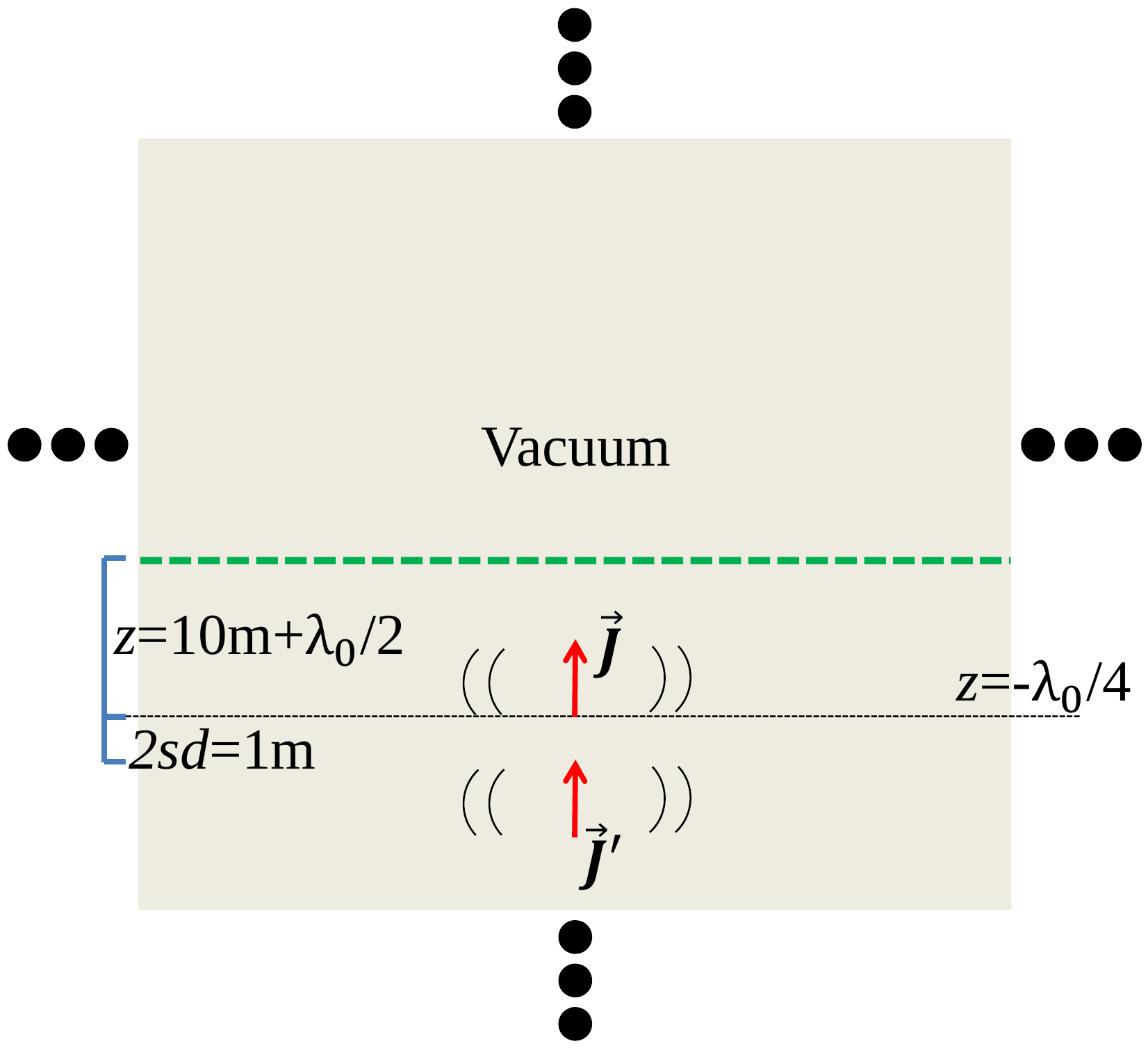}}
\caption{\small \label{geo2} (Color online) $xz$ plane view of the three problem geometries leading to an identical field distribution in the region $z\geq 0$. Physical arguments grounded in Transformation Optics theory and the form invariance of Maxwell's Equations \cite{pendry1,teixeiraJEWA} lead to equivalence in the field distributions between the first two sub-figures (for $z\geq -\lambda_0/4$). On the other hand, image theory-based considerations lead to equivalence in the field distributions between the latter two sub-figures (again, for $z\geq -\lambda_0/4$). We plot the field distribution ($E_z$) on a flat observation plane, residing at $z=10$m, occupying the region $\{-5\leq x \leq 5, -5 \leq y \leq 5\}$ [m] (i.e., at the elevation of the dashed green line seen in the above three $xz$ plane views).}
\end{figure}

To validate the algorithm's capability to simulate fields radiated by wire antennas in planar-layered, anisotropic media, consider a vertically-oriented wire antenna which radiates at $f=10$MHz, has vertical length $L=\lambda_0/2\sim 15$m, is centered at $\bold{r}_o=(0,0,0)$, and supports an electric current filament with distribution $\bm{\mathcal{J}}(\bold{r})=\bold{\hat{z}}\cos(\pi z/L)\mathrm{rect}\left(z/L\right)\delta(x)\delta(y)$. The wire antenna resides in the top vacuum layer $z\geq -\lambda_0/4$, the PEC ground plane\footnote{Akin to the image theory study done in Section \ref{ScatVal}, for the sole purpose of facilitating the present study we coerce, within the code, the TE$_z$ and TM$_z$ reflection coefficients concerning down-going fields impinging upon the ground plane.} half-space occupies the region $z\leq -(d+\lambda_0/4)$ (the longitudinal spacing between the wire antenna and ground plane is $d=5$m), and a ground plane-coating substrate with material properties $\boldsymbol{\bar{\epsilon}}_r=\boldsymbol{\bar{\mu}}_r=\mathrm{diag}\left[s,s,1/s\right]$ ($s=1/10$) occupies the region $-(d+\lambda_0/4) \leq z \leq -\lambda_0/4$. See Figure \ref{geo2}.

The use of this coating layer has the special properties of being perfectly impedance-matched to free space for all plane wave incidence angles and polarizations (hence the name ``isoimpedance" medium \cite{teixeira1}), as well as (equivalently) effecting the metric expansion of space within the isoimpedance layer by a factor of $s$ \cite{teixeiraJEWA,pendry1}. As a consequence, no reflections arise at the isoimpedance/vacuum interface, while plane waves traversing this $d$-meter thick layer exit it having accumulated a (in general complex-valued) phase commensurate with having traversed a $sd$-meter thick region of free space \cite{pendry1,teixeira1}. Equivalently (for our problem), the field solution at some point in the vacuum region $\bold{r}=(x,y,z\geq -\lambda_0/4)$ is \emph{exactly identical} to the field solution at the same $\bold{r}$ but in a two-layer vacuum/PEC ground environment where the semi-infinitely thick PEC ground half-space occupies instead the region $z \leq -(sd+\lambda_0/4)$ (see Figure \ref{geoB2}). Subsequently recalling image theory, we conclude that the solution (in the vacuum/PEC geometry) at $\bold{r}=(x,y,z\geq -\lambda_0/4)$ in turn is identical to the solution at the same $\bold{r}$ in a homogeneous vacuum but with two sources \cite{balanis1}: the original wire antenna along with the image wire source $\bm{\mathcal{J}}'(\bold{r})=\bold{\hat{z}}\cos(\pi (z+d')/L)\mathrm{rect}\left((z+d')/L\right)\delta(x)\delta(y)$, where $d'=\lambda_0/2+2sd$; see Figure \ref{geoC2}. Alternatively stated, setting $s<1$ makes the wire antenna radiate in the region $z\geq -\lambda_0/4$ as if the ground plane were moved closer to the base of the antenna while setting $s>1$ causes the wire antenna to radiate in the region $z\geq -\lambda_0/4$ as if the ground plane were moved further downward in the $z$ direction. Furthermore, the ``scattered" field comprises the direct field distribution, within the region $z\geq -\lambda_0/4$, established by the image source $\bm{\mathcal{J}}'(\bold{r})$.
\begin{figure}[H]
\centering
\subfloat[\label{f0a}]{\includegraphics[width=3in]{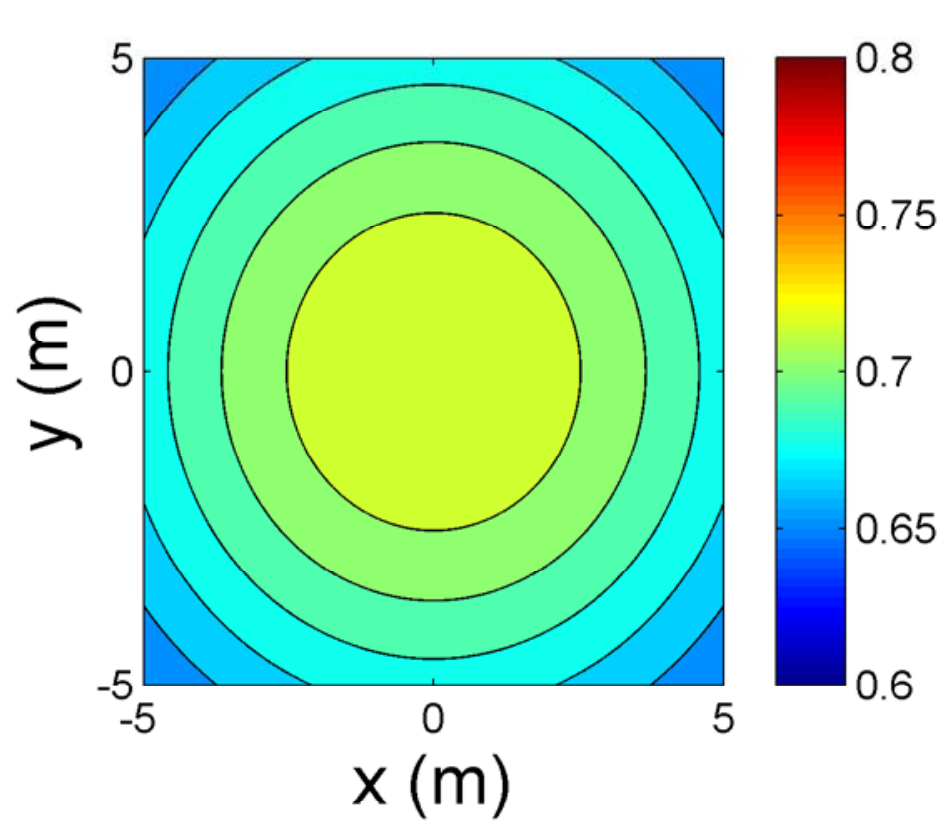}}
\subfloat[\label{f0b}]{\includegraphics[width=3in]{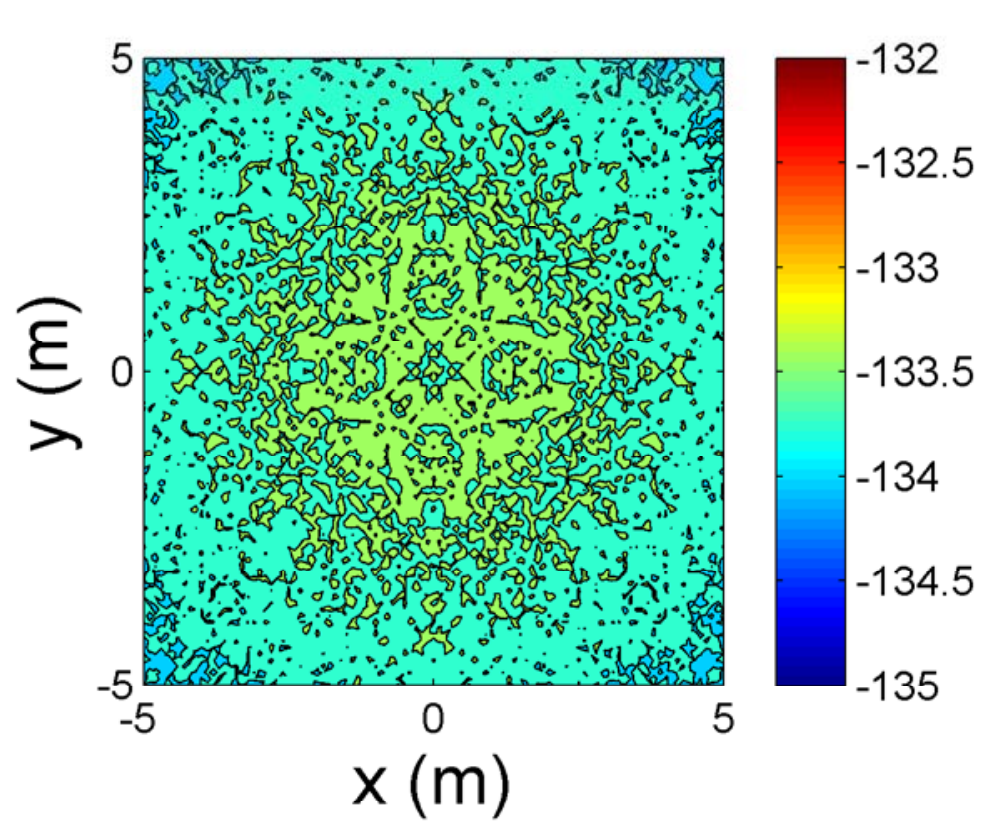}}
\caption{\small \label{field} (Color online) Algorithm-computed electric field $E_z$ distribution (Figure \ref{f0a}) and relative error $10\mathrm{log}_{10}|(E_z-E_z^{\mathrm{val}})/E_z^{\mathrm{val}}|$~[dB] (Figure \ref{f0b}) in the region $\{-5\leq x \leq 5, -5 \leq y \leq 5,z=10\}$~[m]. $E_z^{\mathrm{val}}$ is the closed-form, scattered-field result comprising the image wire current source's radiated field.}
\end{figure}
\subsection{\label{ApAntVal}Validation Results: Aperture Antennas}
In this subsection we first exhibit validation results concerning rectangular aperture antennas radiating at $f$=30MHz in free space. We set the aperture antenna's dimensions along the principal directions as ($L=L_x=L_y=\lambda_0/2$), partition free space into three fictitious layers with the antenna in the central layer of 2m thickness, set the observation point always either in the top or bottom layer to compute the direct field, restrict attention to the $r$ modal index value $r=$1 when cosinusoidal electric current variation along the $\dot{x}$ direction is present (and likewise, set $q$=1 for cosinusoidal electric current variation along $\dot{y}$), and assume cosinusoidal current variation along the direction of current flow.

\begin{figure}[H]
\centering
\subfloat[\label{ApE2a}]{\includegraphics[width=2.8in]{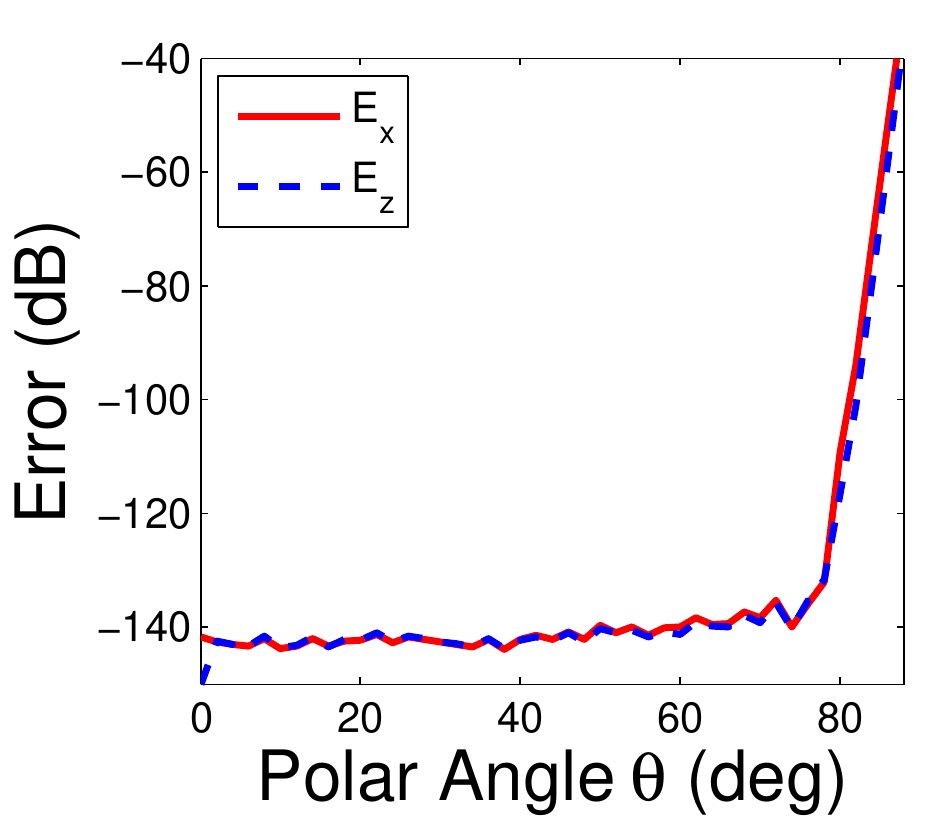}}
\subfloat[\label{ApE2b}]{\includegraphics[width=2.8in]{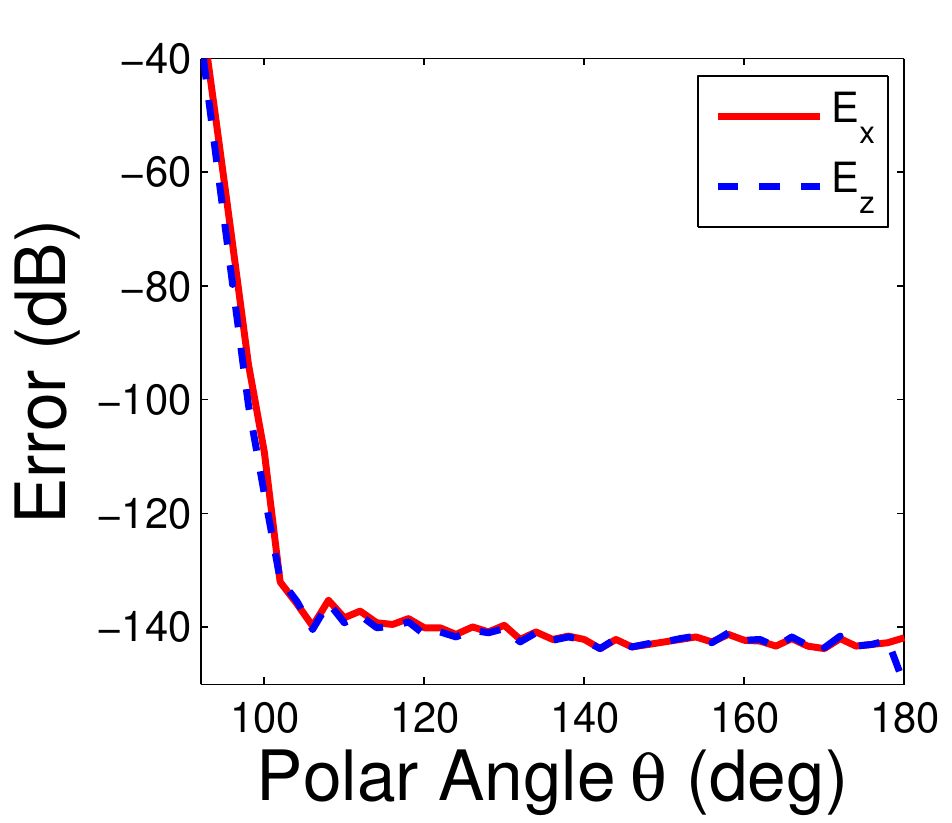}}

\subfloat[\label{ApE2c}]{\includegraphics[width=2.8in]{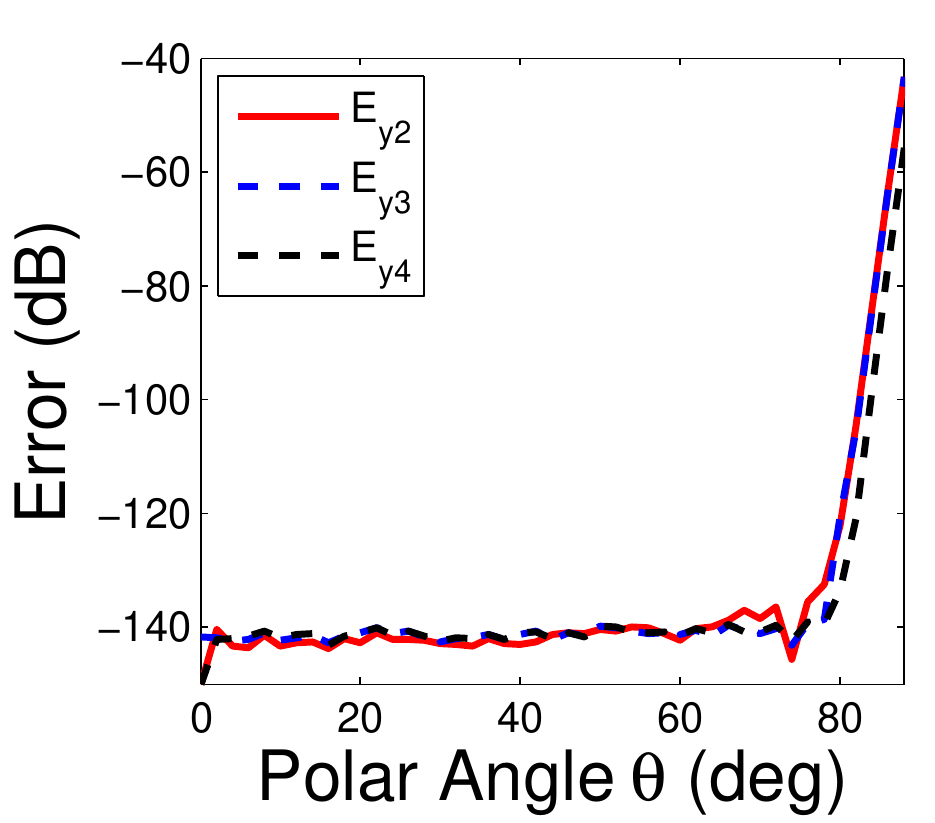}}
\subfloat[\label{ApE2d}]{\includegraphics[width=2.8in]{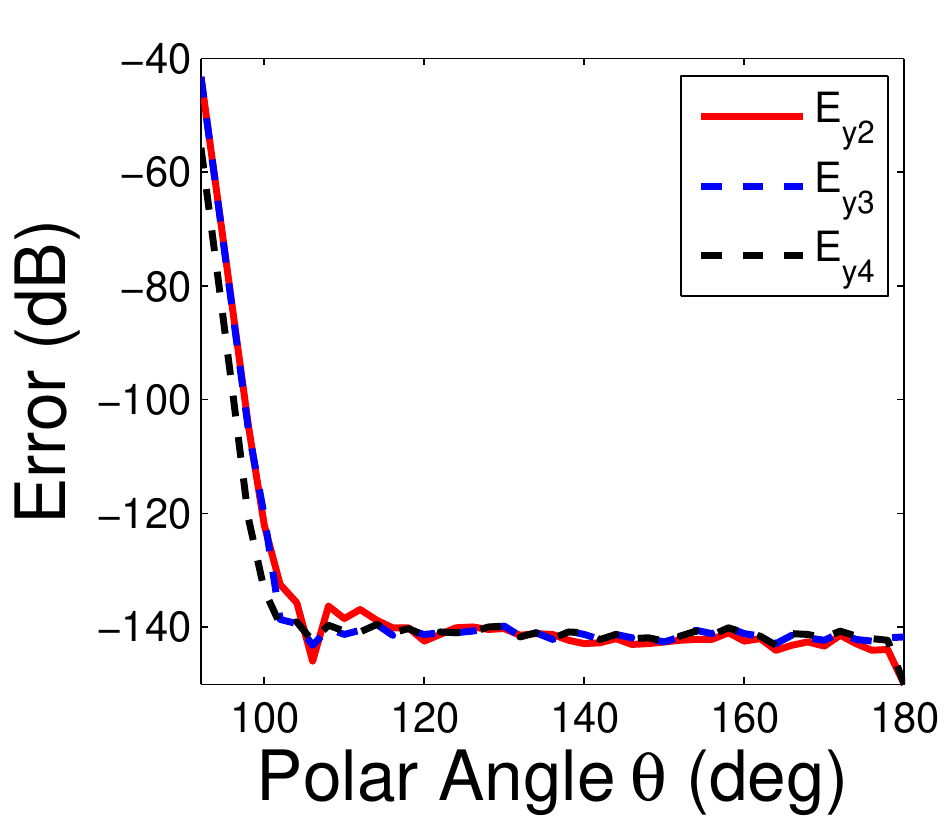}}
\caption{\small \label{ApE2}(Color online) $\epsilon(\bm{\mathcal{E}})$ versus $\theta$ for the aperture antenna. Reference results computed using expressions from \cite{balanis1}.}
\end{figure}

Figure \ref{ApE2} shows the accuracy of the electric field, radiated by an aperture antenna with area normal  $\bold{\hat{n}}=\bold{\hat{z}}$ and centered at $\bold{r}_0'=\bold{0}$, versus polar angle $\theta$ for a fixed azimuthal observation angle $\phi=0^{\circ}$. The radial distance between the center of the aperture and all observation points is held fixed at $|\bold{r}-\bold{r}'|=50$m, while the adaptive integration error tolerance was set to be $1.2 \times 10^{-4}$. Figures \ref{ApE2a}-\ref{ApE2b} concern an aperture with $\bold{\hat{x}}\cos(\pi x/L)\cos(\pi y/L)\mathrm{rect}(x/L)\mathrm{rect}(y/L)\delta (z)$ current amplitude pattern, while the line plots of $E_{y2}$, $E_{y3}$, and $E_{y4}$ in Figures \ref{ApE2c}-\ref{ApE2d} concern an aperture with $\bold{\hat{x}}\cos(\pi x/L)\sin(2\pi y/L)\mathrm{rect}(x/L)\mathrm{rect}(y/L)\delta (z)$, $\bold{\hat{y}}\cos(\pi x/L)\cos(\pi y/L)\mathrm{rect}(x/L)\mathrm{rect}(y/L)\delta (z)$, and $\bold{\hat{y}}\sin(2\pi x/L)\cos(\pi y/L)\mathrm{rect}(x/L)\mathrm{rect}(y/L)\delta (z)$ surface current distributions (resp.). Note that the sampled polar angle $\theta=90^{\circ}$ corresponds to an observation point lying within the central free space layer and thus yields a null scattered-field result. Thus, the polar angle sweep data (both for the first, as well as the latter three, current distribution cases) are shown sub-divided into two plots to remove the artificial discontinuity in the data (versus $\theta$). From Figures \ref{ApE2a}-\ref{ApE2b} we see that for $\theta \in [0^{\circ},78^{\circ}]$ and $\theta \in [102^{\circ},180^{\circ}]$, one realizes an accuracy of between thirteen to fourteen digits in $E_x$. An analogous statement applies for the error in $E_z$ excepting at $\theta=0^{\circ}$ and $\theta=180^{\circ}$, where the algorithm's computed solution (to within machine precision) and closed form solution yield answers for $E_z$ having magnitude equal to zero. We notice that the accuracy degrades as the polar observation angle tends towards horizon, but the algorithm still manages to yield results accurate to approximately three to four digits. Observing the three current cases in Figures \ref{ApE2c}-\ref{ApE2d} leads to similar conclusions for the only non-trivial electric field component $E_y$: accuracy for $E_{y2}$ and $E_{y3}$ is between thirteen to fourteen digits in the polar range $\theta \in [0^{\circ},78^{\circ}]$ and $\theta \in [102^{\circ},180^{\circ}]$, accuracy for $E_{y4}$ is between thirteen to fourteen digits in the polar range $\theta \in [0^{\circ},80^{\circ}]$ and $\theta \in [100^{\circ},180^{\circ}]$, all three accuracies degrade for observation points near to the horizon, and for $E_{y2}$ and $E_{y4}$ the accuracy results at $\theta=0^{\circ}$ and $\theta=180^{\circ}$ are coerced to -150dB since the computed results (to within machine precision) and validation results were of zero magnitude.
Akin to the wire antenna case, extensive error studies were performed to better characterize the algorithm's performance concerning rectangular aperture sources, which consisted of all the following parameter permutations: $(E_{\dot{x}},E_{\dot{y}},E_{\dot{z}},H_{\dot{y}},H_{\dot{z}})\times(\alpha=0^{\circ},\alpha=45^{\circ},\alpha=90^{\circ})\times (\theta=45^{\circ},\theta=135^{\circ})\times (\phi=0^{\circ},\phi=45^{\circ},\phi=90^{\circ},\phi=135^{\circ},\phi=180^{\circ},\phi=225^{\circ},\phi=270^{\circ},\phi=315^{\circ})$. The error results in all these permutations are between -130dB to -140dB. Again, the dot over the field component directions denotes components expressed with respect to the antenna's local (rotated) coordinate system. Furthermore, a baseline computational efficiency study was performed which is identical to the wire antenna study except we observe the electric field $E_x$ due to the source distribution $\bm{\mathcal{J}}(\bold{r})=\bold{\hat{x}}\cos(\pi x/L)\cos(\pi y/L)\mathrm{rect}(x/L)\mathrm{rect}(y/L)\delta (z)$. Analogous conclusions hold, except that now one realizes \emph{two orders of magnitude} in solution speed acceleration due to only one sinusoidal current harmonic required to represent a current sheet otherwise requiring (on the order of) $N_{\mathrm{avg}}=10 \times 10$ Hertzian dipole sampling points for comparable accuracy.

\begin{figure}[H]
\centering
\subfloat[\label{geoA}]{\includegraphics[width=2.9in]{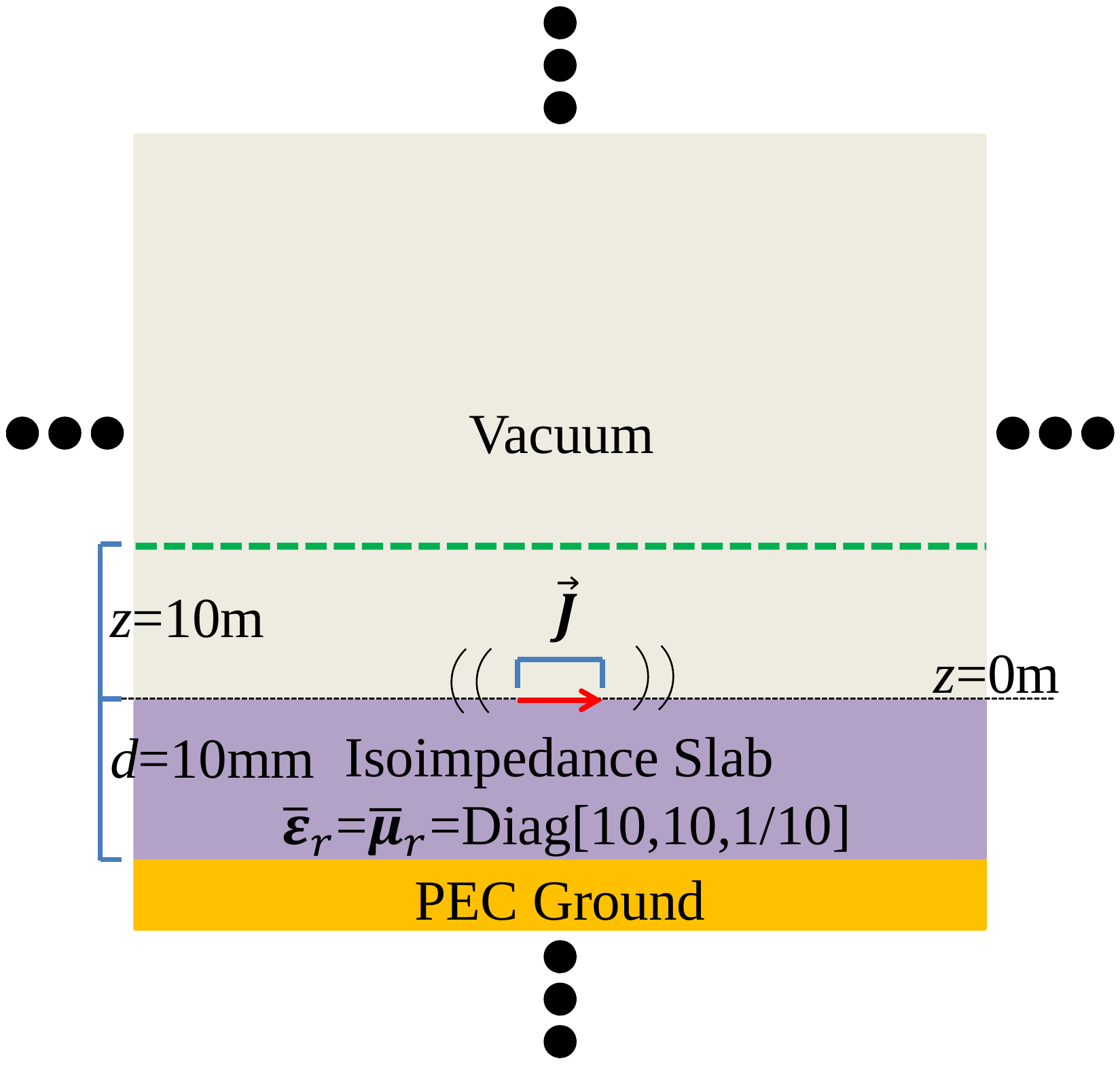}}
\subfloat[\label{geoB}]{\includegraphics[width=2.9in]{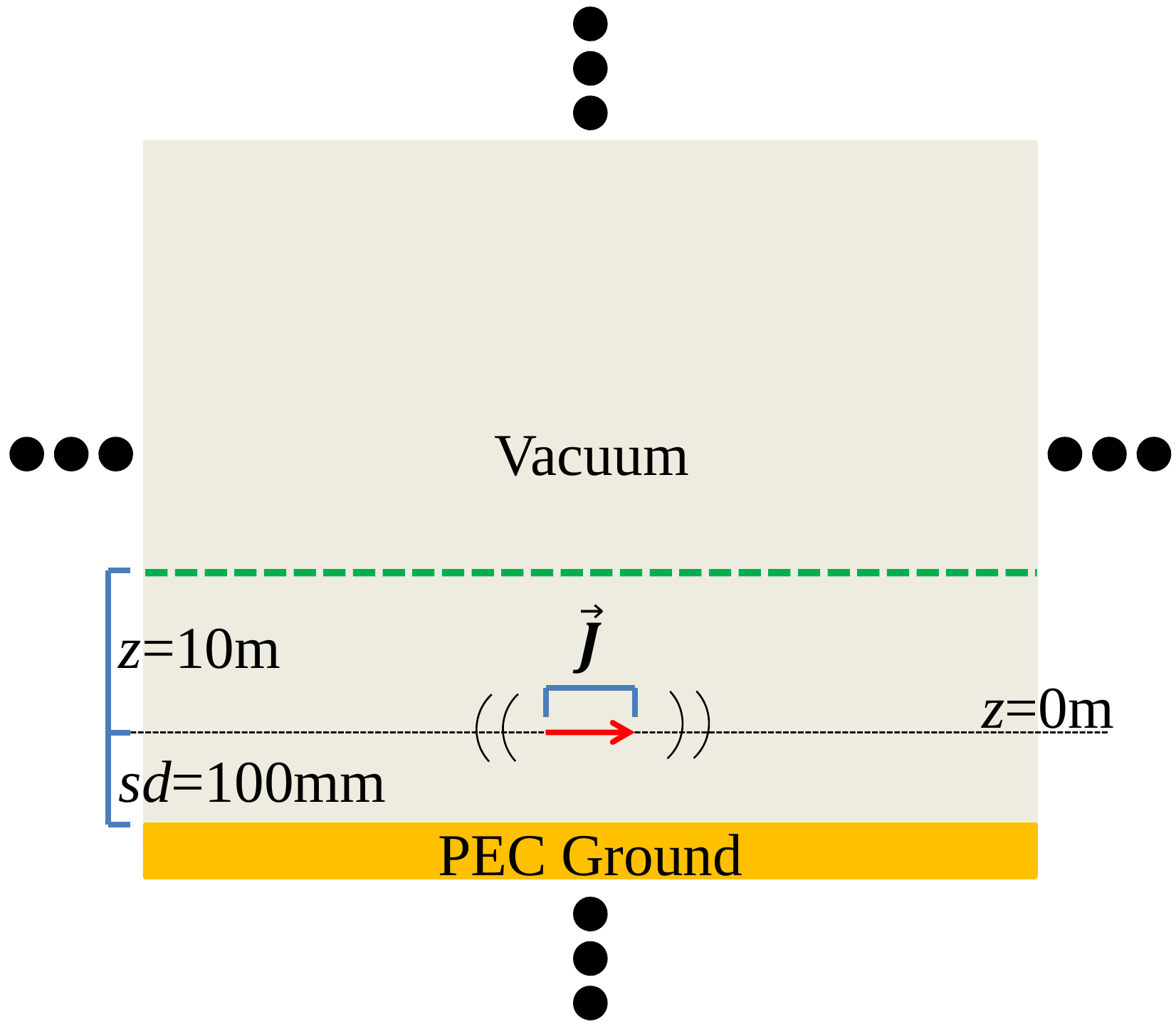}}

\subfloat[\label{geoC}]{\includegraphics[width=2.9in]{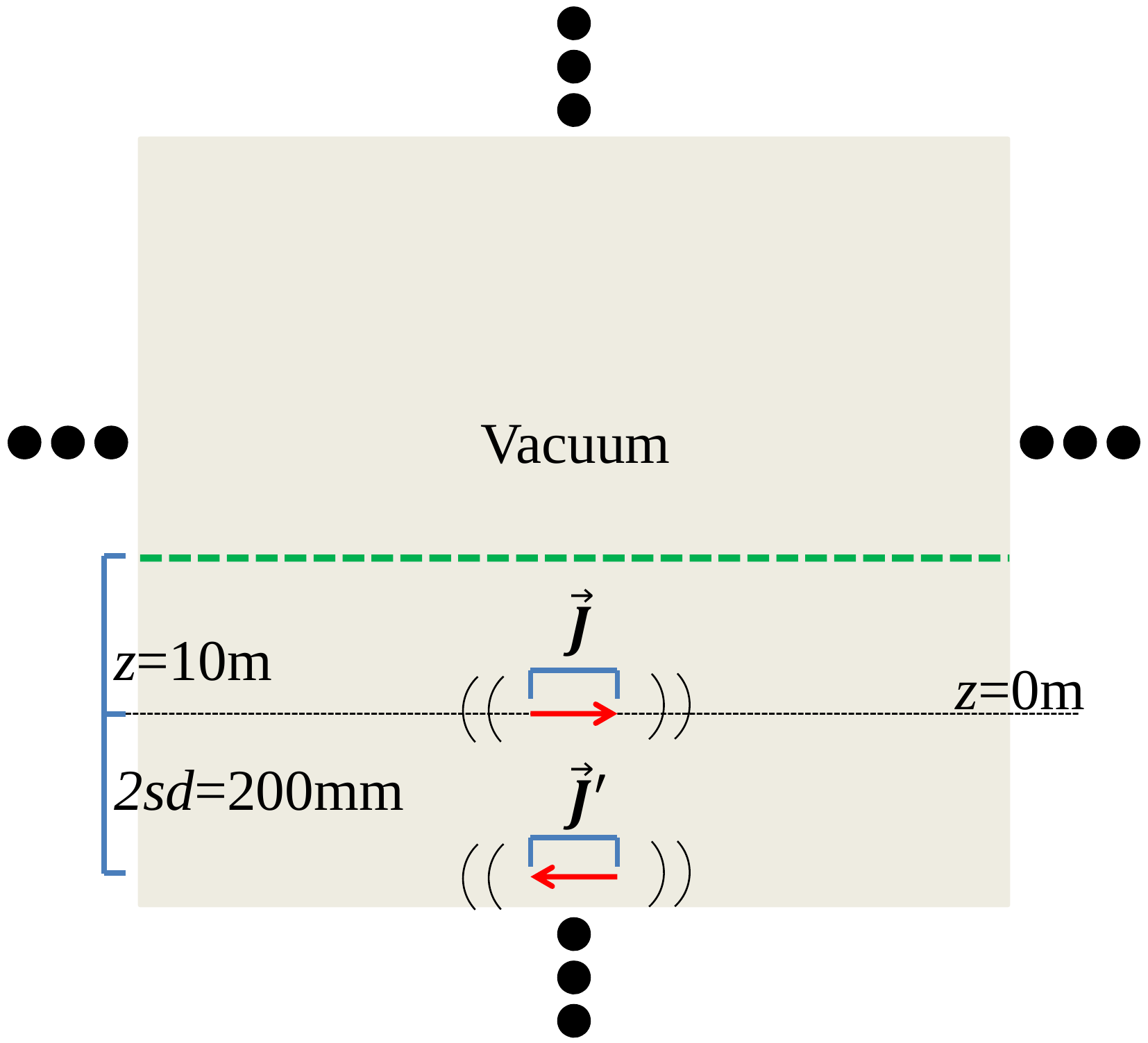}}
\caption{\small \label{geo} (Color online) $xz$ plane view of the three problem geometries leading to an identical field distribution in the region $z\geq 0$. Physical arguments grounded in Transformation Optics theory and the form invariance of Maxwell's Equations \cite{pendry1,teixeiraJEWA} lead to equivalence in the field distributions between the first two sub-figures (for $z\geq 0$). On the other hand, image theory-based considerations lead to equivalence in the field distributions between the latter two sub-figures (again, for $z\geq 0$). We plot the field distribution ($E_x$) on a flat observation plane, residing 10m above the aperture source, occupying the region $\{-10\leq x \leq 10, -4 \leq y \leq 4\}$ [m] parallel to the $xy$ plane (i.e., at the elevation of the dashed green line seen in the above three $xz$ plane views).}
\end{figure}

To validate the algorithm's capability to simulate fields radiated by aperture antennas in planar-layered, anisotropic media, consider a flat, rectangular-shaped aperture antenna radiating at $f=30$MHz, having dimensions $L_x=\lambda_0/2\sim5$m and $L_y=1$m, oriented such that it is parallel to the $xy$ plane, centered at $\bold{r}_o=(0,0,0)$, and supporting an electric current sheet with distribution $\bm{\mathcal{J}}(\bold{r})=\bold{\hat{x}}\cos(\pi x/L_x)\cos(\pi y/L_y)\mathrm{rect}(x/L_x)\mathrm{rect}(y/L_y)\delta(z)$. The aperture antenna resides in the top vacuum layer $z\geq 0$, the PEC ground plane\footnote{Akin to the image theory study done in Section \ref{ScatVal}, for the sole purpose of facilitating the present study we coerce, within the code, the TE$_z$ and TM$_z$ reflection coefficients concerning down-going fields impinging upon the ground plane.} half-space occupies the region $z\leq -d$ (the longitudinal spacing between the aperture antenna and ground plane is $d=10$mm), and a ground plane-coating layer with material properties $\boldsymbol{\bar{\epsilon}}_r=\boldsymbol{\bar{\mu}}_r=\mathrm{diag}\left[s,s,1/s\right]$ ($s=10$) occupies the region $-d \leq z \leq 0$ (see Figure \ref{geoA}). With $s=10$, the aperture will radiate into the region $z\geq 0$ as if the aperture-ground separation were in fact $sd$=100mm (see Figure \ref{geo}). In particular, the ``scattered" field comprises the field distribution, within the region $z\geq 0$, radiated by the image source $\bm{\mathcal{J}}'(\bold{r})=-\bold{\hat{x}}\cos(\pi x/L_x)\cos(\pi y/L_y)\mathrm{rect}(x/L_x)\mathrm{rect}(y/L_y)\delta(z+2sd)$.
\begin{figure}[H]
\centering
\subfloat[\label{f1}]{\includegraphics[width=3in]{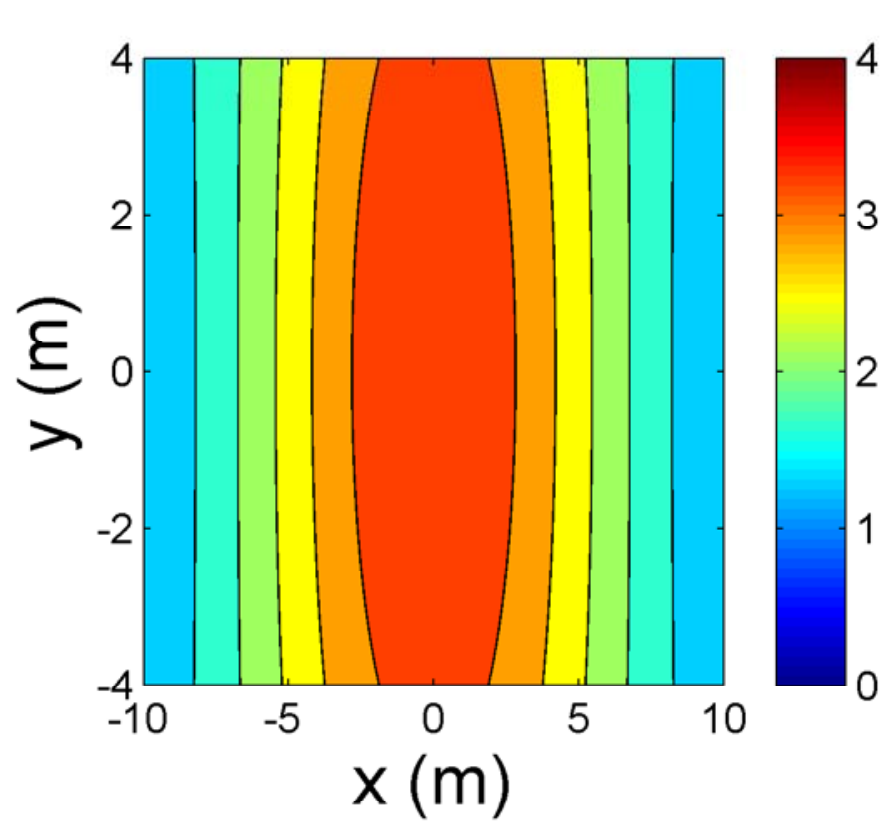}}
\subfloat[\label{f2}]{\includegraphics[width=3in]{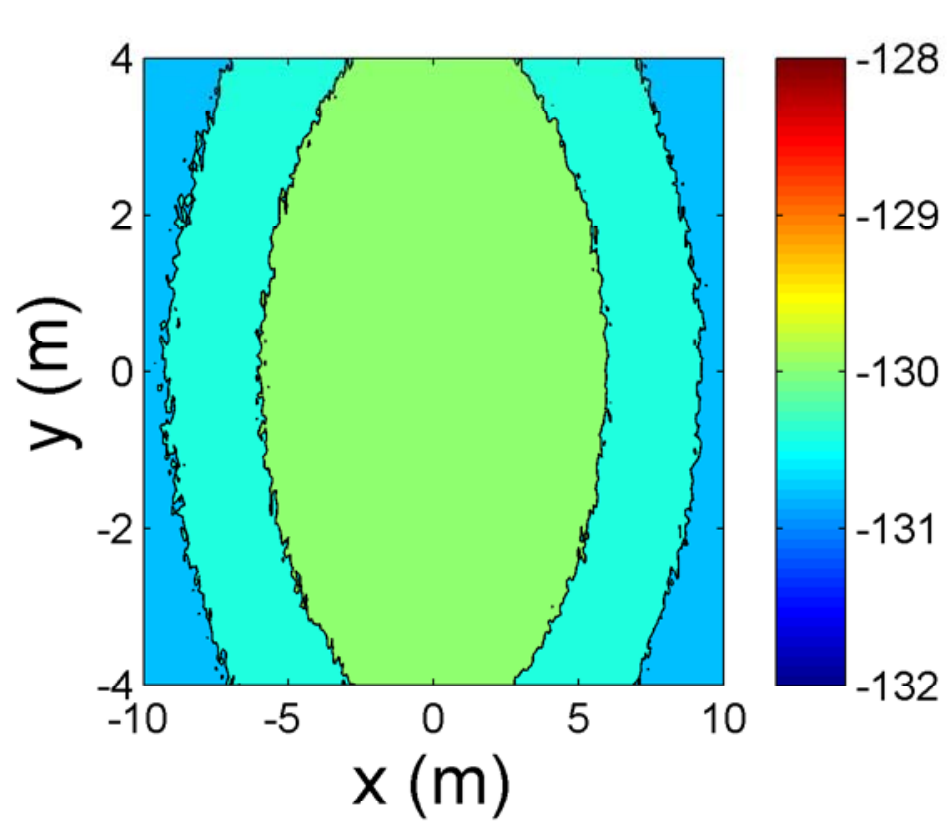}}
\caption{\small \label{field2} (Color online) Algorithm-computed electric field $E_x$ distribution (Figure \ref{f1}) and relative error $10\mathrm{log}_{10}|(E_x-E_x^{\mathrm{val}})/E_x^{\mathrm{val}}|$~[dB] (Figure \ref{f2}) in the region $\{-10\leq x \leq 10, -4 \leq y \leq 4,z=10\}$~[m]. $E_x^{\mathrm{val}}$ is the closed-form scattered-field result comprising the image aperture current source's radiated direct field.}
\end{figure}
\section{\label{hydro}Case Study: Marine Hydrocarbon Exploration}
\begin{figure}[H]
\centering
\subfloat[\label{CSEMa}]{\includegraphics[width=2.5in]{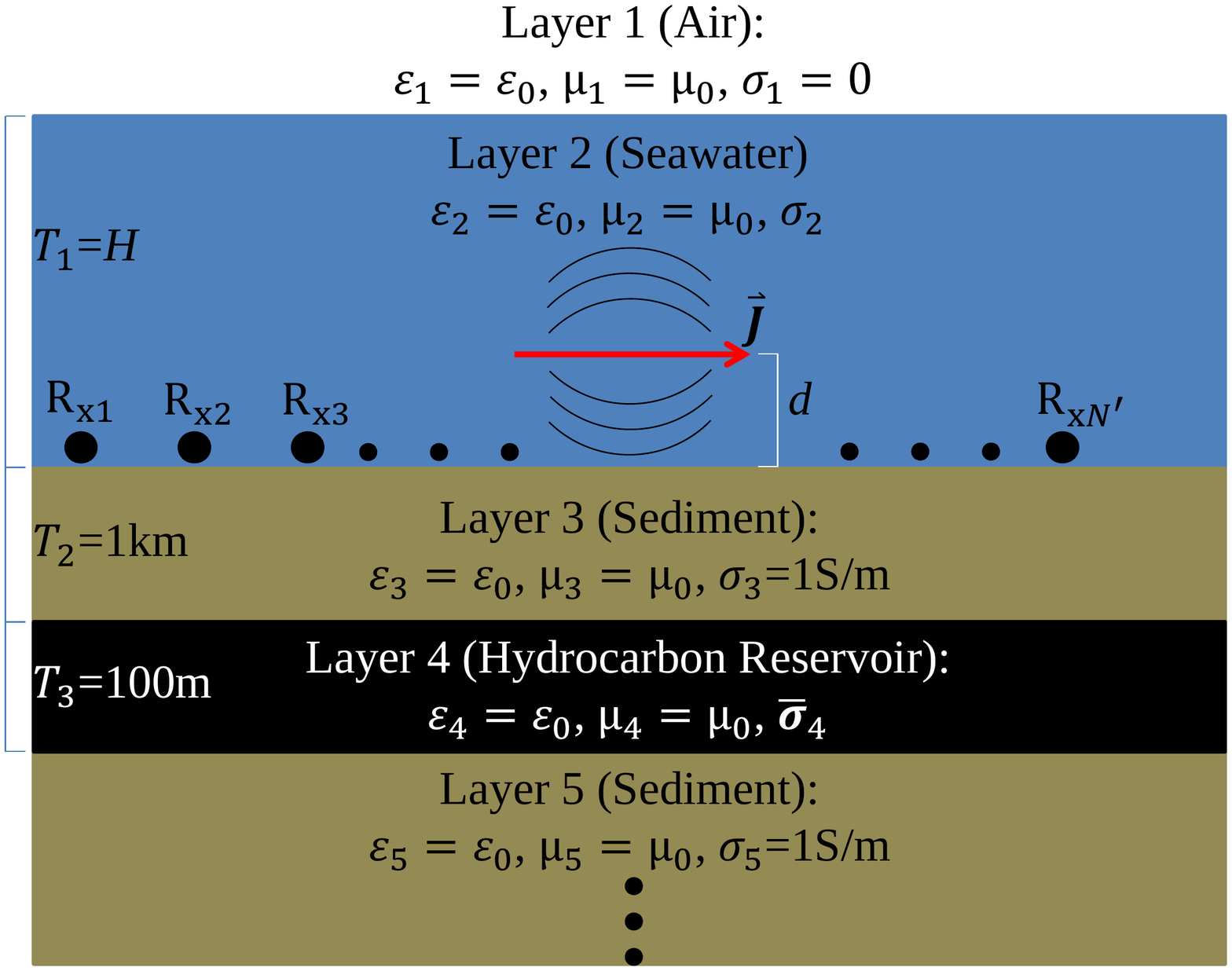}}
\subfloat[\label{CSEMb}]{\includegraphics[width=2.5in]{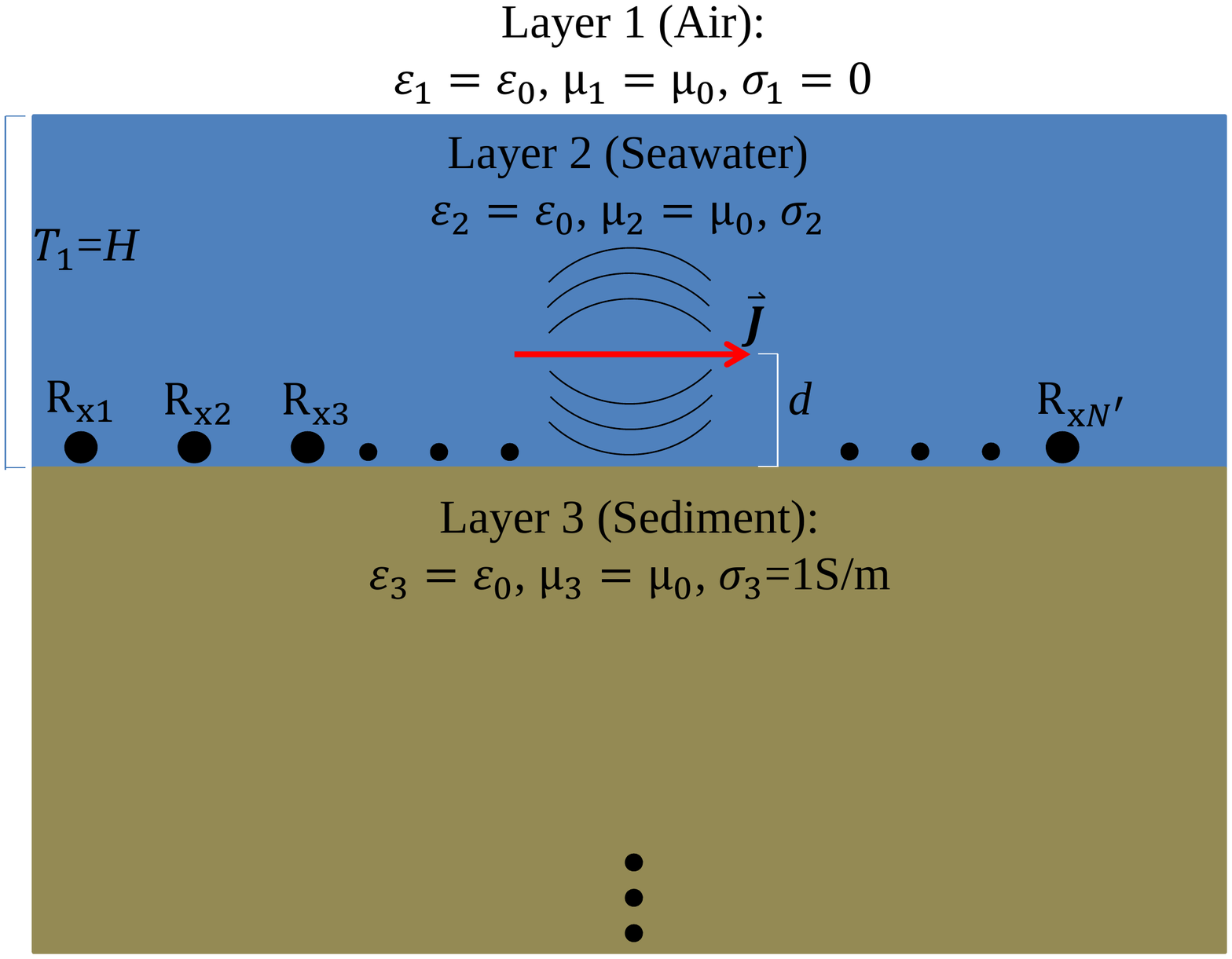}}
\caption{\small \label{CSEMApp2}(Color online) The two contrasting environment geometries with (Figure \ref{CSEMa}) and without (Figure \ref{CSEMb}) the embedded hydrocarbon reservoir. The observation points, mimicking the receiver instruments, lie at the seafloor in the $xz$ plane.}
\end{figure}

\begin{figure}[H]
\centering
\subfloat[\label{MagCase1}]{\includegraphics[width=2in]{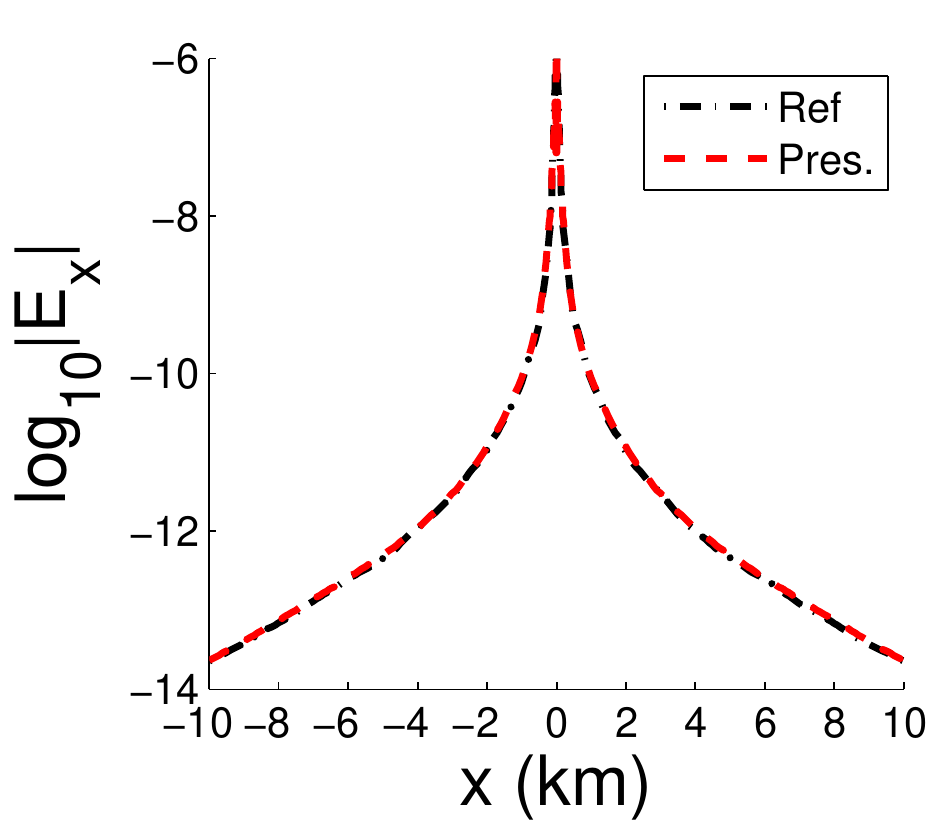}}
\subfloat[\label{PhaseCase1}]{\includegraphics[width=2in]{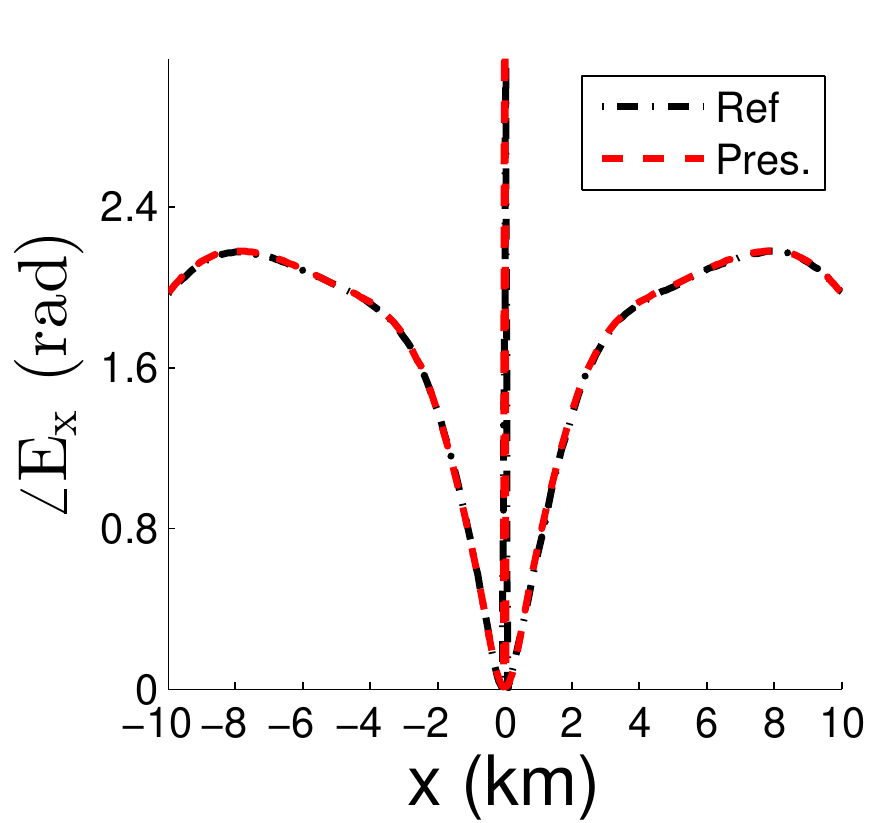}}

\subfloat[\label{MagCase2}]{\includegraphics[width=2in]{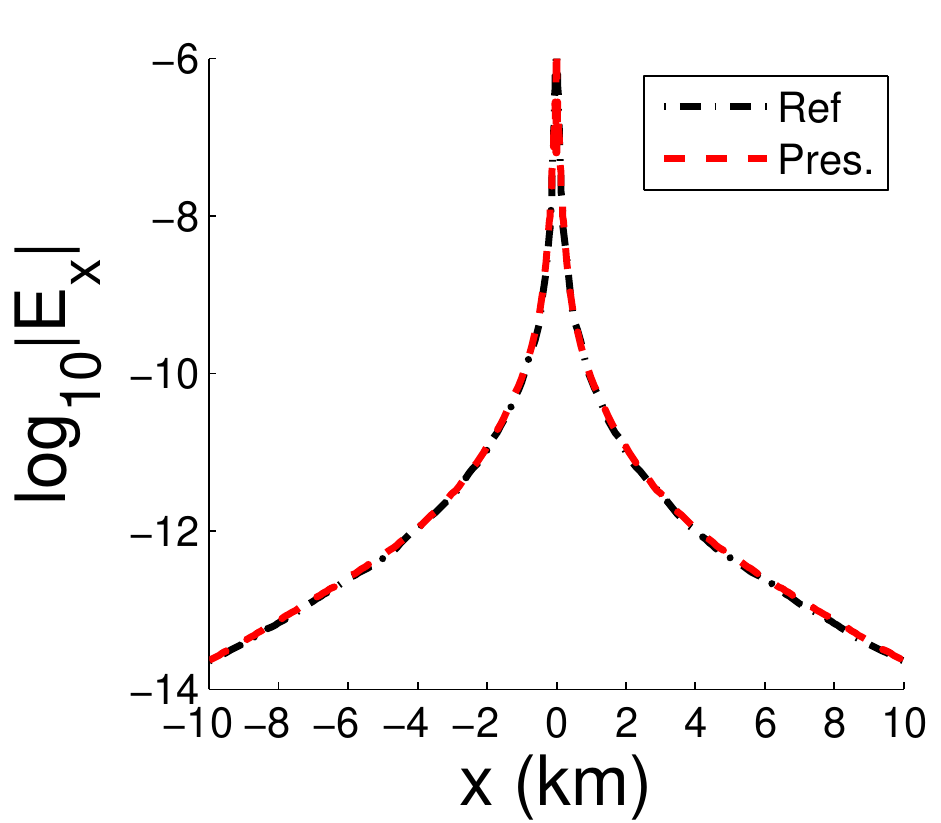}}
\subfloat[\label{PhaseCase2}]{\includegraphics[width=2in]{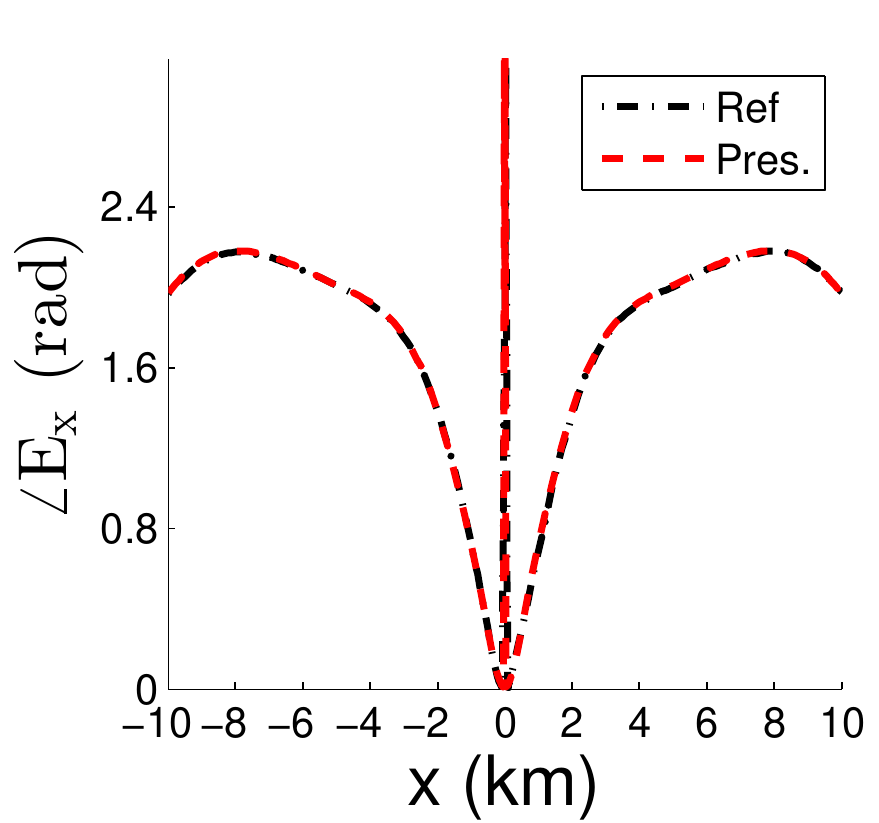}}

\subfloat[\label{MagCase3}]{\includegraphics[width=2in]{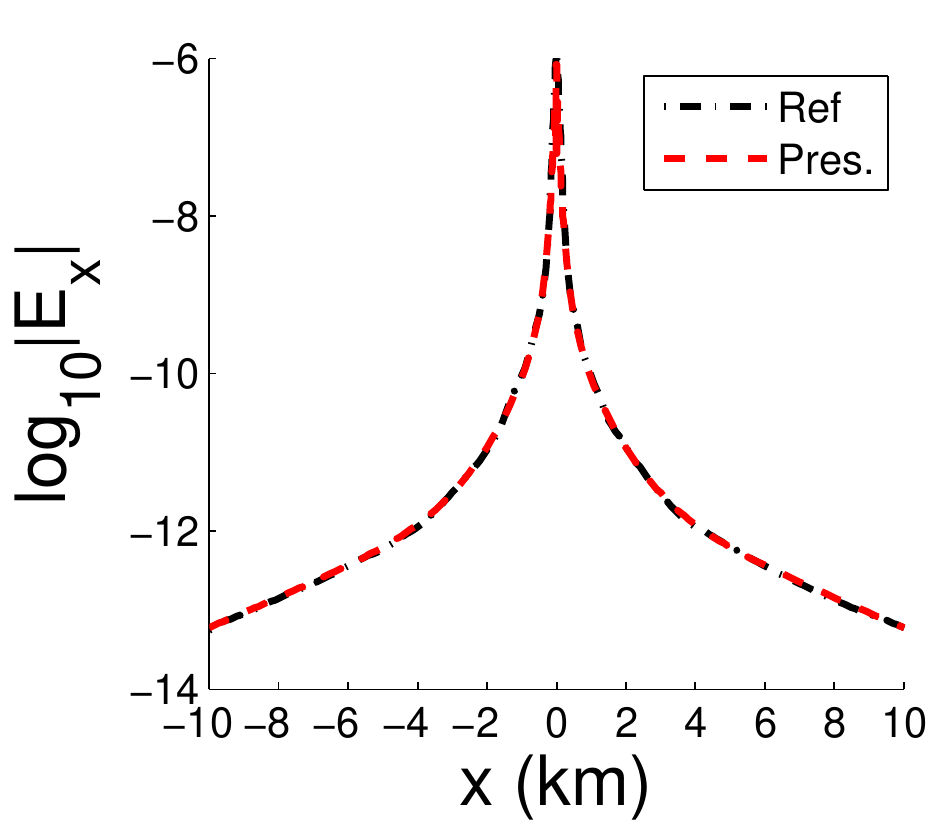}}
\subfloat[\label{PhaseCase3}]{\includegraphics[width=2in]{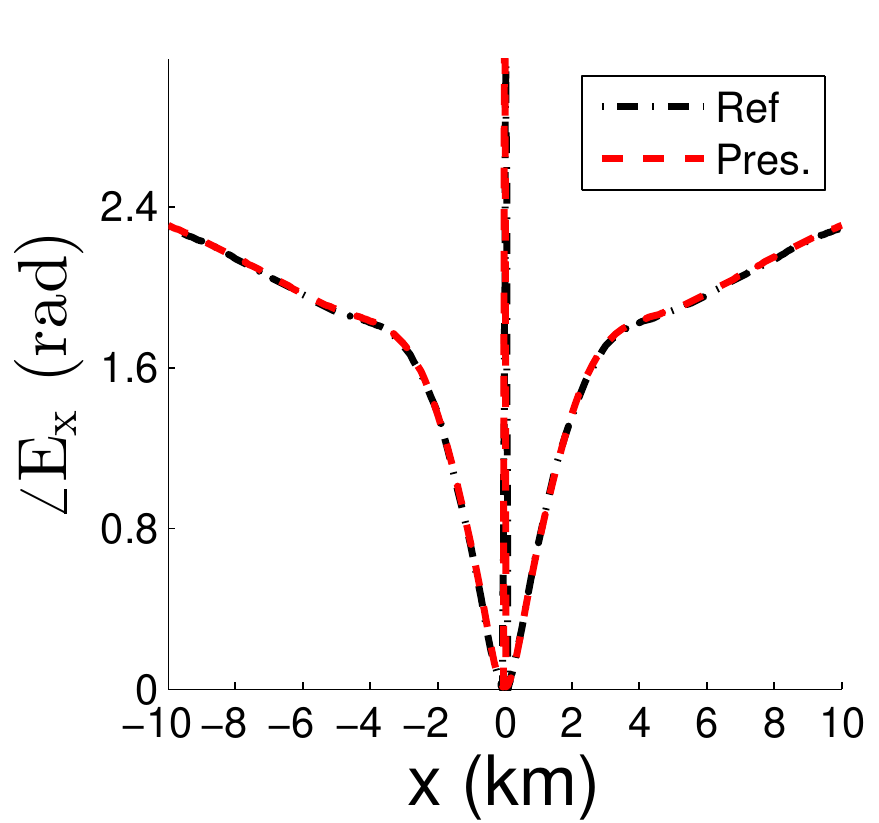}}
\caption{\small \label{Figs6c6d}(Color online) Anisotropic resistive reservoir, with $\sigma_h=10$mS/m and $\sigma_v$=2.5mS/m. Figures \ref{MagCase1}, \ref{MagCase2}, and \ref{MagCase3} show the magnitude of the observed electric field versus $x-x'$ for the isotropic case, intermediate dipping anisotropy case $\{\alpha_4=30^{\circ},\beta_4=0^{\circ}\}$, and fully dipping anisotropy case $\{\alpha_4=90^{\circ},\beta_4=15^{\circ}\}$, respectively. Figures \ref{PhaseCase1}, \ref{PhaseCase2}, and \ref{PhaseCase3} indicate the phase of $E_x$ in these three anisotropy cases, respectively. The curve ``Pres." is our algorithm's result while the curve ``Ref." is the reference result from \cite{loseth}.}
\end{figure}

\begin{figure}[H]
\centering
\subfloat[\label{MagCase4}]{\includegraphics[width=2in]{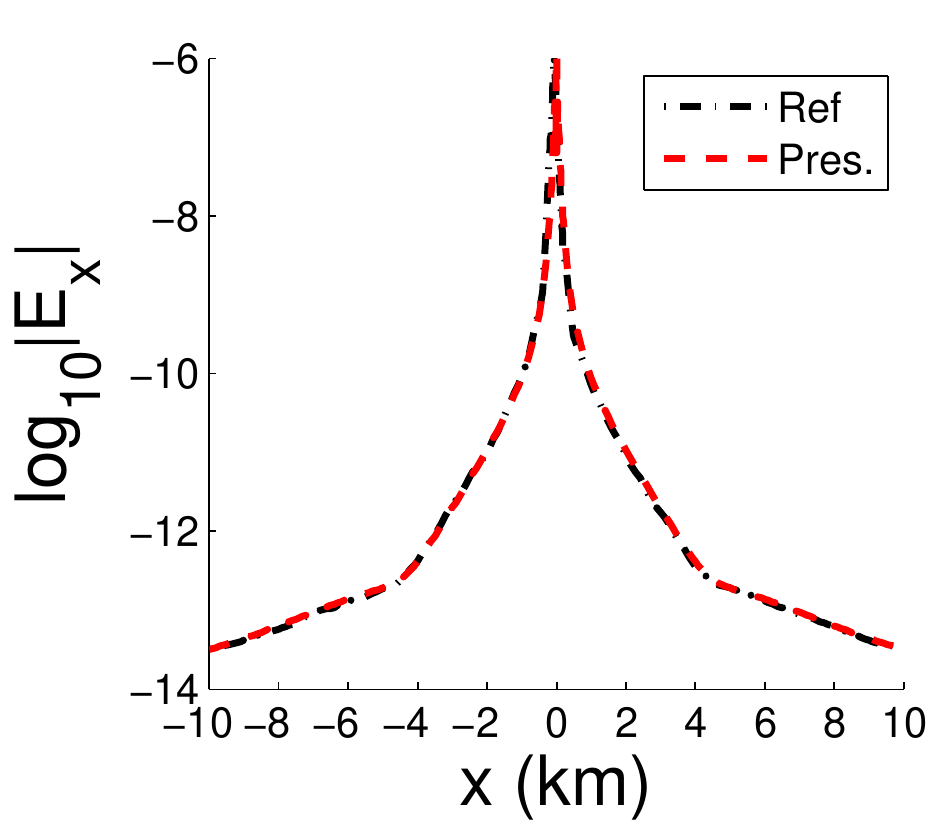}}
\subfloat[\label{PhaseCase4}]{\includegraphics[width=2in]{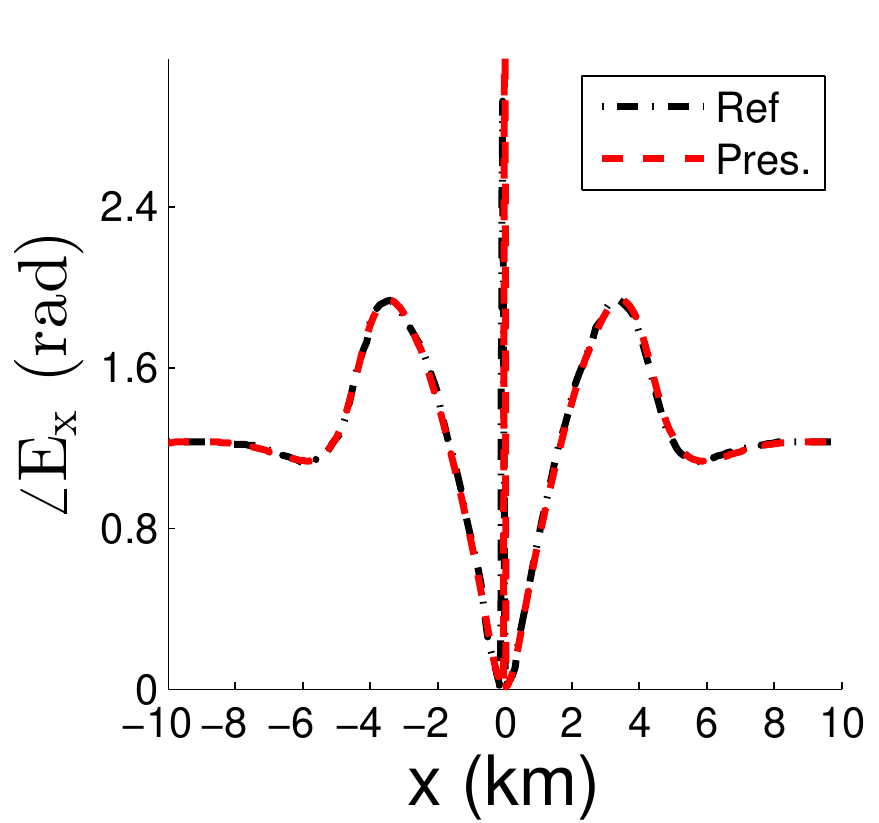}}

\subfloat[\label{MagCase5}]{\includegraphics[width=2in]{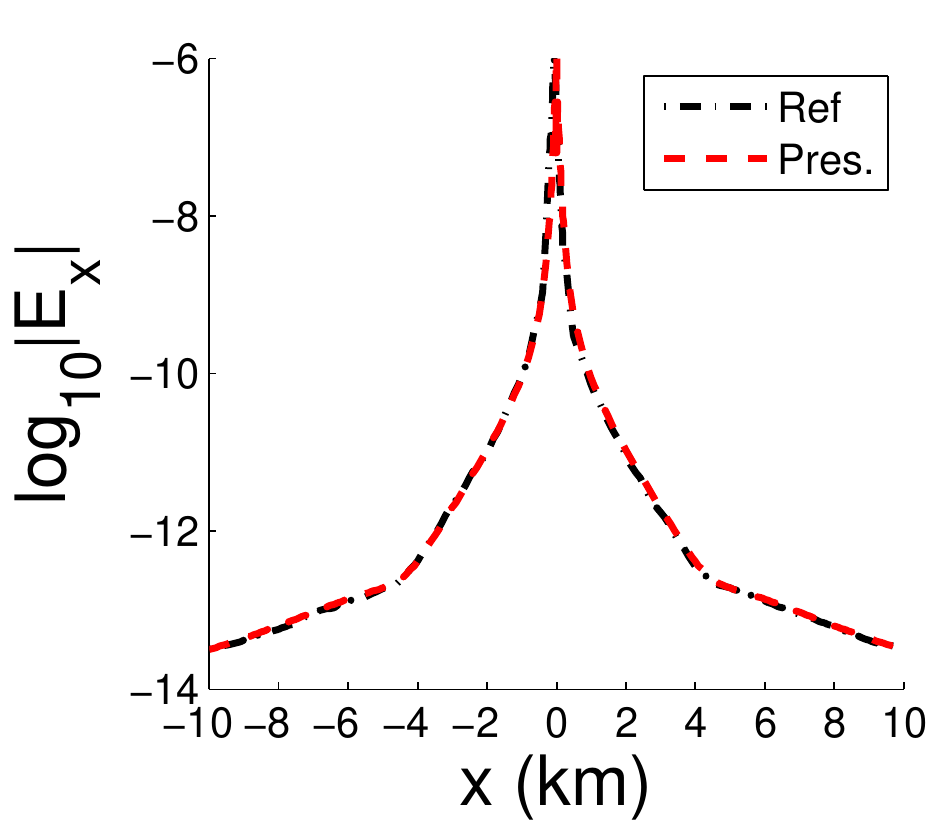}}
\subfloat[\label{PhaseCase5}]{\includegraphics[width=2in]{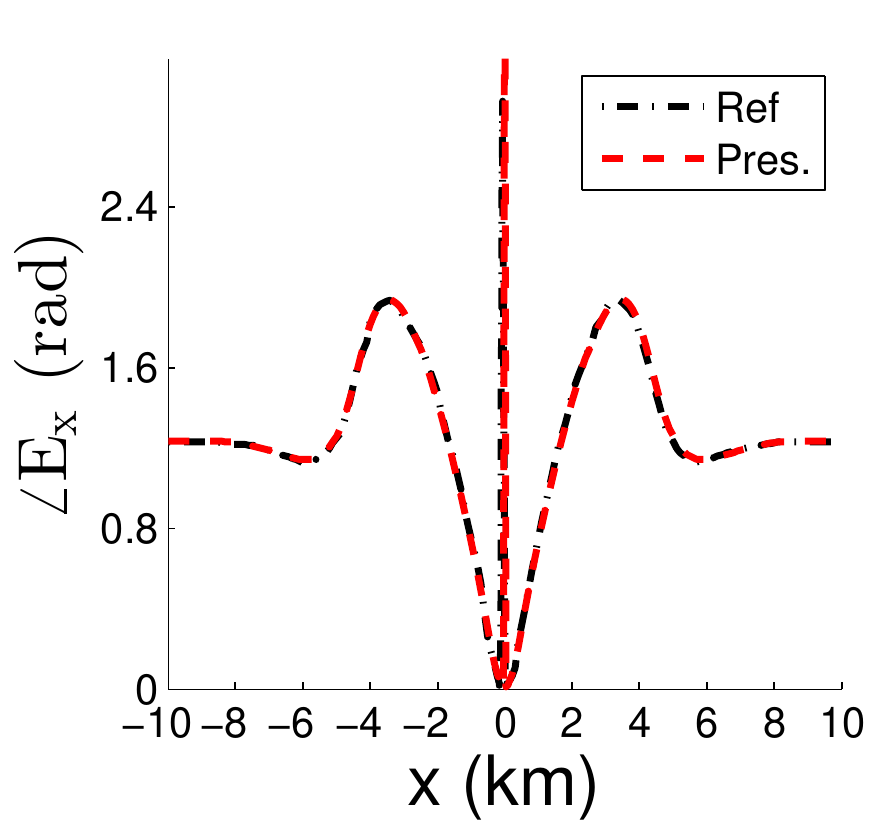}}

\subfloat[\label{MagCase6}]{\includegraphics[width=2in]{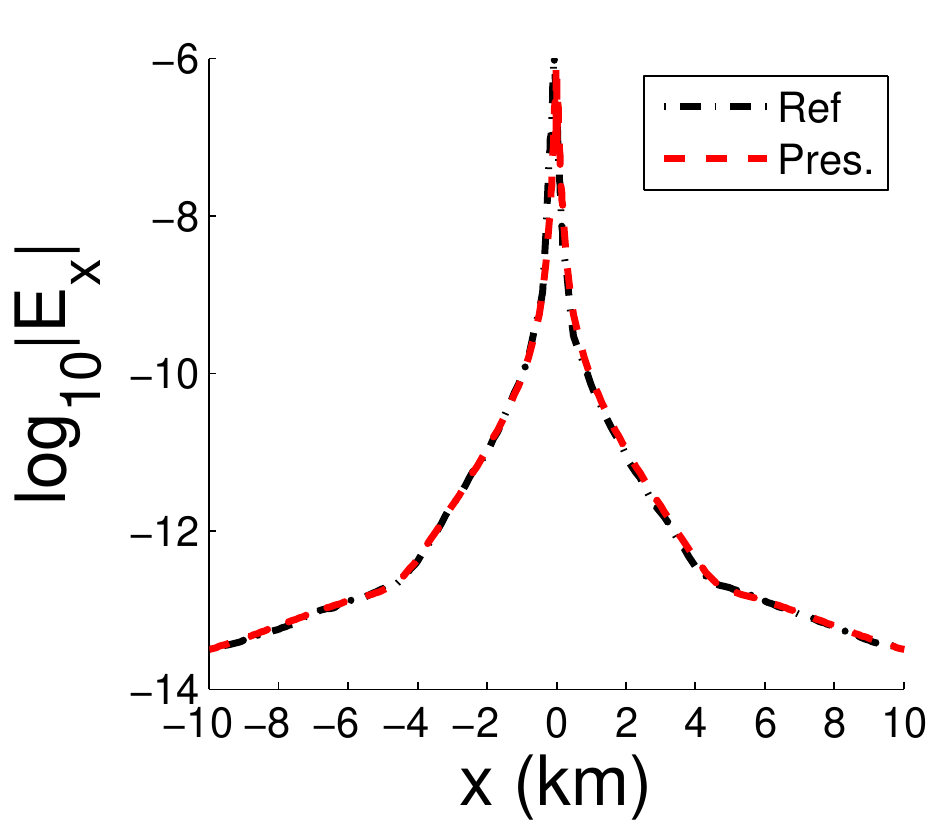}}
\subfloat[\label{PhaseCase6}]{\includegraphics[width=2in]{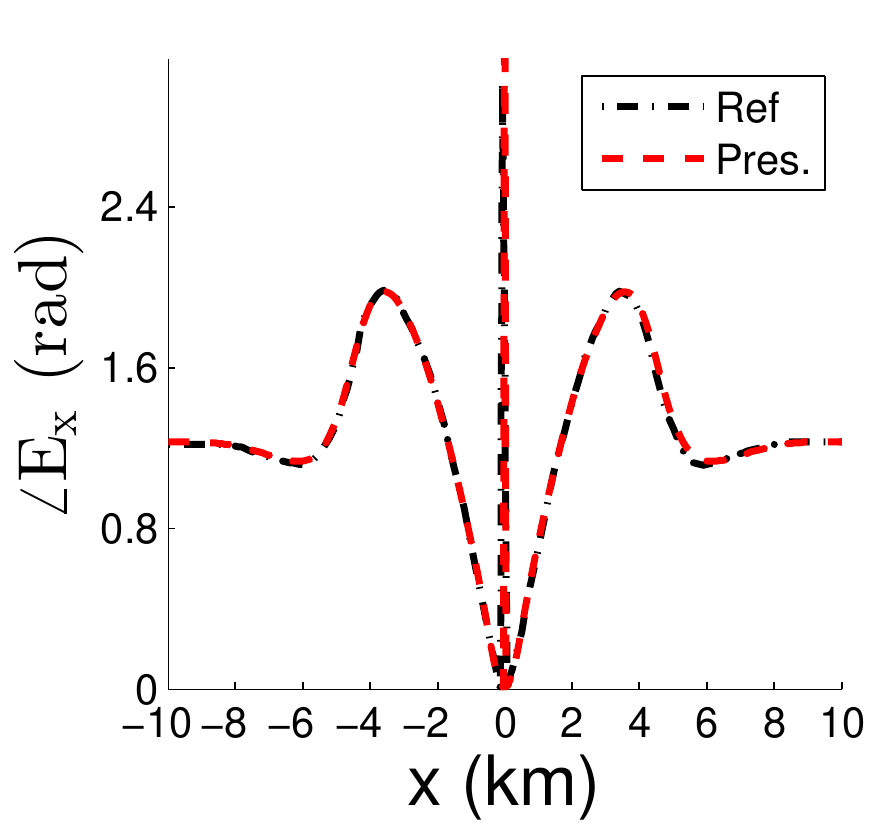}}
\caption{\small \label{Figs7c7d}(Color online) Anisotropic conductive reservoir, with $\sigma_h=500$mS/m and $\sigma_v$=125mS/m. Figures \ref{MagCase4}, \ref{MagCase5}, and \ref{MagCase6} show the magnitude of the observed electric field versus $x-x'$ for the isotropic case, intermediate dipping anisotropy case $\{\alpha_4=30^{\circ},\beta_4=0^{\circ}\}$, and fully dipping anisotropy case $\{\alpha_4=90^{\circ},\beta_4=15^{\circ}\}$, respectively. Figures \ref{PhaseCase4}, \ref{PhaseCase5}, and \ref{PhaseCase6} indicate the phase of $E_x$ in these three anisotropy cases, respectively. The curve ``Pres." is our algorithm's result while the curve ``Ref." is the reference result from \cite{loseth}.}
\end{figure}

\begin{figure}[H]
\centering
\subfloat[\label{E1Arg}]{\includegraphics[width=3.5in]{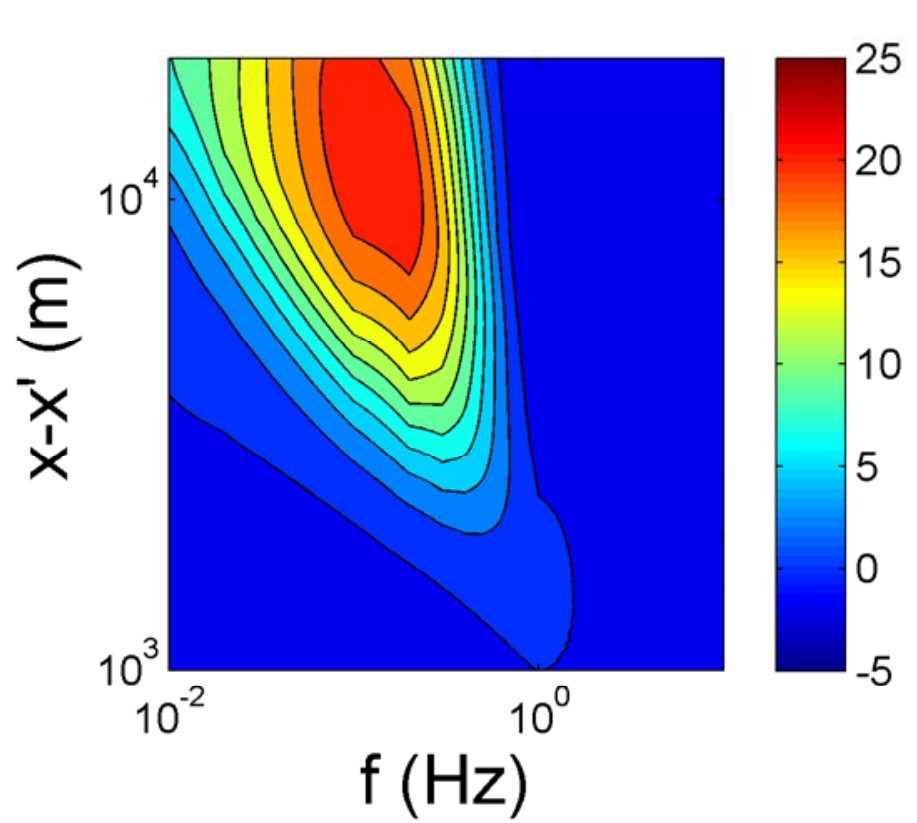}}
\subfloat[\label{E1Mag}]{\includegraphics[width=3.5in]{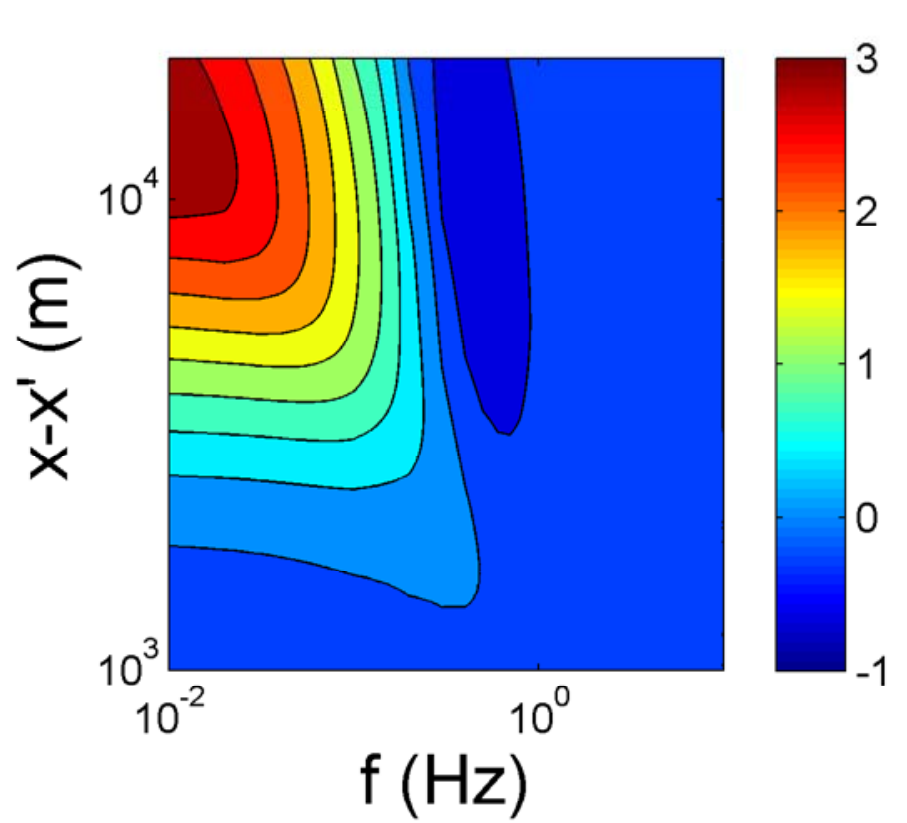}}

\subfloat[\label{E2Arg}]{\includegraphics[width=3.5in]{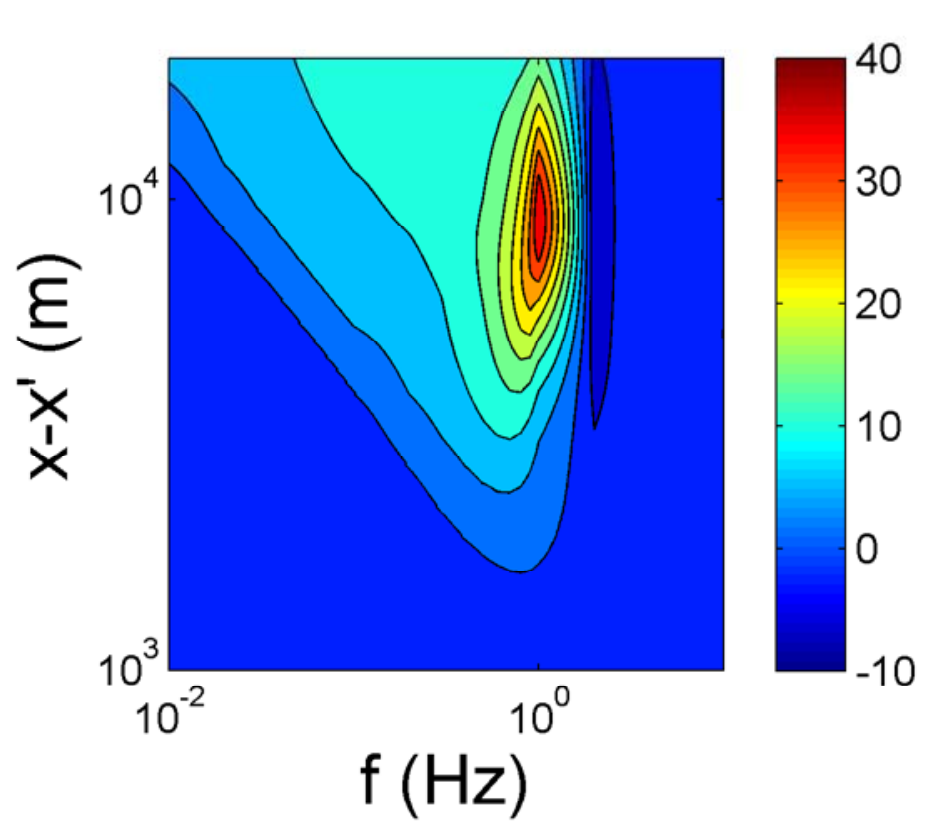}}
\subfloat[\label{E2Mag}]{\includegraphics[width=3.5in]{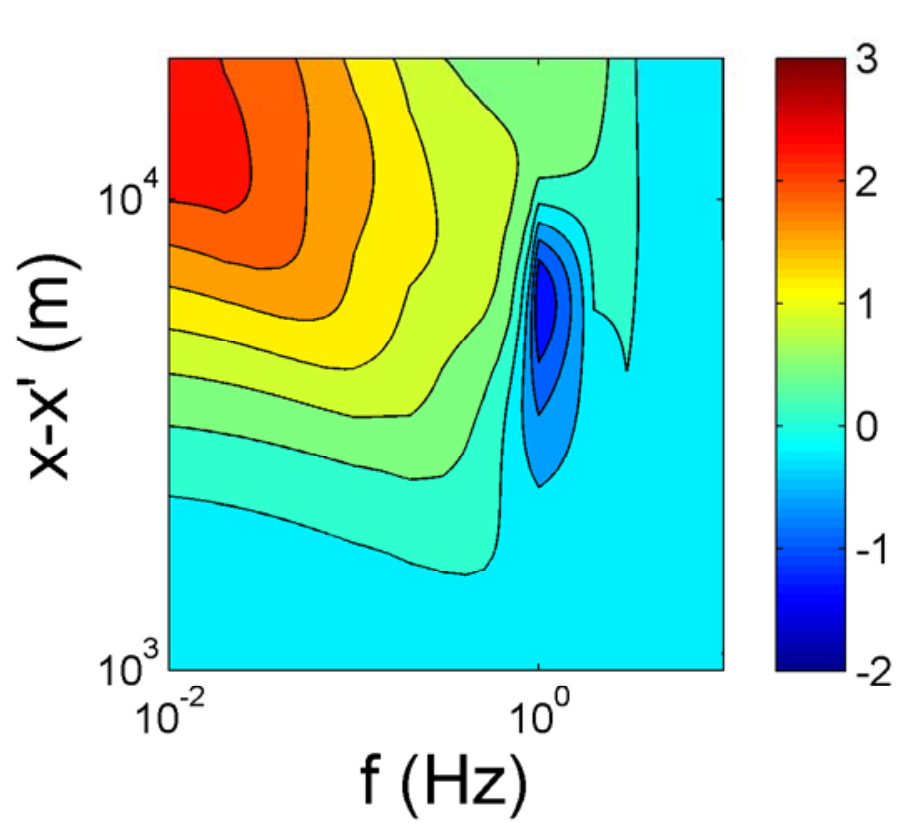}}
\caption{\small \label{CSEM1}(Color online) Figures \ref{E1Arg} and \ref{E2Arg} denote the phase (degrees) of $E_{xr}^s$  when the transmitter operates (resp.) in either shallow water ($H$=100m) or deep water ($H$=500m), while Figures \ref{E1Mag} and \ref{E2Mag} denote the magnitude [dB] of $E_{xr}^s$ when the transmitter operates (resp.) in either shallow water ($H$=100m) or deep water ($H$=500m).}
\end{figure}

\begin{figure}[H]
\centering
\subfloat[\label{H1Arg}]{\includegraphics[width=3.5in]{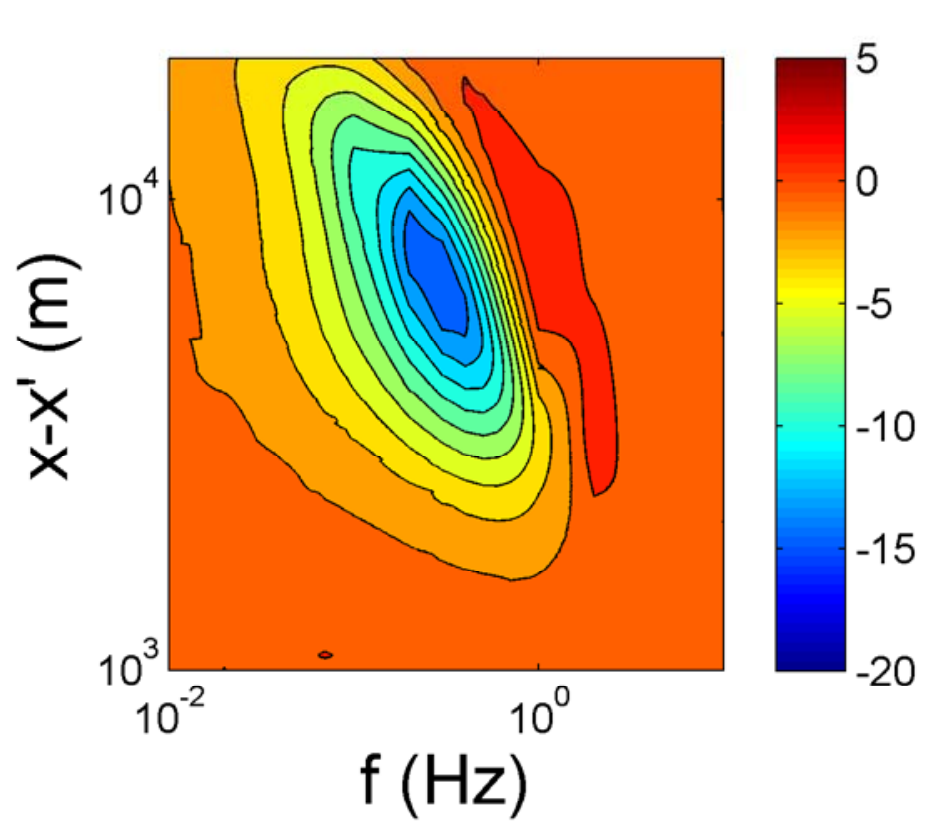}}
\subfloat[\label{H1Mag}]{\includegraphics[width=3.5in]{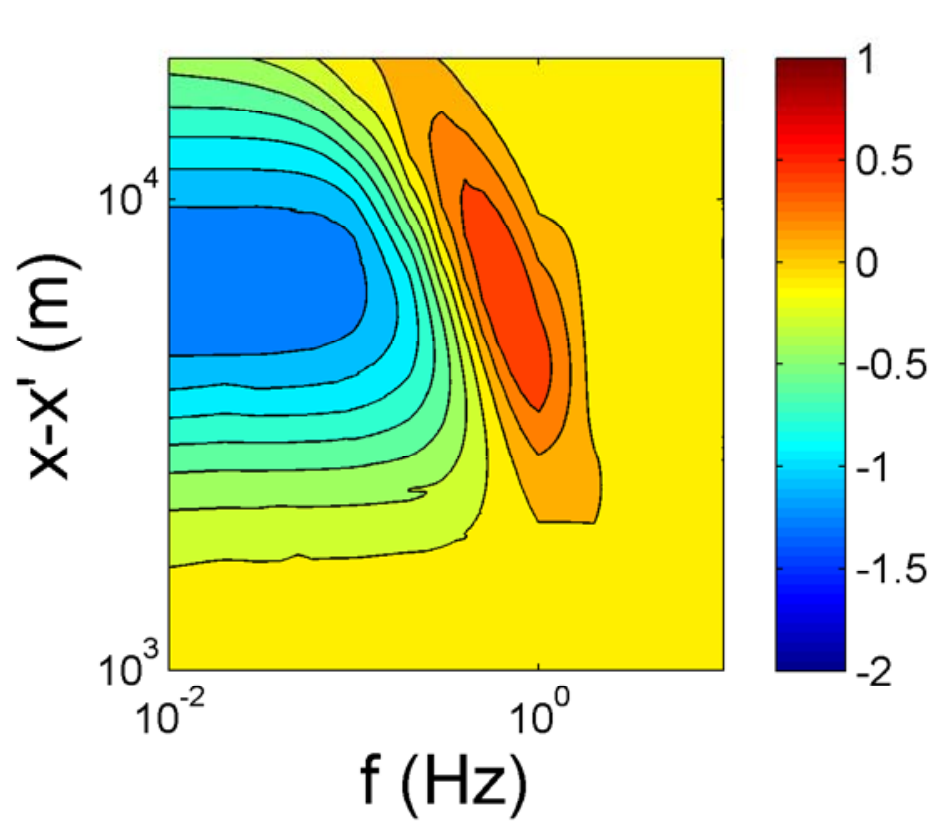}}

\subfloat[\label{H2Arg}]{\includegraphics[width=3.5in]{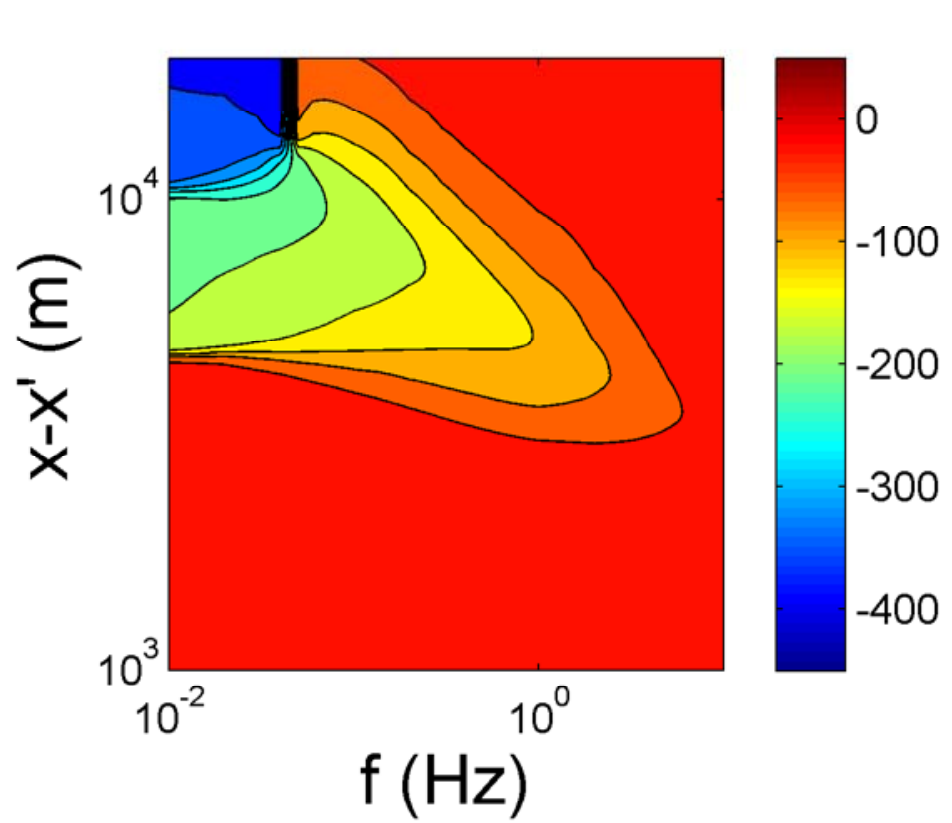}}
\subfloat[\label{H2Mag}]{\includegraphics[width=3.5in]{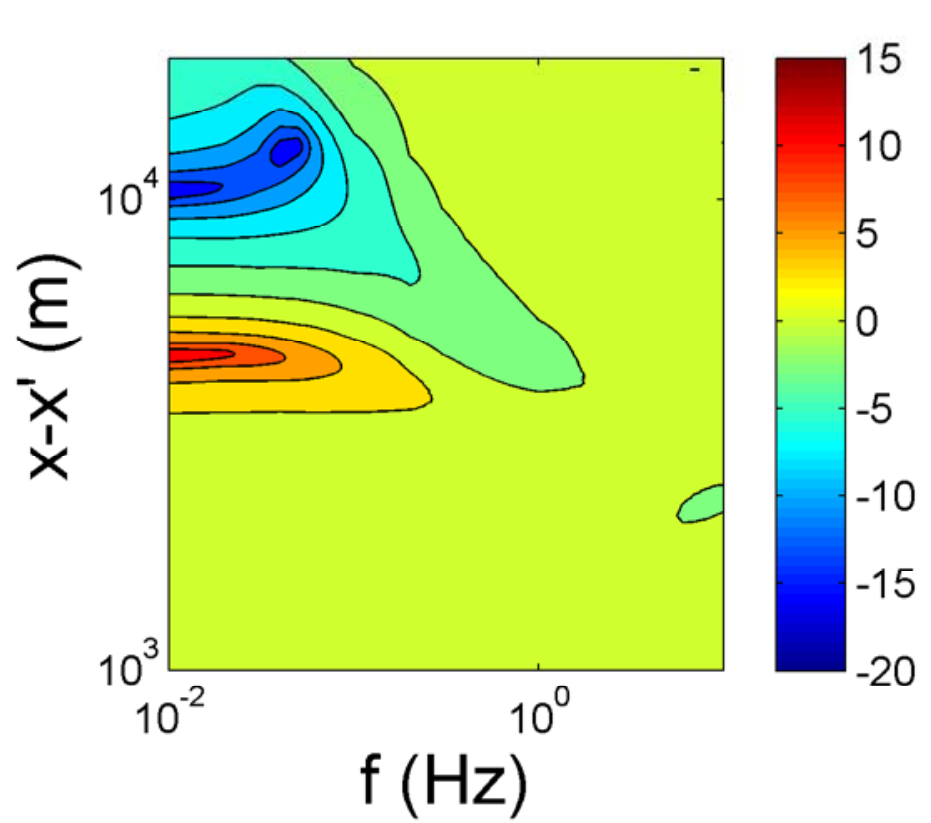}}
\caption{\small \label{CSEM2}(Color online) Figures \ref{H1Arg} and \ref{H2Arg} denote the phase (degrees) of $H_{xr}^s$  when the transmitter operates (resp.) in either shallow water ($H$=100m) or deep water ($H$=500m), while Figures \ref{H1Mag} and \ref{H2Mag} denote the magnitude [dB] of $H_{xr}^s$ when the transmitter operates (resp.) in either shallow water ($H$=100m) or deep water ($H$=500m).}
\end{figure}

\begin{figure}[H]
\centering
\subfloat[\label{E1ArgZ}]{\includegraphics[width=3.5in]{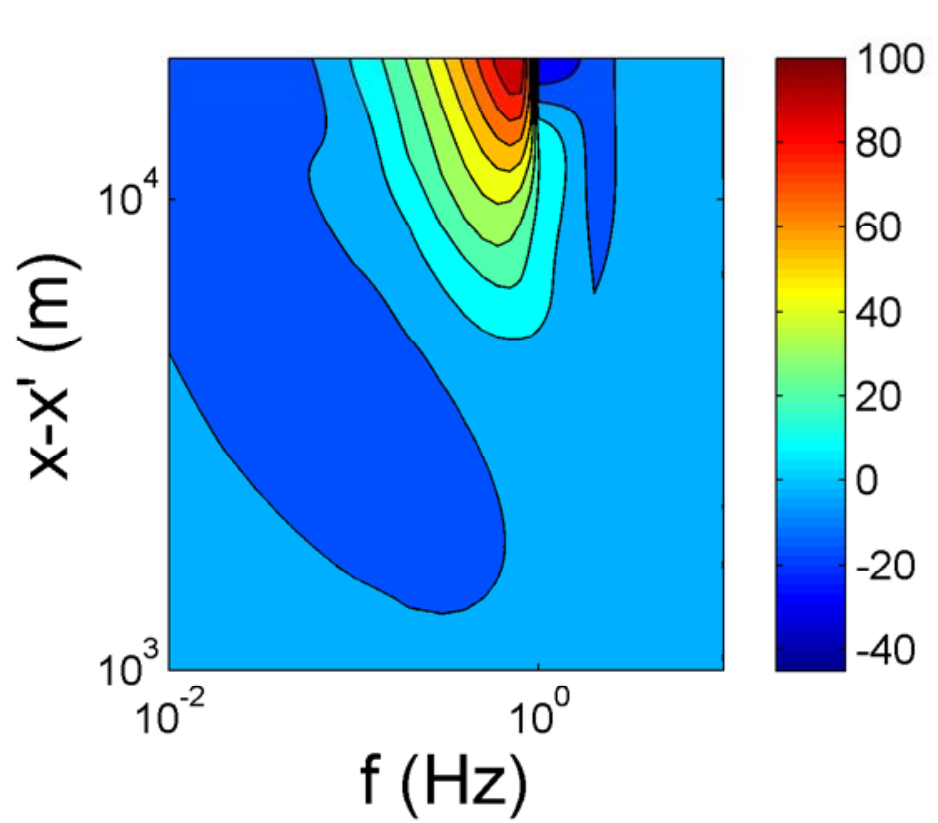}}
\subfloat[\label{E1MagZ}]{\includegraphics[width=3.5in]{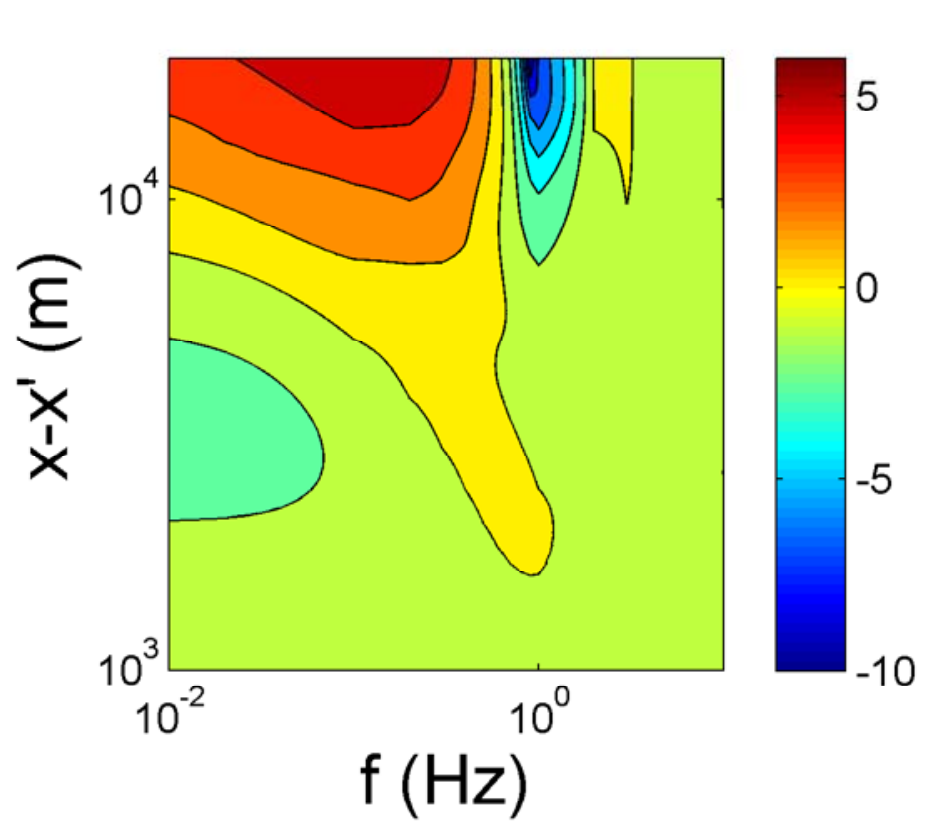}}

\subfloat[\label{E2ArgZ}]{\includegraphics[width=3.5in]{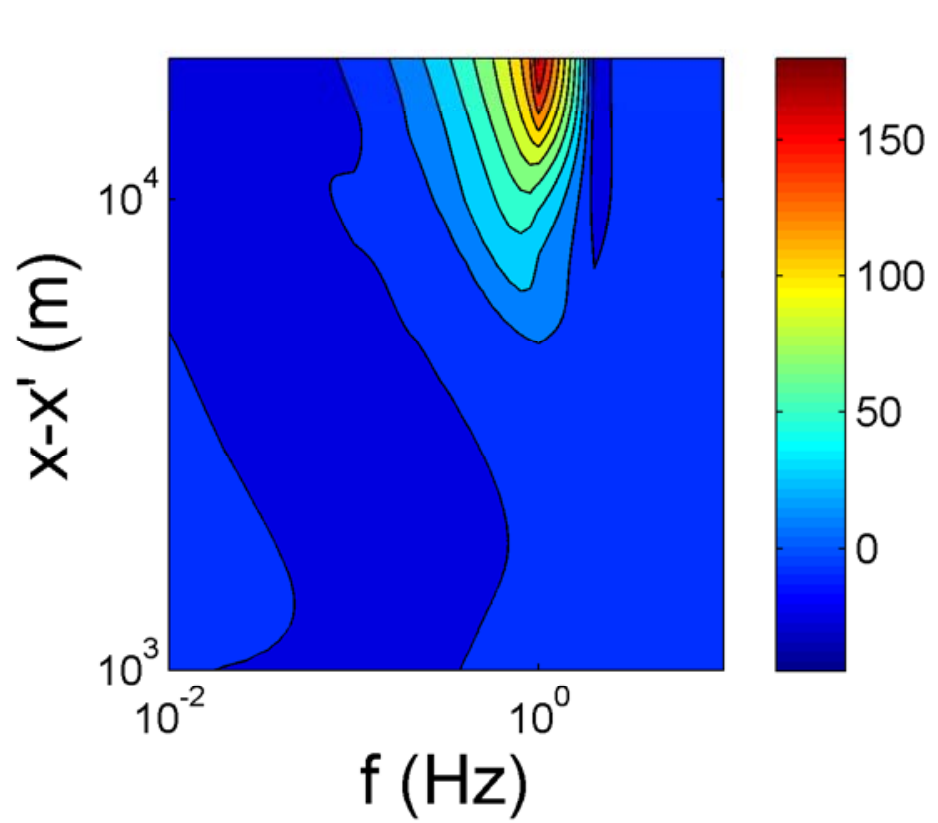}}
\subfloat[\label{E2MagZ}]{\includegraphics[width=3.5in]{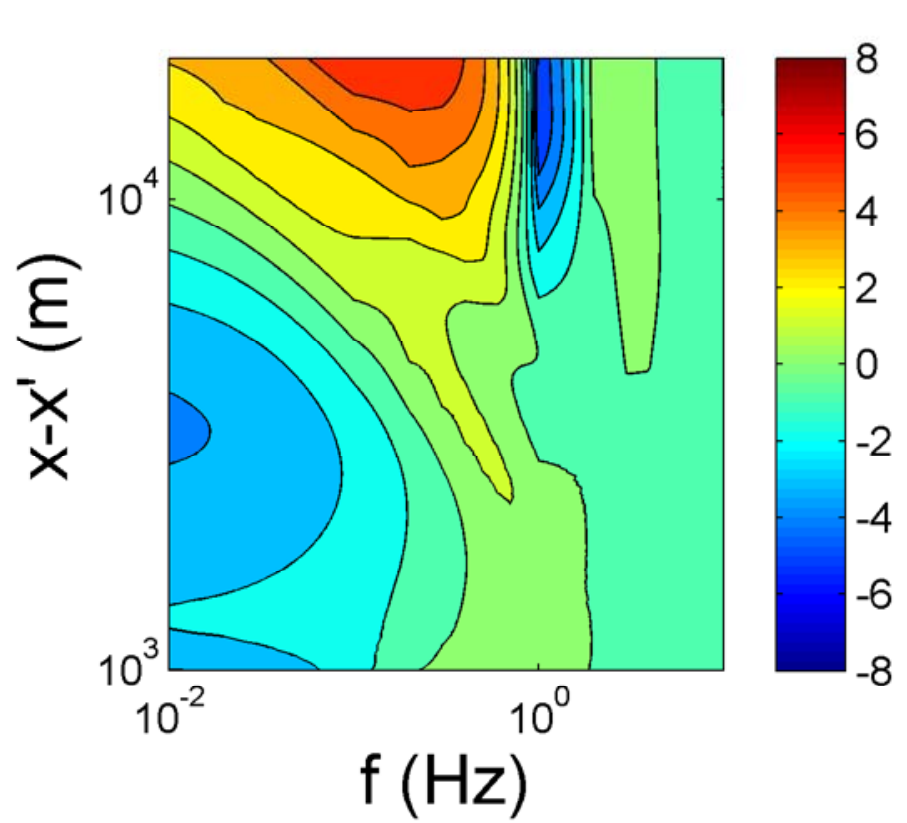}}
\caption{\small \label{CSEM3}(Color online) Figures \ref{E1ArgZ} and \ref{E2ArgZ} denote the phase (degrees) of $E_{zr}^s$  when the transmitter operates (resp.) in either shallow water ($H$=100m) or deep water ($H$=500m), while Figures \ref{E1MagZ} and \ref{E2MagZ} denote the magnitude [dB] of $E_{zr}^s$ when the transmitter operates (resp.) in either shallow water ($H$=100m) or deep water ($H$=500m).}
\end{figure}

Next, we validate and then illustrate one application of the proposed algorithm: Facilitating computation of the fields excited by transmitters operating in highly inhomogeneous and absorptive marine environments characterizing typical controlled source EM (CSEM) operational scenarios. CSEM transmitters, typically radiating in the frequency range 0.01Hz-10Hz \cite{macgregor1}, serve as active illuminators to facilitate detection and characterization of thin, highly resistive hydrocarbon-bearing formations embedded deep under the ocean, which can complement data from magnetotelluric (MT) sounding-based methods \cite{constable2}. Indeed, use of an active source allows one to also exploit galvanic, in addition to inductive, generation of the scattered field that arises from ``blockage", due to a highly resistive layer (e.g., of hydrocarbons), of what was (with the resistive layer absent) a dominantly normally-directed vector current field \cite{um1}. Figure \ref{CSEMApp2} describes the geometry of the problem considered;  note that a hydrocarbon-bearing formation buried at 1km under the sea and having 100m thickness is a typical case study found in the related literature~\cite{constable2,loseth}.
For the first study, we compare our code's results against those found in Figures 6c, 6d, 7c, and 7d of \cite{loseth} involving a Hertzian electric dipole. To this end, we set the $x$-directed source's position to be $d$=30m above the sea floor, the sea depth $H$=300m, the sea water's conductivity $\sigma_2=3.2$S/m, and the antenna's radiation frequency as $f=$0.25Hz. The sub-plots in Figure \ref{Figs6c6d} refer to when the uniaxial-anisotropic hydrocarbon reservoir has in-plane conductivity $\sigma_h=10$mS/m and orthogonal conductivity $\sigma_v=2.5$mS/m,\footnote{\label{dip}That is, $\sigma_h$ and $\sigma_v$ are the principal conductivity components corresponding to applying an electric field either parallel or perpendicular to the reservoir's principal bedding plane, respectively \cite{anderson1}. The representation of the reservoir's (the fourth layer in Figure \ref{CSEMa}) conductivity tensor $\boldsymbol{\bar{\sigma}}_4$ can then be found using the reservoir's conductivity bedding plane polar and azimuthal orientation angles ($\alpha_4$ and $\beta_4$, resp. \cite{anderson1}).} while the sub-plots in Figure \ref{Figs7c7d} refer to when $\sigma_h=500$mS/m and $\sigma_v=125$mS/m. We examine the ``in-line", $x$-directed total electric field $E_x$ observed at the receivers positioned at the sea floor for different source-receiver separations $x-x'$. As can be seen in both Figures \ref{Figs6c6d} and \ref{Figs7c7d}, there is very good agreement observed for all three anisotropy cases exhibited: The isotropic reservoir case (Figures \ref{MagCase1}, \ref{PhaseCase1}, \ref{MagCase4}, and \ref{PhaseCase4}), the intermediate dipping (``cross-bedding" \cite{anderson1}) anisotropy case $\{\alpha_4=30^{\circ},\beta_4=0^{\circ}\}$ (c.f. footnote \ref{dip}) seen in Figures \ref{MagCase2}, \ref{PhaseCase2}, \ref{MagCase5}, and \ref{PhaseCase5}, and finally the fully dipping anisotropy case $\{\alpha_4=90^{\circ},\beta_4=15^{\circ}\}$ seen in Figures \ref{MagCase3}, \ref{PhaseCase3}, \ref{MagCase6}, and \ref{PhaseCase6}.\footnote{For brevity, we omit results from the transverse-isotropic case $\alpha_4=0^{\circ}$ since the closed-form validation results, in Figures \ref{field} and \ref{field2}, adequately demonstrate the algorithm's performance when media with this orientation of principal material axes are present.}
For the second study, we position a 100m long, $x$-directed wire antenna $d$=50m above the ocean-sediment interface while maintaining again the depth of the observation points (``receivers") at the seafloor.\footnote{The length, orientation, and depth above the seafloor of the transmitter antenna, as well as the receiver positions, lead to a case study qualitatively following, and is primarily inspired from, the CSEM field campaign reported in \cite{constable3}.} Furthermore, both the transmitter antenna and observation points are confined to the $xz$ plane while $\sigma_2\sim$3.33S/m and the isotropic reservoir has conductivity $\sigma_4=$10mS/m. Since the field strength can vary significantly over the transmitter-receiver separation distances (taken with respect to the wire's center) $x-x'$ considered herein (1-20km) \cite{constable2}, we plot the magnitude and phase of the \emph{ratio} of the scattered fields received in the two geometries considered in Figures \ref{CSEMa} and \ref{CSEMb}: For example, in the case of $E_x^s$, we observe the phase and magnitude of the received scattered field ratio $E_{xr}^s=E_{x1}^s/E_{x2}^s$, where $E_{x1}^s$ and $E_{x2}^s$ are the scattered fields observed at a particular receiver in the geometries described by Figures \ref{CSEMa} and \ref{CSEMb} (resp.). As a result, a measurement's responsiveness to the hydrocarbon formation's presence is indicated by the extent of phase deviation from $0^{\circ}$ (in the phase plots) and the extent of magnitude swing from 0dB (in the magnitude plots), as observed in Figures \ref{CSEM1}-\ref{CSEM3}.

Figures \ref{CSEM1}, \ref{CSEM2}, and \ref{CSEM3} illustrate the phase and magnitude of the scattered field ratios concerning $E_x^s$, $H_x^s$, and $E_z^s$ (resp.), both for the shallow water ($H$=100m) and deep water ($H$=500m) cases.\footnote{In the top-left corner of Figure \ref{H2Arg}, note the vortex-like behavior of the phase. The seemingly solid vertical black strip corresponds to closely spaced (black) contour lines that, upon zooming in at high resolution, do in fact illustrate the locally rapid variation of phase.} From these Figures, we notice that for both sea water depth scenarios the electric field ratios $E_x^s$ and $E_z^s$ (but particularly $E_z^s$, which corresponds to a pure Transverse-Magnetic to $z$ mode [$\mathrm{TM}_z$] \cite{felsen,chew}) exhibit strong responsiveness to the presence of the deeply buried hydrocarbon bed. On the other hand, in shallow water $H_x^s$ provides little useful information, as can be seen by its relatively poor response to the presence of the resistive hydrocarbon formation compared to the electric field measurements. However, upon increasing the water depth to 500m, both the phase and magnitude of $H_{xr}^s$ show a very high response to the presence of the hydrocarbon bed. By contrast, the phase and magnitude of $H_{zr}^s$ (not shown here), corresponding to a pure Transverse-Electric to $z$ mode ($\mathrm{TE}_z$) \cite{felsen,chew}, fails to yield significant responsiveness to the resistive formation even when the water depth is increased to 500m. These results qualitatively corroborate prior studies indicating that the sea-air interface can significantly dampen instrument sensitivity to deeply buried hydrocarbon reservoirs~\cite{key2,macgregor2}. However, the sensitivity reduction effect is strongly dependent on the field type (electric versus magnetic) and component ($x,y,z$), with the dampening effect much more pronounced in measurements derived from the $\mathrm{TE}_z$ modes as compared to the $\mathrm{TM}_z$ modes ~\cite{key2,macgregor2}.

\section{Concluding Remarks}
 We have introduced and validated numerical algorithms performing two functions widely applicable to myriad applications ranging from radar, antenna, and microwave circuit modeling to aperture synthesis and geophysical prospecting. First we discussed how to extract the scattered field radiated by sources, both Hertzian and distributed, embedded within planar-stratified environments with generally anisotropic and lossy media based on the spectral-domain/Fourier modal synthesis technique. Some of the key features of the extraction algorithm are: (1) Numerical robustness with respect to large material, source, and observer parameter variations, (2) non-reliance on space domain tensor Green's functions, and (3) no added computational burden versus computing the total field~\cite{sainath}. Second, we discussed how to circumvent the tedious and oftentimes computationally expensive task of evaluating the radiation integral in translation-variant environments via calculating the radiation integral directly in the Fourier domain itself. Beyond exhibiting applicability to general sources admitting a closed form Fourier domain representation, this algorithm also enjoys the benefit (as compared to Hertzian dipole-based sampling) of further acceleration due to efficiently sampling and simulating the dominant spatial amplitude distribution of many commonly encountered source current profiles. To validate the algorithms, numerical results were compared against closed-form field solutions in both free space and layered, anisotropic medium radiation scenarios. To illustrate the applicability of the algorithm in geophysical prospecting, we applied it to the modeling of CSEM-based sensing of sub-oceanic hydrocarbon deposits using active wire antenna transmitters.
\section{Acknowledgments}
This work was primarily supported by a NASA Space Technology Research Fellowship (NSTRF). We acknowledge the reviewers for their helpful and insightful commentary concerning this manuscript, as well as Dr. Anthony Freeman of the NASA/JPL. We also acknowledge partial support from the Ohio Supercomputer Center under Grants PAS-0061 and PAS-0110.
\section{References}
\bibliography{reflist}
\bibliographystyle{apsrev4-1}
\end{document}